\documentclass[11pt,a4paper]{article}
\usepackage{jheppub}
\pdfoutput=1

\usepackage{anyfontsize}

\usepackage{hyperref}
\usepackage[mode=buildnew]{standalone}
\usepackage{comment}
\usepackage{caption}
\usepackage{subcaption}
\usepackage{amsmath,amstext,amsgen,amsbsy,amsopn,amsfonts,amsthm}
\usepackage{tikz}
\usetikzlibrary{shapes,snakes,arrows}
\usetikzlibrary{shapes.misc}
\usetikzlibrary{decorations.pathreplacing}
\usepackage{array}
\usetikzlibrary{patterns}
\usepackage{booktabs}
\usepackage{bm}
\usepackage{graphicx}

\tikzset{cross/.style={cross out, draw=black}}

\newtheorem*{constraint}{Constraint}

\usetikzlibrary{arrows,shapes,snakes,automata,backgrounds,petri}
\usepackage[latin1]{inputenc}

\newcommand{\M}{\mathcal{M}}
\newcommand{\Or}{\bar{\mathcal{O}}}

\preprint{Imperial/TP/16/AH/05}

\title{Branes and the Kraft-Procesi Transition}
\author[a]{Santiago Cabrera}
\author[a]{and Amihay Hanany}
\affiliation[a]{Theoretical Physics, The Blackett Laboratory\\
Imperial College London\\ SW7 2AZ United Kingdom}
\emailAdd{santiago.cabrera13@imperial.ac.uk}
\emailAdd{a.hanany@imperial.ac.uk}

 \abstract{The Coulomb and Higgs branches of certain $3d\ \mathcal{N}=4$ gauge theories can be understood as closures of nilpotent orbits. Recently, a new theorem by Namikawa suggests that this is the simplest possible case, thus giving this class a special role. In this note we use branes to reproduce the mathematical work by Kraft and Procesi. It studies the classification of all nilpotent orbits for classical groups and it characterizes an inclusion relation via minimal singularities. We show how these minimal singularities arise naturally in the Type IIB superstring embedding of the $3d$ A-type theories. The Higgs mechanism can be used to   \emph{remove} the minimal singularity, corresponding to a transition in the brane configuration that induces a new effective $3d$ theory. This reproduces the Kraft-Procesi results, endowing the family of gauge theories with a new underlying structure. We provide an efficient procedure for computing such brane transitions.}
\begin{document}

\maketitle
\flushbottom

\section{Introduction}

The study of $3d$ $\mathcal{N}=4$ quantum field theories, their vacua, string embeddings and mirror dualities is today a robust field of theoretical physics. Mirror duality was first discussed in \cite{Intriligator1996}, and its Type IIB superstring realization was proposed very soon after in \cite{Hanany1996}. This analysis has been able to thrive thanks to the use of quivers, the structure of hyperk\"ahler manifolds, 3D mirror symmetry, understanding of 't Hooft monopole operators, algebraic counting methods like the Hilbert series of the chiral ring, and the recent introduction of a way of computing Coulomb branches employing monopole operators \cite{Cremonesi2014a}. \\

An analysis of the theory reveals that for any $3d$ $\mathcal{N}=4$ quiver gauge theory there are two distinct vacuum phases: the \emph{Coulomb branch} and the \emph{Higgs branch}. The Coulomb branch corresponds to the phase where only scalar fields that belong to vector multiplets admit nonzero VEVs. Similarly, in the Higgs branch only scalar fields from the hypers admit nonzero VEVs. Each phase is a different hyperk\"ahler singular variety. Understanding the geometry of these two spaces is crucial to discern various physical properties like the structure of BPS states. The study also provides invaluable help in the endeavor to characterize different families of QFTs.\\

In recent years, a new kind of hyperk\"ahler singular spaces has gained relevance: the closures of nilpotent orbits of Lie algebras. Given any nilpotent element\footnote{An element $X$ of a Lie algebra $\mathfrak{g}$ is considered nilpotent if the operator related to it via any representation map $\rho(X)$ is nilpotent \cite{Collingwood1993}.} in a Lie algebra $\mathfrak{g}$ over the complex numbers $\mathbb{C}$, the orbit of this element under the action of the corresponding group is a geometric space $\mathcal{O}$. The closure of this space $\Or$ is a hyperk\"ahler singularity. Texts in nilpotent orbits are \cite{spaltenstein1982classes,carter1985finite,Collingwood1993,birula2002algebraic}.\\

Nilpotent orbits appear each time there is a problem which involves an embedding of SU(2) into some group\footnote{In fact, we very recently realized that the calculation of the Witten index as Kac and Smilga did for an arbitrary Lie group \cite{Kac1999} is equivalent to the counting of distinguished nilpotent orbits of the group's algebra.} . For example, in the Nahm equations\footnote{Nahm equations first appeared in the study of BPS monopoles \cite{Nahm1979}. For a review on this topic the reader is directed to \cite{Weinberg2006}. } nilpotent orbits arise in a natural way. In Fuzzy spheres there is another natural appearance of nilpotent orbits, etc. Some selected but not complete set of examples can be found in \cite{Bachas2000,Gaiotto2008,Gukov2008,Kim2010,Chacaltana2013,Bourget2015,Heckman}.\\

Furthermore, Namikawa's recent work \cite{Namikawa2016} provides a new source of motivation. Nami-kawa's theorem states that if a Coulomb or Higgs branch is finitely generated by operators with spin $s=1$ under the $SU(2)$ R-symmetry it has to be the closure of a nilpotent orbit of the isometry group's algebra.  Following this theorem, closures of nilpotent orbits represent the simplest non-trivial families of hyperk\"aler singularities that are Higgs and Coulomb branches. Spaces with generators of the chiral ring with spin $s>1$ can be thought as deformations of closures of nilpotent orbits.\\

In this paper we study one aspect of the nilpotent orbits of classical algebras that has been known for some decades among mathematicians \cite{Kraft1982} and is starting to appear in physics \cite{Heckman}. We call it the \emph{Kraft-Procesi transition}. We produce a systematic study of the brane realization of the phenomenon and recover one of its mathematical features: the minimal singularities that characterize each transition.\\

In doing so we hope to bring into physics a new approach on the way of understanding geometrical spaces. This is the idea of Brieskorn \cite{Brieskorn1970} that the structure of a variety can be understood by \emph{slicing it transversally} to a maximal subvariety. We show how this can be realized as a Higgs mechanism from the point of view of quantum field theory. From this point of view, the Hasse diagram gives an interesting view on the set of all possible mixed branches of a given quiver theory.\\

In Section \ref{sec:S} we summarize the results of the present work. Sections \ref{sec:M} and \ref{sec:B} of the paper aim to serve as an introduction to the main discussion. Section \ref{sec:M}  contains an overview of the basic mathematical concepts that are needed: hyperk\"ahler singularities and nilpotent orbits. Section \ref{sec:B} summarizes the required brane dynamics and quiver gauge theory. The reader familiarized with either of these subjects is encouraged to skip those sections and go directly to the new material in Section \ref{sec:K}. In Section \ref{sec:K} we develop the physical interpretation of the Kraft-Procesi transition. Section \ref{sec:F} introduces a formalism which allows to perform the required computations in an efficient way. Section \ref{sec:R} displays the results of such computations. Section \ref{sec:C} contains some conclusions.\\

\section{Summary}\label{sec:S}

The main motivation behind this paper is the discovery of a brane realization for the transition between nilpotent orbits described in \cite{Kraft1982}. The brane configurations and quivers corresponding to $3d\ \mathcal{N}=4$ gauge theories with closures of nilpotent orbits as their Higgs and Coulomb branches are currently known. We want to describe a new physical phenomenon, a Higgsing mechanism, that establishes a relation among them. This phenomenon produces a transition between different theories, and this is precisely the transition developed by Kraft and Procesi.\\

Each of the transitions in \cite{Kraft1982} is characterized by a singularity. We consider remarkably interesting that these singularities arise naturally in the brane configurations for the quiver gauge theories. They are the moduli generated by a minimal set of threebranes that can be \emph{Higgsed away}. For example, in the brane configuration corresponding to the Coulomb branch of a theory, this is the minimal set of D3-branes that can be aligned with D5-branes, split, and taken to the Higgs branch.\\

The Kraft-Procesi transition consists on Higgsing away these minimal singularities until they are no longer part of the configuration. In the previous example where the D3-branes are taken to the Higgs branch this corresponds to taking their coordinates in the Higgs branch to infinity, fully removing the minimal threebranes from the brane system. \\

The result of this transition is a new brane configuration that corresponds to a new quiver gauge theory. If the Coulomb branch of the old theory is the closure of a nilpotent orbit, the Coulomb branch of the new theory would be the closure of a nilpotent orbit as well. They are both connected in the Hasse diagram of their algebra created in \cite{Kraft1982}. The link that connects them is labeled with the minimal singularity that is Higgsed away during the transition.\\

In the present paper we go over many examples of these transitions. We also provide a general description of the process. At the end, we develop a formalism that allows very fast computations of the transitions. With our method, given as an input the quiver corresponding to the closure of the maximal nilpotent orbit, the quivers for the closures of all nilpotent orbits of the same algebra can be obtained, together with the minimal singularities involved in each transition among them, and the superstring embedding of both quivers and transitions.

\section{Mathematical Prelude}\label{sec:M}

In this section we review mathematical concepts that are essential to our discussion. The topics are: \emph{the ring of holomorphic functions over hyperk\"ahler varieties} and \emph{nilpotent orbits of the $\mathfrak{sl}_n$ algebra}. The reader familiarized with either of those concepts is encouraged to move directly to the next section. Our aim in this section is merely to point out some key mathematical features. For a rigorous study on the first subject the reader is directed to \cite{Harris1995}   or any other text in elementary algebraic geometry. On the second subject, \cite{Collingwood1993} is normally the preferred source and is the one we employ here.

\subsection{Hyperk\"ahler Singularities and Their Hilbert Series}


A hyperk\"ahler singularity is a type of affine variety that arises naturally in the study of moduli spaces. Kronheimer \cite{Kronheimer1988} describes it as: a hyperk\"ahler manifold $M$ with three complex structures $I,J,K$ that satisfy quaternionic relations, together with a Riemannian metric $h$ which is K\"ahler with respect to each of the complex structures. Out of the three K\"ahler forms $\omega_I$, $\omega_J$ and $\omega_K$ we can focus in one of them, say $\omega_I$, and the other two combine into a holomorphic (2,0)-form under the complex structure $I$:

\begin{align}
\omega_c=\omega_J+i\omega_K
\end{align} 

In all the cases encountered in physics, there is an $SU(2)$ symmetry acting in the hyperk\"ahler variety that corresponds to the R-symmetry of the quantum field theory\footnote{From the physics point of view, focussing in one complex form $\omega_I$ corresponds to the choice of a subgroup $U(1)_R\subset SU(2)_R$, and therefore the selection of a subalgebra with $3d$ $\mathcal{N}=2$ out of the $3d$ $\mathcal{N}=4$ supersymmetry algebra.} and it is denoted by $SU(2)_R$. The variety can be analyzed using the techniques of algebraic geometry, by studying the ring of holomorphic functions with respect to $\omega_I$. Let us illustrate these concepts with some examples.\\

\subsubsection{Example: $\mathbb{R}^4$}

Let us consider the affine variety $\mathbb{R}^4$. We say that there are four real coordinates:
\begin{align}
\{x_1,x_2,x_3,x_4\}
\end{align} 

There are three different ways of establishing a complex structure by picking the three possible pairs: $(x^1,x^2)(x^3,x^4)$, $(x^1,x^3)(x^2,x^4)$ or $(x^1,x^4)(x^2,x^3)$. We choose the first one of them that sets the following complex coordinates:
\begin{align}
\begin{aligned}
z_1&:=x_1+ix_2\\
z_2&:=x_3+ix_4
\end{aligned}
\end{align}

The variety can be thought now as $\mathbb{C}^2$. We call $\mathbb{C}[z_1,z_2]$ the ring of all holomorphic functions that exist in $\mathbb{C}^2$. This will be the set of all polynomials of the variables $z_1$ and $z_2$ with complex coefficients.\\

To determine the holomorphic ring, it is enough to find all linearly independent homogeneous polynomials. They can be graded according to their degree $d$. In this case there is one polynomial of degree zero, the constant function. There are two polynomials of degree $d=1$: $z_1$ and $z_2$ (note that we could have chosen $z_1$ and $z_1+z_2$ as the two linearly independent ones). There are three with $d=3$: $z_1^2$, $z_2^2$ and $z_1z_2$, for example. We can characterize the variety by stating:

\begin{align}
m_d=d+1
\end{align}

where $m_d$ is the number of linearly independent polynomials of degree $d$ that can be constructed in the variety. $m_d$ is called the \emph{Hilbert function} of the variety $\mathbb{C}^2$.\\

 We say that the two polynomials of degree one, $z_1$ and $z_2$ are the generators of the holomorphic ring, since a generic linearly independent homogeneous polynomial of degree $d$ will have the form:

\begin{align}
z_1^az_2^b\ \ \ \ \ s.t. \ \ a+b=d
\end{align}

The \emph{Hilbert series} $H(t)$ is defined as a power series in the variable $t$ with coefficients determined by the Hilbert function:

\begin{align}
	H(t)=\sum_{d=0}^\infty m_d t^d
\end{align}

The relation between $H(t)$ and $m_d$ is a discrete form on the Legendre transform. In this sense, $\log(t)$ and $d$ are conjugate variables, and one can characterize the system by either a function of $t$ or of $d$, according to convenience, or to the physics of the problem.
This is the source of the name \emph{fugacity} to $t$, if one identifies $d$ as a conserved charge and $\log(t)$ as its chemical potential.\\

In the case of $\mathbb{C}^2$ we have:

\begin{align}
\begin{aligned}
	H_{\mathbb{C}^2}(t)&=\sum_{d=0}^\infty (d+1) t^d\\
					&=\frac{1}{(1-t)^2}
\end{aligned}
\end{align}

where the two terms $(1-t)$ in the denominator correspond to the two generators of degree $d=1$.

\subsubsection{Example: $\mathbb{C}^2/\mathbb{Z}_2$}

Let us consider the singular variety $\mathbb{C}^2/\mathbb{Z}_2$ with the action $(1,1)$ of the finite group $\mathbb{Z}_2$ on the variables $\{z_1,z_2\}$ of $\mathbb{C}^2$. This means that the non identity element of the finite group acts on both variables at the same time, in this case multiplying it by the number $-1$. Therefore, only polynomials of $\mathbb{C}^2$ invariant under $\{z_1,z_2\}\rightarrow\{-z_1,-z_2\}$ are part of the holomorphic ring of $\mathbb{C}^2/\mathbb{Z}_2$. There is one polynomial with $d=0$, the constant function. There are no polynomials with $d=1$, since $z_1\rightarrow-z_1$ and  $z_2\rightarrow-z_2$. There are three polynomials with $d=2$:
\begin{align}
	\begin{aligned}
		p&:=z_1^2\\
		q&:=z_2^2\\
		r&:=z_1z_2
	\end{aligned}
\end{align}

We see that $p$, $q$ and $r$ can generate all other possible polynomials that are invariant under $\mathbb{Z}_2$. Hence they are the generators of the holomorphic ring. To fully characterize the space we need to mention that the three generators satisfy a relation at degree $d=4$:

\begin{align}
pq=r^2
\end{align}

To find the Hilbert function we see that all polynomials of even degree that are present in the holomorphic ring of $\mathbb{C}^2$ are also present in the ring of $\mathbb{C}^2/\mathbb{Z}_2$. No polynomials of odd degree are allowed in the case of $\mathbb{C}^2/\mathbb{Z}_2$. Therefore its Hilbert function is just:

\begin{align}
\begin{aligned}
	m_{d}&=d+1\ \ \  &\text{for}\ d&=2n\ & n&\in \mathbb{N}\\
	m_{d}&=0\ \ \  &\text{for}\ d&=2n+1\ & n&\in \mathbb{N}
\end{aligned}
\end{align}

Hence, the Hilbert series takes the form:

\begin{align}
	\begin{aligned}
H_{\mathbb{C}^2/\mathbb{Z}_2}(t)	&= \sum_{n=0}^\infty (2n+1)t^{2n} \\
							&=\frac{1-t^4}{(1-t^2)^3}
	\end{aligned}
\end{align}

The three terms in the denominator of $H_{\mathbb{C}^2/\mathbb{Z}_2}(t)$ correspond to the three generators with $d=2$, the numerator corresponds to the relation of degree $d=4$. These identifications are  always possible when the variety is a \emph{complete intersection}.\\

\subsubsection{Classification According to the $SU(2)_R$ Spin of the Generators}

For every hyperk\"ahler variety of the type we are considering,  the linearly independent polynomials of the holomorphic ring  can be embedded into multiplets of the symmetry group $SU(2)_R$. They are always assigned the highest weight in the $SU(2)_R$ multiplet. The other weights in the multiplet are normally assigned to non-holomorphic polynomials.\\

For example, in the variety $\mathbb{C}^2$ discussed before, the chosen polynomial with $d=0$ can be just the constant function $f=1$. This does not transform, so it constitutes a singlet of $SU(2)_R$. The generator $z_1$ will be rotated into $\bar{z}_2$, together they constitute the multiplet $(z_1,\bar{z}_2)$ under $SU(2)_R$, with the respective weights $(1,-1)$. This is a representation with spin $s=1/2$. Similarly the generator $z_2$ will be embedded in the spin $s=1/2$ multiplet $(z_2,\bar{z}_1)$ with respective weights $(1,-1)$.\\

Tensor products of the two representations can be taken in order to obtain multiplets containing all other holomorphic polynomials of the form $z_1^az_2^b$.  For example, we can take symmetric product of the multiplet $\gamma_1=(z_1,\bar{z}_2)$ to find:

\begin{align}
Sym^2(\gamma_1)=(z_1^2,z_1\bar{z}_2,\bar{z}_2^2)
\end{align}

with weights $(2,0,-2)$. Therefore it is an irreducible representation of $SU(2)_R$ with highest weight 2, which is equivalent to saying that has spin $s=1$. Hence, we say that the holomorphic polynomial $z_1^2$ is in the spin 1 representation.\\

Similarly, we could take representation $\gamma_2=(z_2,\bar{z}_1)$ and compute:

\begin{align}
Sym^2(\gamma_2)=(z_2^2,z_2\bar{z}_1,\bar{z}_1^2)
\end{align}

with weights $(2,0,-2)$. Therefore we say that also $z_2^2$ is in the spin 1 representation.\\

For $z_1z_2$ we can take the tensor product:

\begin{align}
\gamma_1\otimes\gamma_2=(z_1z_2,z_1\bar{z}_1,\bar{z}_2z_2,\bar{z}_2\bar{z}_1)
\end{align}

In Dynkin labels, this corresponds to:

\begin{align}
[1]\otimes[1]=[2]\oplus[0]
\end{align} 

Therefore, the result is a reducible representation with weights $(2,0,0,-2)$. Since we assign the holomorphic polynomial to the highest weight, in this case we say that $z_1z_2$ carries weight 2. This means that it forms part of the irrep $[2]$ in the RHS of the tensor product, i.e. of the irrep with spin $s=1$. Hence we have seen that all linearly independent holomorphic polynomials that can be constructed in $\mathbb{C}^2$ of degree $d=2$ have spin $s=1$ under $SU(2)_R$.\\

Let us examine the case of the variety $\mathbb{C}^2/\mathbb{Z}_2$. We still have the polynomial of degree zero, chosen to be the constant function $f=1$, with spin $s=0$ with respect to $SU(2)_R$. The generators $p$, $q$ and $r$ all carry spin $s=1$, since they  are inherited from the holomorphic ring of $\mathbb{C}^2$.\\

In general, for every hyperk\"ahler variety with an $SU(2)_R$ symmetry, the generators of the holomorphic ring always carry a highest weight $w$ inside an irreducible representation with spin $s=w/2$. There is a classification for this kind of  varieties whose holomorphic ring is finitely generated. It sorts them according to the spin that is carried by their generators under $SU(2)_R$. The classification is:
\begin{itemize}
	\item  For every variety there is only one object that carries spin $s=0$ and it is always the constant function $f=1$.
	\item If some the generators carry spin $s=1/2$, say $2n$ of them for some $n$, they generate a product of $n$ copies of $\mathbb{C}^2$.
	\item If the generators carry spin $s=1$, they all transform under the adjoint representation of an isometry group of the variety, see for example \cite{Brylinski1994}.
	\item Generators of spin higher than $s=1$ may be called \emph{baryons}, and their intuitive role is to increase the order of the singularity.
\end{itemize}
	
In the two cases above there is an isometry group $SU(2)$ that acts on the variables ${z_1, z_2}$ in the natural way. By this we mean that $\zeta=(z_1,z_2)$ is a doublet of $SU(2)$. We see by taking second symmetric product of this representation that we obtain:

\begin{align}
Sym^2(\zeta)=(z_1^2,z_1z_2,z_2^2)
\end{align}

In Dynkin labels this is written as irreducible representations:

\begin{align}
Sym^2[1]=[2]
\end{align}

Therefore, the generators of the holomorphic ring for the variety $\mathbb{C}^2/\mathbb{Z}_2$ transform under the adjoint representation of spin $s=1$ of the $SU(2)$ isometry group, in addition to spin $s=1$ under $SU(2)_R$.\\

\subsubsection{Classification from the Point of View of Quantum Field Theory}

The classification above is inherited by the set of moduli spaces of quantum field theories with hyperk\"{a}hler moduli spaces. This is due to the existence of a one to one correspondence\footnote{We do assume that this is a one to one correspondence, as we are not aware of any counter example.} between operators in the chiral ring of the theory and polynomials in the holomorphic ring of the affine variety.  In the language of quantum field theory, the classification takes the following form:

\begin{itemize}
	\item  For every moduli space there is only one operator that carries spin $s=0$ under $SU(2)_R$: the identity operator. 
	\item If some of the operators that generate the chiral ring carry spin $s=1/2$, say $2n$ of them for some $n$, they are free fields and form a decoupled sector in the theory. In total there are $n$ free hyper multiplets. Without loss of generality we can proceed by assuming the remaining interacting theory has generators of the chiral ring with spin  $s > 1/2$.
	\item If some of the generators of the chiral ring carry spin $s=1$, they all transform under the adjoint representation of a flavor symmetry group acting on the moduli space \cite{Gaiotto2008}.
	\item Generators of spin higher than $s=1$ may be called \emph{baryons}, and their intuitive role is to increase the order of the singularity of the moduli space. It is interesting to study their role and this is left for future study.
\end{itemize}

\subsection{Namikawa's Theorem}

Namikawa's theorem \cite{Namikawa2016} further restricts the kind of hyperk\"ahler variety that can have generators with spin 1 under $SU(2)_R$. The theorem can be understood as:\\

\emph{
	If all the generators of a hyperk\"ahler singularity with $SU(2)_R$ symmetry have spin $s=1$ under the $SU(2)_R$ group, the variety is the closure of a nilpotent orbit of the Lie algebra of its isometry group.}\\
	
Therefore, any attempt to understand hyperk\"{a}hler moduli spaces with generators with spin $s=1$ under $SU(2)_R$ should always be founded upon an understanding of the geometry of nilpotent orbits. These singularities will either be the closure of a nilpotent orbit, or a deformation of one, in the case when they also contain other generators of spin higher than $s=1$. From this perspective, closures of nilpotent orbits constitute the basis of all hyperk\"{a}hler singularities that exhibit an isometry.  \\

\subsection{Nilpotent Orbits}

As in \cite{Kraft1982}, we want to think of the spaces that are related to an element of a Lie algebra via the adjoint action of the corresponding group. The word \emph{nilpotent} stresses the fact that we are only interested in orbits of the algebra where all the elements are nilpotent. An element $X$ of a complex semisimple Lie algebra $\mathfrak{g}$ is said to be nilpotent if $\rho(X)^m=\rho(X)\circ \dots \circ \rho(X) = 0$ for some $m>0$ and $\rho:\mathfrak{g}\mapsto \text{End}(V)$ is the adjoint representation\footnote{It can be shown that a definition with a different finite representation of the algebra is equivalent to this definition \cite{Collingwood1993}.} of the algebra acting on a complex vector space $V$ \cite{Collingwood1993}.\\

\subsubsection{Definition for $\mathfrak{sl}_n$ Algebra}

In the following paragraphs we present the definition given by \cite{Collingwood1993}, \emph{Section 3.1  Type A}. The first observation is that nilpotent orbits of the algebra $\mathfrak{g}=\mathfrak{sl}_n$ are in one to one correspondence with partitions of $n$. We can define a partition $\lambda$ of $n$ as a tuple $(\lambda_1,\lambda_2,\dots,\lambda_k)$ of integer numbers with properties:

\begin{align}
	\begin{aligned}
	&\lambda_1\geq\lambda_2\geq\dots\geq\lambda_k>0\ \  \text{and}\\ 
	&\sum_{i=1}^k\lambda_i=n
	\end{aligned}
\end{align}
		
\emph{Exponential notation} can be introduced. For example $(3^2,2,1^5)=(3,3,2,1,1,1,1,1)$ is a partition of $n=13$. We  denote $\mathcal{P}(n)$ the set of all partitions of $n$, for example $\mathcal{P}(3)=\{(3),(2,1),(1^3)\}$.\\

We remember that an \emph{elementary Jordan block} of order $i\in\mathbb{Z}^+$ is defined as the $i\times i$ matrix:
\begin{align}
J_i:=\left(\begin{array}{cccccc}
		0&1&0&\dots&0&0\\
		0&0&1&\dots&0&0\\
		\vdots&\vdots&\vdots&\ddots&\vdots&\vdots\\
		0&0&0&\dots&0&1\\
		0&0&0&\dots&0&0\\
		\end{array}\right)
\end{align}\\

Given a partition $\lambda=(\lambda_1,\dots,\lambda_k)$ of $n$ we can form a nilpotent endomorphism of $\mathbb{C}^n$ as:
\begin{align}
		X_\lambda=\left(\begin{array}{cccc}
		J_{\lambda_1}&0&\dots&0\\
		0&J_{\lambda_2}&\dots&0\\
		\vdots&\vdots&\ddots&\vdots\\
		0&0&\dots&J_{\lambda_k}\\
		\end{array}\right)
\end{align}
	
Hence, $X_\lambda$ is a nilpotent element of the algebra $\mathfrak{sl}_n$. \\
		
The $n\times n$ matrix $X_\lambda$ is in the adjoint representation of $PSL(n)$ group\footnote{In \cite{Collingwood1993} the adjoint group that defines the action of elements of $\mathfrak{sl}_n$ on the algebra itself is $PSL(n)=SL(n)/Z$ where $Z$ is the center of $SL(n)$, this is the group that generates the nilpotent orbits acting on the right and on the left on the matrix $X_\lambda$.}. It generates an orbit under the group called the nilpotent orbit:
\begin{align}
\mathcal{O}_\lambda:=PSL(n)\cdot X_\lambda
\end{align}

 Note also that two different partitions give rise to two disjoint nilpotent orbits by the uniqueness of the Jordan normal form. Therefore, for every different partition of $n$ there is a different nilpotent orbit of $\mathfrak{sl}_n$. Furthermore, a generic nilpotent element $X\in\mathfrak{sl}_n$ has a Jordan normal form $X_\lambda$ for some $\lambda\in\mathcal{P}(n)$, i.e. it is $PSL(n)$-conjugate to $X_\lambda$. Therefore it belongs to the nilpotent orbit $\mathcal{O}_\lambda$.\\

\subsubsection{Example: Non-trivial Orbit of $\mathfrak{sl}_2$}

Let us study a specific example for the algebra $\mathfrak{g}=\mathfrak{sl}_2$. The set of all partitions of $n=2$ is 
\begin{align}
\mathcal{P}(2)=\{(2),(1^2)\}
\end{align}

The partition $\lambda=(1^2)$ corresponds to the trivial orbit, this is the orbit of the zero element. The corresponding Jordan matrix is 

\begin{align}
X_{(1^2)}=\left(\begin{array}{cc}
			0&0\\
			0&0\\
		\end{array}\right)
\end{align}.

Therefore, $X_{(1^2)}$ is actually the only element in the orbit. The closure of the orbit is equivalent to the orbit itself and defines an affine variety that only contains one point. This is called the \emph{trivial nilpotent orbit}.\\

The partition $\lambda=(2)$ corresponds to the non-trivial orbit. Its Jordan matrix is:

\begin{align}
X_{(2)}=\left(\begin{array}{cc}
			0&1\\
			0&0\\
		\end{array}\right)
\end{align}

In order to obtain the orbit we define the action of the group. A generic element $S\in SL(2)$ is:
\begin{align}
S=\left(\begin{array}{cc}
			a&b\\
			c&d\\
		\end{array}\right),\ \ \ ab-cd=1
\end{align}
where $a,b,c,d\in\mathbb{C}$.\\

We can define the nilpotent orbit  as
\begin{align}
\mathcal{O}_{(2)}:=\{M=S \cdot X_{(2)}\cdot S^{-1}|S\in SL(2)\}
\end{align}

We see that the action of $S$ on $X_{(2)}$ in this way gives the same element of the orbit than the action of $-S$, so the group that is acting on the nilpotent element to generate the orbit is actually $PSL(2)=SL(2)/Z$ with $Z=\{I,-I\}$, where $I$ is the identity matrix.\\

Any element $M\in \mathcal{O}_{(2)}$ can be written explicitly as:
\begin{align}
M=\left(\begin{array}{cc}
			-ac&a^2\\
			-c^2&ac\\
		\end{array}\right)
\end{align}

We can check that all matrices in $\mathcal{O}_{(2)}$ are nilpotent, since $M^2=0$. Note that the matrix with all zero entries is not included in the orbit, since that would imply $a=c=0$ and would contradict the condition $ab-cd=1$. If we take the set of all matrices $M\in \mathcal{O}_{(2)}$ together with the  matrix with all zero entries we obtain an affine variety. This set corresponds to $\bar{\mathcal{O}}_{(2)}$, the \emph{closure} of the nilpotent orbit $\mathcal{O}_{(2)}$.\\

The closure of the nilpotent orbit corresponding to partition $\lambda=(2)$ is then a variety with 3 generators of degree $d=2$:
\begin{align}
	\begin{aligned}
		p&:=a^2\\
		q&:=c^2\\
		r&:=ac
	\end{aligned}
\end{align}
and one relation of degree $d=4$:
\begin{align}
	p q = r^2
\end{align}

This defines the polynomial ring for the variety $\mathbb{C}^2/\mathbb{Z}_2$. Hence:
\begin{align}
\bar{\mathcal{O}}_{(2)}=\mathbb{C}^2/\mathbb{Z}_2
\end{align}

We can see that to obtain the closure we take the union with all nilpotent orbits with lower dimension, in particular $\mathcal{O}_{(1^2)}$, obtaining:

\begin{align}
\bar{\mathcal{O}}_{(2)}=\mathcal{O}_{(2)}\cup\mathcal{O}_{(1^2)}
\end{align}

This will always be the case in general for other closures of nilpotent orbits. Notice that this is also the case for the closure of the trivial nilpotent orbit:

\begin{align}
\bar{\mathcal{O}}_{(1^2)}=\mathcal{O}_{(1^2)}
\end{align}

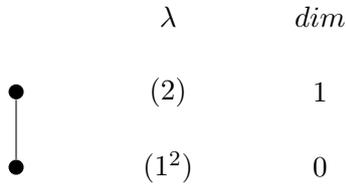
\begin{figure}[h]
	\centering
	\begin{tikzpicture}
		\tikzstyle{hasse} = [circle, fill,inner sep=2pt];
		\node [hasse] (r) [] {};
		\node [hasse] (m) [below of=r] {};
		\draw (r)--(m);	
		\node (er) [right of=r] {};
		\node (dr) [right of=er] {$(2)$};
		\node (cr) [right of=dr] {};
		\node (br) [right of=cr] {1};
		\node (em) [right of=m] {};
		\node (dm) [right of=em] {$(1^2)$};
		\node (cm) [right of=dm] {};
		\node (bm) [right of=cm] {0};
		\node (d) [above of=dr] {$\lambda$};
		\node (b) [above of=br] {$dim$};
	\end{tikzpicture}
	\caption{Hasse diagram with the partial order of all closures of nilpotent orbits of the algebra $\mathfrak{sl}_2$. The numbers $dim$ refer to the quaternionic dimension of the variety.}
	\label{fig:HasseSU2}
\end{figure}

From the previous results we can infer an inclusion relation in the closures of both nilpotent orbits:

\begin{align}
\bar{\mathcal{O}}_{(1^2)}\subset\bar{\mathcal{O}}_{(2)}
\end{align}
where $\subset$ denotes  that the LHS is a subvariety of the RHS variety. This relation induces a partial ordering in the set of all closures of nilpotent obits of $\mathfrak{sl}_2$. A Hasse diagram can be plotted to represent such ordering. In this case there are only two varieties and said diagram results extremely simple. However, we want to include it here, in fig. \ref{fig:HasseSU2}, since it constitutes the first step towards the characterization of the inclusion relation of closures of nilpotent orbits for algebras of the form $\mathfrak{g}=\mathfrak{sl}_n$.\\

\section{Branes}\label{sec:B}

Now that some of the key mathematical aspects for the present discussion have been revisited we would like to go over some physical arguments that lie at the core of our problem. In this section we review the Type IIB superstring embedding for $3d\ \mathcal{N}=4$ effective gauge theories from \cite{Hanany1996}. If the reader is already familiar with this description we direct them to subsection \ref{sec:HBC}, where the connection with closures of nilpotent orbits is presented.\\

\subsection{Brane Configurations and $3d$ $\mathcal{N}=4$ Quiver Gauge Theories}

Let us start by constructing the most simplest possible models. Out of the three introductory examples discussed in this subsection, the first has the variety $\mathbb{C}^2$ as its Coulomb branch and the trivial variety as its Higgs branch. The second corresponds to the mirror dual model of the first one. Hence,  its Higgs branch is $\mathbb{C}^2$ and its Coulomb branch is trivial. The third model is self-mirror: both its Coulomb and its Higgs branches are described by $\mathbb{C}^2/\mathbb{Z}_2$.

\subsubsection{First Example: $3d$ $\mathcal{N}=4$ SQED with Zero Flavours}

In this Type IIB superstring configuration there are only D3-branes, D5-branes and NS5-branes. The D3-brane spans directions $x^1$ and $x^2$ and stretches between two fivebranes along the $x^6$ direction. In a vacuum configuration it has constant positions along the remaining space directions $\{x^3,x^4,x^5,x^7,x^8,x^9\}$. We call $\vec{x}$ the position of the D3-brane along $\{x^3,x^4,x^5\}$ directions and  $\vec{y}$ the position along $\{x^7,x^8,x^9\}$.\\

D5-branes span directions $\{x^1,x^2,x^7,x^8,x^9\}$, and have positions along coordinates $\vec{m}=(x^3,x^4,x^5)$ and $x^6$. We denote by $\vec{m}_i$ and $z_i$ the position of the $i^{th}$ D5-brane along the directions $\vec{m}$ and $x^6$ respectively.\\

Similarly, NS5-branes span directions $\{x^1,x^2,x^3,x^4,x^5\}$, and have positions along coordinates $\vec{w}=(x^7,x^8,x^9)$ and $x^6$. We denote by $\vec{w}_j$ and $t_j$ the position of the $j^{th}$ NS5-brane along the directions $\vec{w}$ and $x^6$ respectively.\\

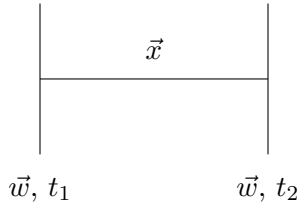
\begin{figure}[h]
	\centering
		\begin{tikzpicture}
			\draw 	(1,0)--(1,2)
					(4,0)--(4,2);
			\draw	(1,1)--(4,1);
		\draw 	(2.5,1) node[label=above:{$\vec{x}$}]{}
				(1,0) node [label=below:{$\vec{w}$, $t_1$}]{}
				(4,0) node [label=below:{$\vec{w}$, $t_2$}]{};
		\end{tikzpicture}
	\caption{In this picture vertical lines correspond to NS5-branes. The vertical direction corresponds to directions $\vec{m}$, spanned by the NS5-branes. The horizontal direction corresponds to $x^6$, so the different positions of $t_i$ of the two NS5-branes along this direction are evidenced in this way. The third axis, perpendicular to the paper, would correspond to directions $\vec{w}$, in this case both NS5-branes are in the picture since their $\vec{w}_i$ position coincides and a D3-brane with the same position $\vec{y}=\vec{w}_i$ can be stretched between them.}
	\label{fig:U1}
\end{figure}

This kind of configuration preserves 8 out of the 32 initial supercharges \cite{Hanany1996}. In the first example we set two NS5-branes at positions $t_1\neq t_2$ along direction $x^6$ and same value of $\vec{w}_1=\vec{w}_2$. This allows a D3-brane with the coordinates $\vec{y}=\vec{w}_1=\vec{w}_2$ to be stretched between them. Hence, there is a continuous and infinite set of positions $\vec{x}$ that the D3-brane can have. The brane configuration is sketched in fig. \ref{fig:U1}.\\

The low energy physics of this configuration is described by a $3d\ \mathcal{N}=4$ effective gauge theory living in the worldvolume of the threebrane. The effective gauge group is $G=U(1)$. There is one vector multiplet and no hypermultiplets. The gauge coupling of the effective theory is proportional to the distance between the NS fivebranes. Up to a universal multiplicative constant we have:

\begin{align}
	\frac{1}{g^2}=|t_1-t_2|
\end{align}

The three scalars on the vector multiplet correspond to the three real coordinates of the position of the D3-brane, $\vec{x}$. Since the vector field in the multiplet has only one degree of freedom, it can be dualized into a real scalar field $a$ that admits non zero vacuum expectation value. Due to the boundary conditions imposed by the NS5-branes, it can take any value on the circle $S^1$. The radius $R$ of the circle is proportional to the gauge coupling. We recover the non-compact variety $\mathbb{R}^4$ in the infrared, where all couplings are taken to infinity.  Therefore, in the low energy limit, the moduli space is $(\vec{x},a)\in \mathbb{R}^4$. We write:

\begin{align}
	\mathcal{M}_C=\mathbb{C}^2
\end{align}

where $\M_C$ is the Coulomb branch. Since there are no hyper multiplets, there is no Higgs branch, we assign this to the trivial variety, the point.

\subsubsection{Second Example: $3d$ $\mathcal{N}=4$, One Free Massless Hyper}

Let us consider a theory with only one free massless hyper multiplet in three dimensions and with 8 supercharges. There are 4 real scalars in the model, that admit any constant (non space dependent) vacuum expectation value. Therefore the moduli space is the variety $\mathbb{R}^4$ or equivalently $\mathbb{C}^2$. We say that the Higgs branch of this theory is $\mathbb{C}^2$, while the Coulomb branch is the trivial variety.  We write:

\begin{align}
	\mathcal{M}_H=\mathbb{C}^2
\end{align}

where $\M_H$ is the Higgs branch. Therefore we see that this model is \emph{mirror dual} to the previous example.\\

\subsubsection*{Mirror Symmetry}

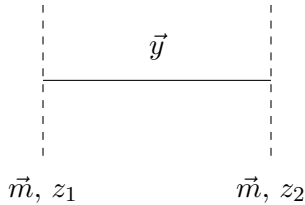
\begin{figure}[h]
	\centering
		\begin{tikzpicture}
			\draw [dashed]	(1,0)--(1,2)
					(4,0)--(4,2);
			\draw	(1,1)--(4,1);
		\draw 	(2.5,1) node[label=above:{$\vec{y}$}]{}
				(1,0) node [label=below:{$\vec{m}$, $z_1$}]{}
				(4,0) node [label=below:{$\vec{m}$, $z_2$}]{};
		\end{tikzpicture}
	\caption{In this picture the dashed vertical lines correspond to D5-branes. The vertical direction corresponds to directions $\vec{w}$, spanned by the D5-branes. The horizontal direction corresponds to $x^6$, so the different positions of $z_i$ of the two D5-branes along this direction are evidenced in this way. The third axis, perpendicular to the paper, would correspond to directions $\vec{m}$, in this case both D5-branes are in the picture since their $\vec{m}_i$ position coincides and a D3-brane with the same position $\vec{x}=\vec{m}_i$ can be stretched between them.}
	\label{fig:U1mirror}
\end{figure}

In terms of brane configurations, mirror duality corresponds to an S-duality that effectively swaps D5-branes and NS5-branes. If we perform this duality on the previous model, fig. \ref{fig:U1}, we obtain a new model, depicted in fig. \ref{fig:U1mirror}. Three out of the four real scalars in the model correspond to the position of the D3-brane $\vec{y}$. To understand the role of the fourth real scalar we need to think of the $4d$ effective theory in the worldvolume of the infinite D3-brane. This theory has a vector multiplet with a four-dimensional vector field living in it. When the theory decomposes into a $3d$ theory the four-dimensional vector multiplet decomposes into a three-dimensional vector multiplet and a three-dimensional hyper multiplet. The four-dimensional vector field also decomposes into a three-dimensional vector field, that lives in the three-dimensional vector multiplet, and a three-dimensional scalar field, that lives inside the three-dimensional hyper multiplet. It is this three-dimensional scalar that does not correspond to the position of the brane $\vec{y}$, but can also admit a nonzero vacuum expectation value in the three-dimensional theory, due to the boundary conditions imposed by the D5-branes \cite{Hanany1996}. 

\subsubsection{Third Example: $3d$ $\mathcal{N}=4$ SQED with 2 Flavours}

In the last example we consider a $3d\  \mathcal{N}=4$ theory with gauge group $G=U(1)$, one vector multiplet transforming in the adjoint of such group, and two hypers transforming in the fundamental of the gauge group. Let the two hyper multiplets transform under the fundamental representation of the flavor group $SU(2)$.\\

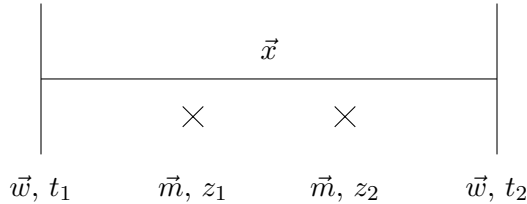
\begin{figure}[h]
	\centering
		\begin{tikzpicture}
			\draw (2,0)--(2,2)
					(8,0)--(8,2);
			\draw	(2,1)--(8,1);
		\draw 	(5,1) node [label=above:{$\vec{x}$}]{}
				(4,.5) node[cross]{}
				(6,.5) node[cross]{}
				(4,0) node[label=below:{$\vec{m}$, $z_1$}]{}
				(6,0) node[label=below:{$\vec{m}$, $z_2$}]{}
				(2,0) node [label=below:{$\vec{w}$, $t_1$}]{}
				(8,0) node [label=below:{$\vec{w}$, $t_2$}]{};
		\end{tikzpicture}
	\caption{In the phase depicted in this figure the two NS5-branes share the same position $\vec{w}$ along directions $\{x^7,x^8,x^9\}$, this makes the existence of a Couolomb branch possible. The two crosses correspond to two D5-branes stretching along the perpendicular direction to the paper, which corresponds to $\{x^7,x^8,x^9\}$. They share the same position $\vec{m}$ along directions $\{x^3,x^4,x^5\}$, which in the diagram is represented by the vertical direction, this makes the existence of a Higgs branch possible. In the special point of the moduli where the D3-brane $\vec{x}$ position coincides with the position $\vec{m}$ of the D5-branes the two hyper multiplets become massless due to fundamental strings of length zero stretching between the D5-branes and the D3-brane.}
	\label{fig:SU2U1}
\end{figure}

In the Coulomb branch of this theory all the hyper multiplets are massive and the vector multiplet is massless. In the singular point where Higgs and Coulomb branch coincide both hypers become massless. In the Higgs branch the vector multiplet becomes massive, \emph{eating} one of the hypers, and leaving one massless hyper multiplet. Hence, the theory has a four dimensional Coulomb branch, and a four dimensional Higgs branch, that intersect in a singular point of the moduli space. The brane description of this theory is shown in fig. \ref{fig:SU2U1}.\\

In the Coulomb branch the D3-brane ends in both NS5-branes. Its position $\vec{x}$ corresponds to the three massless real scalar fields in the vector multiplet, and the boundary conditions on the NS5-branes allow the scalar field $a$ dual to the vector field in the super multiplet to admit a nonzero VEV. The Coulomb branch as seen in the brane picture has four dimensions. The new feature of this model is the existence of a singular point in the branch. The point where $\vec{x}=\vec{m}$. At this point two hyper multiplets become massless, this is the intersection of the Higgs branch with the Coulomb branch. Therefore the Coulomb branch must have four real dimensions and a singular point at which it is connected to the Higgs branch. Any variety of the form $\mathbb{C}^2/\Gamma$ where $\Gamma\subset SU(2)$ is a finite subgroup of $SU(2)$ could be a good candidate. In this case the answer is the simplest nontrivial group $\Gamma=\mathbb{Z}_2$ \cite{Seiberg1996, Hanany1996}.\\

\begin{figure}[h]
	\centering
		\begin{tikzpicture}
			\draw 	[dashed](4,0)--(4,2)
					(6,0)--(6,2);
			\draw	(2,.5)--(4,.5)
					(6,.5)--(8,.5)
					(4,1)--(6,1);
		\draw 	(5,1) node [label=above:{$\vec{y}$}]{}
				(2,.5) node[cross]{}
				(8,.5) node[cross]{}
				(2,.5) node[circle,draw]{}
				(8,.5) node[circle,draw]{}
				(4,0) node[label=below:{$\vec{m}$, $z_1$}]{}
				(6,0) node[label=below:{$\vec{m}$, $z_2$}]{}
				(2,0) node [label=below:{$\vec{w}$, $t_1$}]{}
				(8,0) node [label=below:{$\vec{w}$, $t_2$}]{};
		\end{tikzpicture}
	\caption{The phase depicted in this picture corresponds to the Higgs branch of the $3d\ \mathcal{N}=4$ SQED theory with two flavours. The vertical dashed lines correspond to D5-branes stretching along directions $\{x^7,x^8,x^9\}$. The circled crosses correspond to NS5-branes stretching along the directions $\{x^3,x^4,x^5\}$.}
	\label{fig:SU2U1Higgs}
\end{figure}
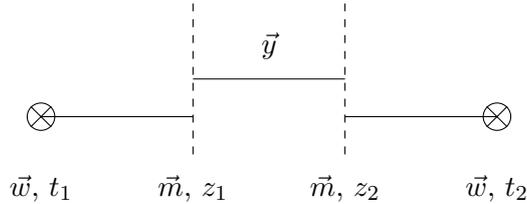

At the point of the singularity we can use the Higgs mechanism to transition to the Higgs branch. In the brane system this is realized in two steps:
\begin{enumerate}
	\item The D3-brane aligns with the two D5-branes.
	\item At that position, the D3-brane can split in three segments, each between two fivebranes. The rightmost and leftmost segments are \emph{frozen}, since they are connecting two fivebranes of different kind, they are indeed fixed at position $(\vec{x},\vec{y})=(\vec{m},\vec{w})$. The segment in the middle stretches between the two D5-branes. Its position along the $\{x^3,x^4,x^5\}$ directions is fixed, $\vec{x}=\vec{m}$, but the position $\vec{y}$ can change freely, a diagram of a phase with this configuration and $\vec{y}\neq \vec{w}$ is displayed on fig. \ref{fig:SU2U1Higgs}.
\end{enumerate}

After the two step transition described above the Higgs branch is reached. As in the case with only one free hyper multiplet, there is a four dimensional moduli given by the position $\vec{y}$ of the D3-brane and  the scalar field $b$, such that $(\vec{y},b)\in \mathbb{R}^4$. The main difference is that there is a singular point, when $\vec{y}=\vec{w}$. The branch has to be four dimensional and has one singularity. The answer in this case is also $\mathbb{C}^2/\mathbb{Z}_2$ \cite{Hanany1996}. The Higgs branch is identical to the Coulomb branch and we write:

\begin{align}
	\M_C=\M_H=\mathbb{C}^2/\mathbb{Z}_2
\end{align}

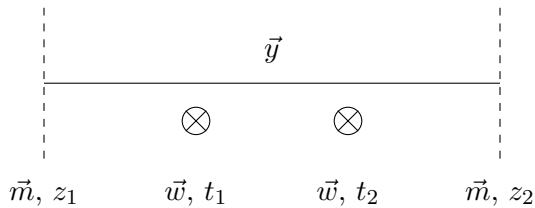
\begin{figure}[h]
	\centering
		\begin{tikzpicture}
			\draw [dashed] (2,0)--(2,2)
					(8,0)--(8,2);
			\draw	(2,1)--(8,1);
		\draw 	(5,1) node [label=above:{$\vec{y}$}]{}
				(4,.5) node[cross]{}
				(6,.5) node[cross]{}
				(4,.5) node[circle,draw]{}
				(6,.5) node[circle,draw]{}
				(2,0) node[label=below:{$\vec{m}$, $z_1$}]{}
				(8,0) node[label=below:{$\vec{m}$, $z_2$}]{}
				(4,0) node [label=below:{$\vec{w}$, $t_1$}]{}
				(6,0) node [label=below:{$\vec{w}$, $t_2$}]{};
		\end{tikzpicture}
	\caption{This is a depiction of the Higgs branch of the $3d\ \mathcal{N}=4$ SQED theory with two flavours, after the frozen D3-branes have been annihilated.}
	\label{fig:SU2U1Higgs2}
\end{figure}

This identity can be made manifest by performing two Hanany-Witten transition. In this phase transition the NS5-branes can go thought the D5-branes, annihilating the frozen D3-branes that connected them. The result is shown in fig. \ref{fig:SU2U1Higgs2}.

Since both the Higgs and the Coulomb branch are $\mathbb{C}^2/\mathbb{Z}_2$ we say that this model is mirror dual to itself. To check this we can perform S-duality in the Coulomb branch brane configuration, fig. \ref{fig:SU2U1}, as we did for the first example. The result is the Higgs branch brane configuration , in fig. \ref{fig:SU2U1Higgs2}.\\ 

\begin{figure}[h]
	\centering
		\begin{tikzpicture}
			\tikzstyle{gauge} = [circle, draw];
			\tikzstyle{flavour} = [regular polygon,regular polygon sides=4,draw];
			\node (g1) [gauge,label=below:{$1$}] {};
			\node (f1) [flavour,above of=g1,label=above:{$2$}] {};
			\draw (g1)--(f1);
		\end{tikzpicture}
	\caption{Quiver of the model with $U(1)$ gauge theory with 2 flavours.}
	\label{fig:SU2U1Quiver}
\end{figure}
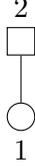

This theory belongs to a family of models that are called \emph{quiver gauge theories}. This means that we can draw a graph (quiver) where nodes and edges represent the different particles in the model. In this case the quiver of the model is depicted in fig. \ref{fig:SU2U1Quiver}. The circular node with the label $n$ symbolizes a gauge group $U(n)$, in this case $n=1$. There are always $n^2$ vector multiplets transforming in the adjoint representation of such nodes. In this case there is one vector multiplet transforming as a singlet. The square node with a label $k$ represents a flavor group $U(k)$, in this case $k=2$. There is a $U(1)$ \emph{center of mass} factor that decouples, so the final flavor group will be $SU(2)$. The edge corresponds to \emph{bifundamental} hyper multiplets. These are hyper multiplets transforming in the fundamental representation of both the gauge node and the flavor node. In this case there are $2\times 1=2$ hyper multiplets.

\subsection{Generic Example of a Brane Configuration for a Quiver Gauge Theory}

\begin{figure}[h]
	\centering
	\begin{subfigure}[t]{1\textwidth}
    \centering
	\begin{tikzpicture}
		\draw (-1,1) node[cross]{};
		\draw (0,1) node[cross]{};
		\draw (1,0)--(1,2);
		\node [cross] at (2,1){};
		\draw (3,0)--(3,2);
		\node [cross] at (4,.8){};
		\node [cross] at (5,.8){};
		\draw (6,0)--(6,2);
		\draw (7,0)--(7,2);
		\draw (8,0)--(8,2);
		\draw (1,.5)--(3,.5)
			(1,1.5)--(3,1.5);
		\draw (3,.25)--(6,.25)
			(3,1.35)--(6,1.35)
			(3,1.75)--(6,1.75);
		\draw (6,.5)--(7,.5)
			(6,1.5)--(7,1.5);
		\draw (7,1)--(8,1);
	\end{tikzpicture}
        \caption{}
    \end{subfigure}
    \hfill
    \begin{subfigure}[t]{1\textwidth}
    \centering
	\begin{tikzpicture}
		\draw[dashed] (-1,0)--(-1,2)
				(0,0)--(0,2)
				(2,0)--(2,2)
				(4,0)--(4,2)
				(5,0)--(5,2);
		\draw 	(1,1.5) node[cross]{}
				(3,.5) node[cross]{}
				(6,1.5) node[cross]{}
				(7,1) node[cross]{}
				(8,.5) node[cross]{}
				(1,1.5) node[circle,draw]{}
				(3,.5) node[circle,draw]{}
				(6,1.5) node[circle,draw]{}
				(7,1) node[circle,draw]{}
				(8,.5) node[circle,draw]{};
		\draw 	(1.05,1.55)--(2,1.55)
				(1.05,1.45)--(4,1.45)
				(3,.5)--(4,.5)
				(5,1.5)--(6,1.5)
				(5,1)--(7,1)
				(5,.5)--(8,.5)
				(2,1)--(4,1)
				(4,1.6)--(5,1.6)
				(4,1.2)--(5,1.2)
				(4,.7)--(5,.7);
	\end{tikzpicture}
        \caption{}
    \end{subfigure}
	\caption{Brane configurations for the theory with linking numbers for the D5-branes $\vec{l}_d=(0,0,1,2,2)$ and linking numbers for the NS5-branes \mbox{$\vec{l}_s=(4,4,4,4,4)$}. (a) shows the Coulomb branch and  (b) depicts the Higgs branch. In (a) the D5-branes are represented by crosses and the NS5-branes by vertical solid lines. In (b) the D5-branes are represented by vertical dashed lines and the NS5-branes by circled crosses. In both figures the D3-branes are represented by horizontal solid lines.}
	\label{fig:firstExample}
\end{figure}
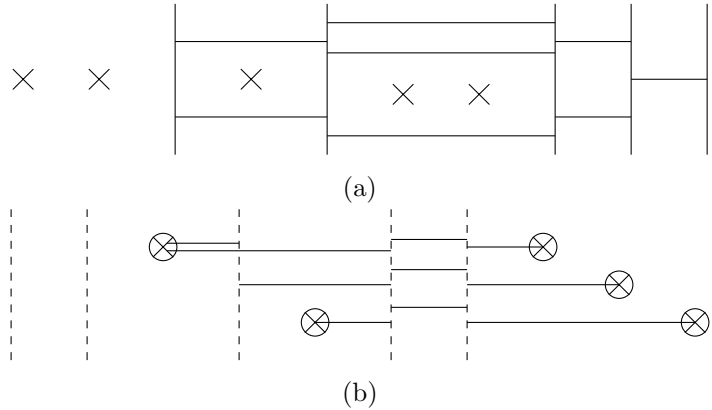

Let us see a more general example of a brane configuration, like the one in fig. \ref{fig:firstExample} (a). In this diagram, as in the previous examples and in the remaining figures of this paper, vertical solid lines correspond to NS5-branes and horizontal solid lines correspond to D3-branes. D5-branes are represented with crosses. The directions each brane spans are the same as in previous examples, hence 8 supercharges are always preserved. Since all D3-branes that are not fixed stretch between NS5-branes, their positions $\vec{x}_i$ along $\{x^3,x^4,x^5\}$ directions, together with the VEVs of the $a_i$ fields constitute the Coulomb branch of the theory. We will call a brane configuration of this kind a Coulomb branch brane configuration, or for short, a \emph{Coulomb brane configuration}.\\

To obtain the Higgs branch brane configuration we can perform a phase transition exactly in the same fashion as in the example of SQED with 2 flavors, we align all D5-branes, so a total higgsing can be achieved, align the D3-branes with the D5-branes, and them perform the splitting. To preserve supersymmetry we need to follow one constraint:

\begin{constraint}
	Given any pair of NS5-brane and D5-brane in the system, there can only be at most one D3-brane stretching between them.
\end{constraint}

After the splitting we have two kinds of D3-branes, those that stretch between an NS5-brane and a D5-brane, whose position is fixed, and those that stretch between two D5-branes, which can move freely along their $\vec{y}_i$ positions. These last ones will generate the Higgs branch of the theory. Therefore we call a brane configuration where there are no D3-branes that can move along their $\vec{x}_i$ the Higgs branch brane configuration, or \emph{Higgs brane configuration} for short. In this case it is depicted in fig. \ref{fig:firstExample} (b).\\

We can see that in this example the Coulomb branch of the model has $8 \times 4=32$ real dimensions, since there are 8 D3-branes that generate the moduli in the Coulomb brane configuration , and for each brane there are four real scalar fields $(\vec{x}_i,a_i)$ that admit nonzero VEVs. On the other hand, the Higgs branch has $4\times 4=16$ real dimensions, since there are only four D3-branes in fig. \ref{fig:firstExample} (b) that are free to move and generate the moduli; for each of them the real scalar fields $(\vec{y}_i,b_i)$ admit nonzero VEVs.\\

The matter content of the model can be obtained from the Coulomb brane configuration. This can be summarized in a quiver. In order to do this, the first step is to perform Hanany-Witten transitions to make sure that all frozen D3-branes have been annihilated, then split the D3-branes that stretch between NS5-branes as much as possible. The elements in the quiver are:

\begin{itemize}
	\item For each interval between two consecutive NS5-branes, there is a gauge node with label $n_i$ corresponding to a factor of the gauge group of the form $U(n_i)$, where $n_i$ is the number of D3-branes stretching between said fivebranes.
	\item Between two consecutive gauge nodes there is one edge corresponding to hyper multiplets transforming in the fundamental representation of each node.
	\item For each interval between two consecutive NS5-branes that contains at least one D5-brane, there is a flavor node with label $k_i$ connected to its respective gauge node by an edge. The edge represents hypermultiplets transforming under the fundamental of the flavor node and the fundamental of the gauge node. $k_i$ is the number of D5-branes in said interval, and the flavor group is $U(k_i)$.
\end{itemize}

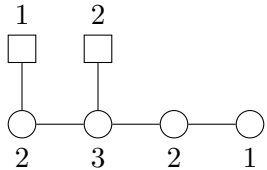
\begin{figure}[h]
	\centering
	\begin{tikzpicture}
	\tikzstyle{gauge} = [circle, draw];
	\tikzstyle{flavour} = [regular polygon,regular polygon sides=4, draw];
	\node (g1) [gauge, label=below:{$2$}] {};
	\node (g2) [gauge,right of=g1, label=below:{$3$}] {};
	\node (g3) [gauge,right of=g2, label=below:{$2$}] {};
	\node (g4) [gauge,right of=g3, label=below:{$1$}] {};
	\node (f1) [flavour, above of=g1, label=above:{$1$}] {};
	\node (f2) [flavour,above of=g2, label=above:{$2$}] {};
	\draw (g1)--(g2)--(g3)--(g4)
		(f1)--(g1)
		(f2)--(g2);
	\end{tikzpicture}
	\caption{Quiver from the model with linking numbers for the D5-branes $\vec{l}_d=(0,0,1,2,2)$ and linking numbers for the NS5-branes \mbox{$\vec{l}_s=(4,4,4,4,4)$}. The circular nodes represent vectorplets in the three-dimensional $\mathcal{N}=4$ gauge theory on the worldvolume of the D3-branes. Each vectorplet with label $i$ on its node transforms on the adjoint representation of a different factor $U(i)$ of the gauge group of the theory $G=\bigotimes_i U(i)$. Edges of the quiver represent hypermultiplets transforming under the bifundamental representation of the group $U(i)\times U(j)$, where $i,j$ are the labels on the nodes connected by the edge. The square nodes with label $k$ represent therefore global symmetries $U(k)$ of the hypermultiplets.}
	\label{fig:firstExampleQuiver}
\end{figure}

The final gauge group is $G=\bigotimes_iU(n_i)$. There is an overall $U(1)$ center of mass factor that decouples from the total flavor symmetry group, $\bigotimes_i U(k_i)$. If there are no flavors, then there is an overall $U(1)$ factor that decouples from the gauge group $G$. For the present model we can read the quiver, it is depicted in fig. \ref{fig:firstExampleQuiver}.\\

\subsubsection*{Conserved Quantities}

Despite the multiplicity of brane configurations that correspond to a single theory, there exist some quantities that are always preserved. Furthermore, specifying these quantities fully characterizes the model and its effective quiver gauge theory. The conserved quantities are: the number of NS5-branes in the system $n_s$, the number of D5-branes $n_d$, and the \emph{linking number} of each of the fivebranes. \\

The linking numbers, using the conventions in \cite{Gaiotto2008}, are just the net number of D3-branes ending on the fivebrane (D3-branes ending on it from the right minus D3-branes ending on it from the left) plus the total number of fivebranes of the opposite kind to its left (for example, if we are computing the linking number of a D5-brane with position $z$ along the $x^6$ direction, we have to add the number of NS5-branes with position $t_i$ along the same direction such that $t_i<z$).\\

For the model in this example we have $n_s=n_d=5$. Let us compute the linking number of the NS5-branes, choosing the Coulomb brane configurationin fig. \ref{fig:firstExample} (a), starting from the leftmost NS5-brane:
\begin{itemize}
	\item The $1^{st}$ NS5-brane: it has 2 D3-branes ending on it from the right and none from the left, so we obtain a factor of 2. Since there are 2 D5-branes to its left we have another factor of 2 and the final linking number is $2+2=4$.
	\item The $2^{nd}$ NS5-brane: There are 3 threebranes ending on it from the right and 2 ending on it from the left, this gives a factor of 1. We need to add 3 D5-branes to its left and obtain a total linking number of $1+3=4$.
	\item The $3^{rd}$ NS5-brane: There are 2 D3-branes ending on it from the right and 3 from the left, giving a total factor of $-1$. We need to add 5 D5-branes that appear at its left, giving a total linking number of $-1+5=4$.
	\item The $4^{th}$ NS5-brane: There is 1 threebrane ending on it from the left and two from the right, giving a total factor of $-1$. We need to add 5 D5-branes to its left, obtaining a linking number of $4$.
	\item The $5^{th}$ NS5-brane: There is only 1 D3-brane ending on it from the left, giving a factor of $-1$. After adding the 5 D5-branes that are at its left we obtain the linking number $-1+5=4$
\end{itemize}

Therefore all NS5-branes have the same linking number, 4. We can arrange all linking numbers in a vector:
\begin{align}
\vec{l}_s=(4,4,4,4,4)
\end{align}

Let us compute the linking number of the D5-branes, starting from the leftmost one and going to the right. None of them have D3-branes ending on them, so we just need to count the number of NS5-branes to their left: the first two have 0 NS5-branes to their left, the third one has 1, and the fourth and fifth ones have 2. Therefore the linking numbers of all of them, ordered in an array, are:

\begin{align}
\vec{l}_d=(0,0,1,2,2)
\end{align}

 We see that $n_s$, $n_d$, $\vec{l}_s$ and $\vec{l}_d$ are preserved in the Higgs brane configuration, fig. \ref{fig:firstExample} (b).\\

\begin{figure}[ht]
	\centering
	\begin{tikzpicture}
	\tikzstyle{gauge} = [circle,draw];
	\tikzstyle{flavour} = [regular polygon,regular polygon sides=4,draw];
	\node (g1) [gauge,label=below:{$1$}] {};
	\node (g2) [gauge,right of=g1,label=below:{$3$}] {};
	\node (f2) [flavour,above of=g2,label=above:{$5$}] {};
	\draw (g1)--(g2)--(f2);
	\end{tikzpicture}
	\caption{Quiver from the model with linking numbers for the NS5-branes \mbox{$\vec{l}_s=(0,0,1,2,2)$} and linking numbers for the D5-branes \mbox{$\vec{l}_d=(4,4,4,4,4)$}.}
	\label{fig:firstExampleQuiverSdual}
\end{figure}
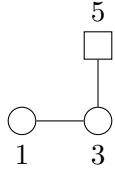

Taking the mirror dual can also be understood as swapping $n_s$ with $n_d$ and  $\vec{l}_s$ with $\vec{l}_d$. We can see the mirror model of the present example in fig. \ref{fig:firstExampleSdual}. The quiver can be read from the Coulomb branch, fig. \ref{fig:firstExampleSdual} (b), after doing Hanany-Witten transitions that annihilate all frozen threebranes. After all these transitions there are one D3-brane in the fourth interval between NS5-branes and 3 D3-branes in the fifth one. All D5-branes are now in the fifth interval. The quiver takes the form of fig. \ref{fig:firstExampleQuiverSdual}.\\

\begin{figure}[ht]
	\centering
	\begin{subfigure}[t]{1\textwidth}
    \centering
	\begin{tikzpicture}
		\draw[dashed] (1,0)--(1,2)
				(3,0)--(3,2)
				(6,0)--(6,2)
				(7,0)--(7,2)
				(8,0)--(8,2);
		\draw (-1,1) node[cross]{};
		\draw (0,1) node[cross]{};
		\node [cross] at (2,1){};
		\node [cross] at (4,.8){};
		\node [cross] at (5,.8){};
		\draw (-1,1) node[circle,draw]{};
		\draw (0,1) node[circle,draw]{};
		\node [circle,draw] at (2,1){};
		\node [circle,draw] at (4,.8){};
		\node [circle,draw] at (5,.8){};
		\draw (1,.5)--(3,.5)
			(1,1.5)--(3,1.5);
		\draw (3,.25)--(6,.25)
			(3,1.35)--(6,1.35)
			(3,1.75)--(6,1.75);
		\draw (6,.5)--(7,.5)
			(6,1.5)--(7,1.5);
		\draw (7,1)--(8,1);
	\end{tikzpicture}
        \caption{}
    \end{subfigure}
    \hfill
    \begin{subfigure}[t]{1\textwidth}
    \centering
	\begin{tikzpicture}
		\draw	(-1,0)--(-1,2)
				(0,0)--(0,2)
				(2,0)--(2,2)
				(4,0)--(4,2)
				(5,0)--(5,2);
		\draw 	(1,1.5) node[cross]{}
				(3,.5) node[cross]{}
				(6,1.5) node[cross]{}
				(7,1) node[cross]{}
				(8,.5) node[cross]{};
		\draw 	(1.05,1.55)--(2,1.55)
				(1.05,1.45)--(4,1.45)
				(3,.5)--(4,.5)
				(5,1.5)--(6,1.5)
				(5,1)--(7,1)
				(5,.5)--(8,.5)
				(2,1)--(4,1)
				(4,1.6)--(5,1.6)
				(4,1.2)--(5,1.2)
				(4,.7)--(5,.7);
	\end{tikzpicture}
        \caption{}
    \end{subfigure}
	\caption{Brane configurations for the theory with linking numbers for the NS5-branes $\vec{l}_s=(0,0,1,2,2)$ and linking numbers for the D5-branes \mbox{$\vec{l}_d=(4,4,4,4,4)$}. (a) shows the Higgs branch and (b) depicts the Coulomb branch. In (a) the D5-branes are represented by vertical dashed lines and the NS5-branes by circled crosses. In (b) the D5-branes are represented by crosses and the NS5-branes by vertical solid lines. In both figures the D3-branes are represented by horizontal solid lines.}
	\label{fig:firstExampleSdual}
\end{figure}
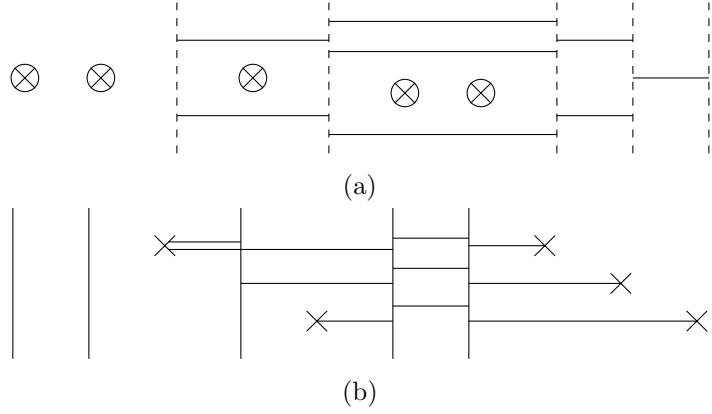

\subsubsection*{Global Symmetries and Linking Numbers}

From this example we see that a very interesting feature of the moduli space of the three-dimensional effective theories is already present in the linking numbers. This is the fact that the global symmetry of the Higgs branch, and the global symmetry of the Coulomb branch are both determined in the linking numbers of the fivebranes.\\

The general statement is as follows: for each integer number different from zero $i\in\mathbb{Z}^+$ that appears in $\vec{l}_d$ (resp. $\vec{l}_s$) there is a factor $U(r_i)$ in the global symmetry group of the Higgs branch (resp. Coulomb branch), where $r_i$ is the the number of times that $i$ appears in $\vec{l}_d$  (resp. $\vec{l}_s$). The global symmetry group of the Higgs branch (resp. Coulomb branch) is:

\begin{align}
	G_F=S\left((U(r_1)\times\dots\times U(r_N)\right)
\end{align}

where the $S(\dots)$ symbol denotes that an overall $U(1)$ factor decouples and is removed from the product.\\

In the current example with $\vec{l}_s=(4,4,4,4,4)$ and $\vec{l}_d=(0,0,1,2,2)$, the global symmetry of the Coulomb branch is $SU(5)$ (there is only one integer number $i=4$ different from zero in $\vec{l}_s$, and it appears five times, hence $r_4=5$). The global symmetry of the Higgs branch is $S(U(1)\times U(2))$, since the number $i=1$ appears once in $\vec{l}_d$ and the number $i=2$ appears twice, giving $r_1=1$ and $r_2=2$.\\

In the quiver, the global symmetry group of the Higgs branch corresponds to the flavor symmetry group of the effective gauge theory, represented in the square nodes. In fig. \ref{fig:firstExampleQuiver} we see two flavor nodes with ranks 1 and 2, corresponding to the global symmetry group in the Higgs branch $S(U(1)\times U(2))$. After performing mirror symmetry, the global symmetry group in the Coulomb branch becomes the flavor symmetry group of the mirror quiver. In fig. \ref{fig:firstExampleQuiverSdual}, the mirror quiver, we see one flavor node with rank 5, corresponding to the global symmetry group $SU(5)$ of the Coulomb branch before mirror symmetry is performed.

\subsection{One Parameter Family of Theories: $3d$ $\mathcal{N}=4$ SQED with N Flavours}

Let us take this section to focus in a very important set of models that play a crucial role in our following discussion of the Kraft-Procesi transitions. Actually the example with $\mathbb{C}^2/\mathbb{Z}_2$ Coulomb and Higgs branches is the first member of this family. As we said above, we can fully characterize a model by specifying the values of $n_s$, $n_d$, $\vec{l}_s$ and $\vec{l}_d$. Table \ref{tab:SUNU1} summarizes these conserved quantities for each element in the family.

\begin{table}[h]
	\centering		
	\begin{tabular}{ l l l }
	\toprule
	 &$\bm{n}$ &\multicolumn{1}{c}{$\vec{\bm{l}}$}\\ 
	\midrule 
	D5 & $N$ & $(1,1,\dots,1)$  \\
	NS5 & $2$ & $(1,N-1)$	\\
	\bottomrule
	\end{tabular}
	\caption{This table fully characterizes all elements of the family of theories with  $U(1)$ gauge group and $N$ flavours. This is a one parameter family, for each value of $N\in \mathbb{Z}^+$ there is a different model.}
	\label{tab:SUNU1}
\end{table}

\begin{figure}[h]
	\centering
	\begin{subfigure}[t]{.49\textwidth}
    \centering
	\begin{tikzpicture}
		\draw (1,0)--(1,2);
		\draw (4,0)--(4,2);
		\draw (1,.5)--(4,.5);
		\draw (2,1) node[cross] {};
		\draw (3,1) node[cross] {};
		\fill (2.5,1) circle [radius=1pt];
		\fill (2.40,1) circle [radius=1pt];
		\fill (2.60,1) circle [radius=1pt];
		\draw [decorate,decoration={brace,amplitude=5pt}](1.8,1.3) -- (3.2,1.3) node [black,midway,xshift=-0.6cm] { };
		\draw node at (2.5,1.7) {\footnotesize $N$};
	\end{tikzpicture}
        \caption{}
    \end{subfigure}
    \hfill
	\begin{subfigure}[t]{.49\textwidth}
    \centering
	\begin{tikzpicture}
	\tikzstyle{gauge} = [circle,draw];
	\tikzstyle{flavour} = [regular polygon,regular polygon sides=4,draw];
	\node (g1) [gauge,label=below:{$1$}] {};
	\node (f1) [flavour,above of=g1,label=above:{$N$}] {};
	\draw (g1)--(f1)
		;
	\end{tikzpicture}
        \caption{}
    \end{subfigure}
    \hfill
 	\caption{(a) Coulomb brane configuration for the model with  $n_s=2$, $n_d=N$, \mbox{$\vec{l}_s=(1,N-1)$} and \mbox{$\vec{l}_d=(1,1,\dots,1)$}. (b) Quiver obtained from brane configuration (a).}
	\label{fig:SUNU1}
\end{figure}
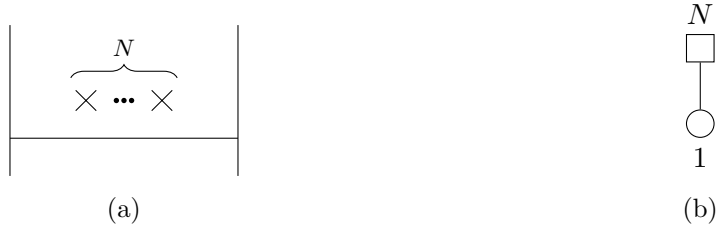

The Coulomb brane configuration for a generic member of this family is depicted in fig. \ref{fig:SUNU1} (a). From the quiver, fig. \ref{fig:SUNU1} (b), we see that we are dealing with the one parameter family of models with $U(1)$ gauge group and $N$ hypermultiplets that transform under an $SU(N)$ flavour symmetry.\\

The Coulomb brane configuration  only has one D3-brane that generates the moduli space. Hence its real dimension $1\times 4=4$. The first step in the analysis of the moduli space is to align all D5-branes positions $\vec{m}_i$, we will always consider this configuration, since it presents the maximum Higgsing. Actually in fig. \ref{fig:SUNU1} (a) the two NS5-branes are also aligned $\vec{w}_1=\vec{w}_2$. From now on we always consider $\vec{m}_1=\dots=\vec{m}_{n_d}$ and $\vec{w}_1=\dots=\vec{w}_{n_s}$. \\

Following the same reasoning as before we see that there is a singular point in the space of positions $\vec{x}$ of the D3-brane. This is when it coincides with the D5-branes:

\begin{align}
\vec{x}=\vec{m}_1=\dots=\vec{m}_N
\end{align}

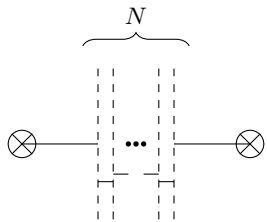
\begin{figure}[h]
	\centering
	\begin{tikzpicture}
		\draw[dashed] 	(2,0)--(2,2)
				(3,0)--(3,2)
				(2.2,0)--(2.2,2)
				(2.8,0)--(2.8,2);
		\draw 	(1,1) node[cross]{}
				(4,1) node[cross]{};
		\draw 	(1,1) node[circle,draw]{}
				(4,1) node[circle,draw]{};
		\draw (2,.5)--(2.2,.5)
				(2.8,.5)--(3,.5)
				(2.2,.6)--(2.4,.6)
				(2.6,.6)--(2.8,.6)
				(1,1)--(2,1)
				(3,1)--(4,1);
		\fill (2.5,1) circle [radius=1pt];
		\fill (2.40,1) circle [radius=1pt];
		\fill (2.60,1) circle [radius=1pt];
		\draw [decorate,decoration={brace,amplitude=5pt}](1.8,2.3) -- (3.2,2.3) node [black,midway,xshift=-0.6cm] { };
		\draw node at (2.5,2.7) {\footnotesize $N$};
	\end{tikzpicture}
	\caption{Higgs branch of the model with $n_s=2$, $n_d=N$, $\vec{l}_s=(1,N-1)$ and $\vec{l}_d=(1,1,\dots,1)$.}
	\label{fig:SUNU1Higgs}
\end{figure}

At this point there are $N$ hyper multiplets that become massless. Therefore this is the point where the Coulomb and the Higgs branches meet. Once in this point we can perform the splitting of the D3-brane to obtain the Higgs brane configuration. This phase of the moduli space is depicted in fig \ref{fig:SUNU1Higgs}. We see that now there are $N-1$ D3-branes that admit nonzero values of $(\vec{y}_i,b_i)$. This corresponds to a Higgs branch where $N-1$ hypermultiplets are massless. This is coherent with the fact that the vectorplet becomes massive, and \emph{eats} one out of the $N$ hypermultiplets. We find  that the Higgs branch has real dimension $(N-1)\times 4= 4N-4$.\\

As before, in this case a good candidate for the Coulomb branch would be the variety $\mathbb{C}^2/\Gamma$, with $\Gamma$ a finite subgroup of $SU(2)$. The final answer was computed by \cite{Seiberg1996} and is that $\Gamma$ is the cyclic group of order $N$, i.e. $\Gamma=\mathbb{Z}_N$. We write:

 \begin{align}
	\mathcal{M}_C=\mathbb{C}^2/\mathbb{Z}_N
\end{align}

The family of varieties $\mathbb{C}^2/\mathbb{Z}_N$ clearly generalizes the Coulomb branch of the self-dual model with gauge group $G=U(1)$ and 2 flavors, $\mathcal{M}_C=\mathbb{C}^2/\mathbb{Z}_2$. However, there is another way of generalizing this variety. As we saw before, $\mathbb{C}^2/\mathbb{Z}_2$ is also the closure of the minimal nilpotent orbit for the $\mathfrak{sl}_2$ algebra. This is the same as the algebra for the flavor group $SU(2)$. The flavor group acts on the hyper multiplets, so it is a symmetry of the Higgs branch. A generalization can be the Higgs branch being the closure of the minimal orbit of the algebra corresponding to the flavor group. For a generic element of the family the flavor group acting on the hyper multiplets would be $SU(N)$, so a candidate for the Higgs branch that generalizes the $SU(2)$ case would be the closure of the minimal nilpotent orbit of $\mathfrak{sl}_N$, corresponding to partition $\lambda=(2,1^{N-2})$. In the next section we show how the Higgs branch can be computed and how this is indeed the correct guess.\\

In the literature the variety $\mathbb{C}^2/\mathbb{Z}_N$ has been labeled as $A_{N-1}$, and the closure of the minimal nilpotent orbit of $\mathfrak{sl}_N$ as $a_{N-1}$. We write that for the model with N flavors:

\begin{align}
	\begin{aligned}
		\M_C&=A_{N-1}\\
		\M_H&=a_{N-1}	
	\end{aligned}
\end{align}

\subsubsection*{Mirror model}

Now we can ask for the mirror model, the one to with $\mathcal{M}_H=A_{N-1}$ and $\mathcal{M}_C=a_{N-1}$. This will be the model described by Table \ref{tab:SUNU1mirror} (we just need to swap the two rows of Table \ref{tab:SUNU1}).

\begin{table}[h]
	\centering
	\begin{tabular}{ l l l }
	\toprule
	 &$\bm{n}$ &\multicolumn{1}{c}{$\vec{\bm{l}}$}\\ 
	\midrule 
	D5 & $2$ & $(1,N-1)$	\\
	NS5 & $N$ & $(1,1,\dots,1)$  \\
	\bottomrule
	\end{tabular}
	\caption{This data characterizes the mirror model of the theory with gauge group $G=U(1)$ and $N$ flavours.}
	\label{tab:SUNU1mirror}
\end{table}

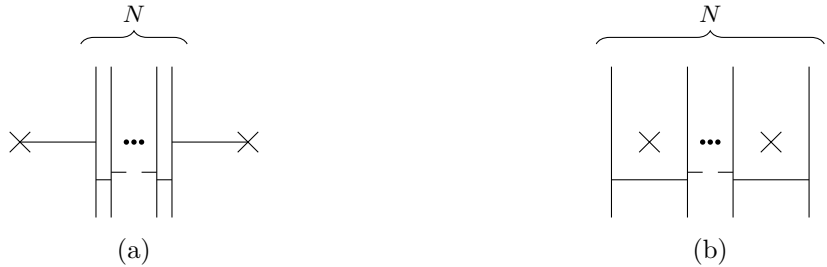
\begin{figure}[h]
	\centering
    \begin{subfigure}[t]{.49\textwidth}
    \centering
	\begin{tikzpicture}
		\draw	(2,0)--(2,2)
				(3,0)--(3,2)
				(2.2,0)--(2.2,2)
				(2.8,0)--(2.8,2);
		\draw 	(1,1) node[cross]{}
				(4,1) node[cross]{};
		\draw (2,.5)--(2.2,.5)
				(2.8,.5)--(3,.5)
				(2.2,.6)--(2.4,.6)
				(2.6,.6)--(2.8,.6)
				(1,1)--(2,1)
				(3,1)--(4,1);
		\fill (2.5,1) circle [radius=1pt];
		\fill (2.40,1) circle [radius=1pt];
		\fill (2.60,1) circle [radius=1pt];
		\draw [decorate,decoration={brace,amplitude=5pt}](1.8,2.3) -- (3.2,2.3) node [black,midway,xshift=-0.6cm] { };
		\draw node at (2.5,2.7) {\footnotesize $N$};
	\end{tikzpicture}
        \caption{}
    \end{subfigure}
    \hfill
    \begin{subfigure}[t]{.49\textwidth}
    \centering
	\begin{tikzpicture}
		\draw	(1.2,0)--(1.2,2)
				(3.8,0)--(3.8,2)
				(2.2,0)--(2.2,2)
				(2.8,0)--(2.8,2);
		\draw 	(1.7,1) node[cross]{}
				(3.3,1) node[cross]{};
		\draw (1.2,.5)--(2.2,.5)
				(2.8,.5)--(3.8,.5)
				(2.2,.6)--(2.4,.6)
				(2.6,.6)--(2.8,.6);
		\fill (2.5,1) circle [radius=1pt];
		\fill (2.40,1) circle [radius=1pt];
		\fill (2.60,1) circle [radius=1pt];
		\draw [decorate,decoration={brace,amplitude=5pt}](1,2.3) -- (4,2.3) node [black,midway,xshift=-0.6cm] { };
		\draw node at (2.5,2.7) {\footnotesize $N$};
	\end{tikzpicture}
        \caption{}
    \end{subfigure}
    \hfill
    \caption{Coulomb brane configuration for the model with  $n_s=N$, $n_d=2$, \mbox{$\vec{l}_s=(1,1,\dots,1)$} and \mbox{$\vec{l}_d=(1,N-1)$}. (a) is the Coulomb brane configuration obtain via mirror duality. (b) is the brane configuration without frozen D3-branes after performing two Hanany-Witten transitions, the quiver can be read more easily from this configuration.}
    \label{fig:SUNU1mirror}
\end{figure}

\begin{figure}[h]
	\centering
	\begin{tikzpicture}
	\tikzstyle{gauge} = [circle, draw];
	\tikzstyle{flavour} = [regular polygon,regular polygon sides=4, draw];
	\node (g1) [gauge,label=below:{$1$}] {};
	\node (g2) [gauge, right of=g1,label=below:{$1$}] {};
	\node (gd) [right of=g2] {$\dots$};
	\node (g3) [gauge, right of=gd,label=below:{$1$}] {};
	\node (g4) [gauge, right of=g3,label=below:{$1$}] {};
	\node (f1) [flavour,above of=g1,label=above:{$1$}] {};
	\node (f4) [flavour,above of=g4,label=above:{$1$}] {};
	\draw (g1)--(g2)
			(g3)--(g4)
			(g1)--(f1)
			(g4)--(f4)
		;
	\draw [decorate,decoration={brace,mirror,amplitude=5pt}](-.5,-.8) -- (4.5,-.8) node [black,midway,xshift=-0.6cm] { };
		\draw node at (2,-1.3) {\footnotesize $N$};
	
	\end{tikzpicture}
    \caption{Quiver for the model with  $n_s=N$, $n_d=2$, \mbox{$\vec{l}_s=(1,1,\dots,1)$} and \mbox{$\vec{l}_d=(1,N-1)$}. The brace indicates that there are $N$ gauge nodes with label $1$ in the sequence.}
    \label{fig:SUNU1mirrorQuiver}
\end{figure}
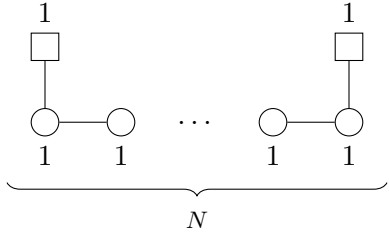

To obtain the Coulomb brane configuration  we take the Higgs brane configuration of the previous model and swap the D5-branes with NS5-branes and vice-versa. The result is given in fig. \ref{fig:SUNU1mirror} (a). To read the quiver we can perform two Hanany-Witten transitions to annihilate the frozen D3-branes: the result is given in fig. \ref{fig:SUNU1mirror} (b). The quiver can now be read, and is in fig. \ref{fig:SUNU1mirrorQuiver}.

\subsection{Higgs Branch Computation}\label{sec:HBC}

Let us take a pause for a moment to review how the Higgs branch computation can be performed for the one parameter family of two node quivers. This can make manifest the relation between the chiral ring and the ring of holomorphic functions in the affine variety. It also shows how the closure of the minimal nilpotent orbit arises as a natural variety in the context of moduli spaces. We follow the instructions by \cite{Benvenuti2010}.\\

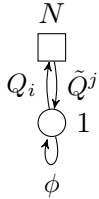
\begin{figure}[ht]
	\centering
	\begin{subfigure}[t]{.45\textwidth}
    \centering
    \begin{tikzpicture}[->,>=stealth',shorten >=0pt,shorten <=0pt,bend angle=10,auto]
	\tikzstyle{gauge} = [circle, draw];
	\tikzstyle{flavour} = [regular polygon,regular polygon sides=4,draw];
	\node (g1) [gauge,label=right:{$1$}] {};
	\node (f1) [flavour,above of=g1,label=above:{$N$}] {};
	\path[every node/.style={font=\sffamily\small}]
			(g1) edge [bend left]node {$Q_{i}$} (f1)
				edge [loop below] node{$\phi$} (g1)
			(f1) edge [bend left] node {$\tilde{Q}^j$} (g1);
	\end{tikzpicture}
    \end{subfigure}
 	\caption{Model with $n_s=2$, $n_d=N$, $\vec{l}_s=(1,N-1)$, $\vec{l}_d=(1,1,\dots,1)$. The figure represents the quiver where particles are shown as representations of the subalgebra generated only by 4 supercharges.}
	\label{fig:U1withNflavours}
\end{figure}

Let us focus on the set of quivers from the previous section, where we want to show that the Higgs branch corresponds to $\mathcal{M}_H=a_{N-1}$, fig. \ref{fig:SUNU1} (b). To compute the Hilbert series of the Higgs branch as in \cite{Benvenuti2010} we count chiral operators. These chiral operators correspond to holomorphic functions in the hyperk\"ahler variety. First we need to identify all scalar fields that admit nonzero VEV and that are contained in hyper multiplets. In order to do this we focus in the description of the model from the point of view of 4 supercharges. For each hyper multiplet with 8 supercharges there will be two chiral multiplets, they will be supersymmetric multiplets under the subalgebra generated only by 4 supercharges. In the quiver this can be realized in the following way: every edge turns into two directed edges, with opposite directions. For every vector multiplet in the 8 supercharges description there is a chiral multiplet and a vector multiplet under the 4 supercharges subalgebra. In the quiver this is realized: every gauge node turns into a gauge node with a directed loop. The 4 supercharges version quiver of the model is shown in fig. \ref{fig:U1withNflavours}.\\

In this case the $N$ hyper multiplets split into $N$ chiral multiplets, with $N$ complex scalars $Q_i$ that transform under the fundamental representation of $SU(N)$, $[1,0,\dots,0]$, and have charge $-1$ under the gauge group $U(1)$, and $N$ chiral multiplets, with $N$ complex scalars $\tilde{Q}^j$, that transform under the antifundamental representation of $SU(N)$, $[0,\dots,0,1]$, and have charge $1$ under the gauge group $U(1)$. There is also a complex scalar in the vectorplet, $\phi$, that transforms under the adjoint representation of $U(1)$, i.e. the singlet of charge $0$.\\

To obtain the Higgs branch of the theory we need to find all operators made with the scalar fields $Q_i$ and $\tilde{Q}^j$ that are gauge invariant and satisfy the zero energy condition. The simplest combination of fields that is gauge invariant would be the combination of a $Q_i$ and a $\tilde{Q}^j$, we can see that the set of all operators of this type generates the rest of all other invariant operators. We denote them with:

\begin{align}
	M_i^j:=Q_i\tilde{Q}^j
\end{align}

$M_i^j$ generates a ring of operators. We can think of this ring as the set of all $N\times N$ complex matrices with rank smaller or equal to 1:

\begin{align}
\{M_{N\times N}|M_i^j\in \mathbb{C},\ rank(M)\leq 1\}
\end{align}
Now the zero energy condition will impose some relations in this ring, giving a quotient ring called \emph{the chiral ring}. Let us see how this takes place.\\

Let $W$ be the superpotential function over the chiral superfields that correspond to each chiral multiplet. Let $W(\phi_A)$ be the same function but this time evaluated in the complex scalar fields $\phi_A:=\{Q_i,\tilde{Q}^j,\phi\}$ that correspond to each chiral multiplet. The zero energy condition is then 

\begin{align}
	\frac{\partial W}{\partial \phi_A}&=0,\ \ \forall A
\end{align}

 Following the prescription in \cite{Benvenuti2010}, in this model the superpotential is 

\begin{align}
	W=\phi\sum_iQ_i\tilde{Q}^i
\end{align} 

The only relevant condition that can be applied to the chiral ring is the one derived from taking partial derivative with respect to $\phi$, this gives:

\begin{align}
	\sum_iQ_i\tilde{Q}^i=0 
\end{align} 

A consequence is: 

\begin{align}
	Tr(M)=0 
\end{align} 

Another consequence is:

\begin{align}\label{eq:nilp}
	M^2=0 
\end{align} 

Therefore the Higgs branch can be characterized as: 
\begin{align}
	\{M_{N\times N}|M_i^j\in \mathbb{C},\ M^2=0,\ Tr(M)=0,\ rank(M)\leq 1\}
\end{align}

The Jordan condition is satisfied:

\begin{align}
	Tr(M^p)=0 \ \ \forall p\in \mathbb{N} \Leftrightarrow \text{all eigenvalues of $M$ are 0}
\end{align} 

M can also be thought as endomorphisms on $\mathbb{C}^N$. In this way the matrices can be identified as nilpotent elements of the adjoint representation of $\mathfrak{sl}_N$. Since all nilpotent elements are classified into nilpotent orbits, we can use their Jordan normal form $X_\lambda$. We see that the only Jordan normal matrix $X_\lambda$ that fulfills 

\begin{align}
	\text{rank}(X_\lambda)\leq 1
\end{align}

is for the trivial partition $\lambda=(1^N)$ and for the minimal partition $\lambda'=(2,1^{N-2})$:

\begin{align}
X_{\lambda'}=\left(\begin{array}{cccccc}
		0&1&0&\dots&0&0\\
		0&0&0&\dots&0&0\\
		\vdots&\vdots&\vdots&\ddots&\vdots&\vdots\\
		0&0&0&\dots&0&0\\
		0&0&0&\dots&0&0\\
		\end{array}\right)
\end{align}

Due to the properties of the trace, the rank and the power of matrices, any element in the nilpotent orbits $\mathcal{O}_{\lambda}=\{S\cdot X_\lambda \cdot S^{-1}|S\in SL(N)\}$ for $\lambda$ trivial and minimal will fulfill the conditions to be part of the Higgs branch. Hence we can write all elements in the Higgs branch as:

\begin{align}
\begin{aligned}
\mathcal{M}_H&=\mathcal{O}_{(1^N)}\cup\mathcal{O}_{(2,1^{N-2})}\\
		&=\bar{\mathcal{O}}_{(2,1^{N-2})}
\end{aligned}
\end{align}\\


\subsection{Brane Systems for Closures of Nilpotent Orbits of $\mathfrak{sl}_N$}

To conclude the present section we introduce the brane models that give rise to $3d$ quiver gauge theories whose Higgs branch is $\Or_{\lambda}$, the closure of a nilpotent orbit of $\mathfrak{sl}_N$ corresponding to partition $\lambda\in\mathcal{P}(N)$. The conserved quantities that define such models are defined as follows. \\



For a given algebra $\mathfrak{sl}_N$, the number of fivebranes of each type is $n_s=n_d=N$. The linking numbers of the D5-branes are all the same, $N-1$. The linking numbers of the NS fivebranes depend on the partition $\lambda\in\mathcal{P}(N)$. In particular, each of the different NS5-branes has a linking number corresponding to the different parts in $\lambda^t$. $\lambda^t$ is the \emph{transpose partition} of $\lambda$.\\

To obtain the transpose partition $\lambda^t$ in its \emph{exponential notation}:

\begin{align}
\lambda^t=(1^{m_1},2^{m_2},\dots,N^{m_N})
\end{align}

where $m_1$ is the number of parts in $\lambda^t$ that are equal to $1$, $m_2$ the number of parts that are equal to $2$, etc., we use the definition:

\begin{align}
	m_i:=\lambda_{i}-\lambda_{i+1}
\end{align}

where $\lambda_i$ are the different parts of  $\lambda=(\lambda_1,\lambda_2,\dots,\lambda_k)$.\\

For example, if $\lambda =(2,1,1)$, we have:

\begin{align}
	\begin{aligned}
		m_1&=2-1=1\\
		m_2&=1-1=0\\
		m_3&=1-0=1
	\end{aligned}
\end{align}

Therefore

\begin{align}
	\begin{aligned}
		\lambda^t&=(3^1,2^0,1^1)\\
		&=(3,1)
	\end{aligned}
\end{align}

To obtain the linking numbers for the NS5-branes we pad $\lambda^t$ with zeroes until it contains $N$ parts. Then, we invert the order of the parts so the order in $\vec{l}_s$ corresponds to linking numbers monotonically increasing from left to right. For example, if $\lambda=(2,1^2)$, then $\lambda^t=(3,1)$ and the linking numbers are $\vec{l}_s=(0,0,1,3).$ This description is summarized in Table \ref{tab:Tsigma}.\\

\begin{table}[ht]
	\centering
	\begin{tabular}{ l l l }
	\toprule
	 &$\bm{n}$ &\multicolumn{1}{c}{$\vec{\bm{l}}$}\\ 
	\midrule 
	D5 & $N$ & $(N-1,N-1,\dots,N-1)$	\\
	NS5 & $N$ & $\lambda^t$  \\
	\bottomrule
	\end{tabular}
	\caption{This table fully characterizes all elements of the family of theories with Higgs branch the closure of the nilpotent orbit denoted by partition $\lambda$ of the algebra $\mathfrak{sl}_N$. $\vec{l}_s$ should be the transpose partition of $\lambda$, padded with zeroes; we also want to invert the order of the parts so the order of the array corresponds to the linking numbers of the fivebranes ordered from left to right. For example, if $N=4$ and $\lambda=(2,1^2)$, then $\lambda^t=(3,1)$ and the linking numbers are $\vec{l}_s=(0,0,1,3)$.}
	\label{tab:Tsigma}
\end{table}

To obtain a $3d$ effective gauge theory with the Coulomb branch being the closure of a nilpotent orbit $\Or_{\lambda}\subset \mathfrak{sl}_N$ for any $\lambda\in\mathcal{P}(N)$ we can perform a mirror symmetry. The result is a model with linking numbers $\vec{l}_s=(N-1,\dots,N-1)$ and $\vec{l}_d$ equal to $\lambda^t$.\\

The $3d\ \mathcal{N}=4$ low energy effective theories that are obtained with these brane constructions are known in the literature with the name $T_{\lambda^t} (SU(N))$, for linking numbers $\vec{l}_d=(N-1,\dots,N-1)$ and $\vec{l}_s$ equal to $\lambda^t$. The mirror theories are denoted by $T^{\lambda^t} (SU(N))$. They belong to a bigger family of theories,  $T_\rho^{\sigma} (SU(N))$, where $\rho,\sigma \in \mathcal{P}(N)$. For more details the reader is directed to \cite{Gaiotto2008,Chacaltana2013}

\subsubsection{Example: Closure of the Maximal Orbit of $\mathfrak{sl}_3$ as Higgs Branch}\label{sec:SU3max}

In the literature the \emph{maximal nilpotent orbit} is the one whose closure has the highest dimension. In the case of $\mathfrak{sl}_N$ this corresponds to the orbit with partition $\lambda=(N)$. This is a very special case since the closure of this orbit is the union of all nilpotent orbits. Therefore all closures of all nilpotent orbits are contained within this variety. Sometimes this variety is also referred to as the \emph{nilpotent cone}. The nilpotent orbit is also referred to as the \emph{regular orbit}.\\

The maximal orbit of $\mathfrak{sl}_3$ corresponds to partition $\lambda=(3)$.   To obtain the transpose partition we realize that the only part of $\lambda$ different from zero is $\lambda_1=3$, therefore 
\begin{align}
\begin{aligned}
m_1&=3-0=3\\ 
m_i&=0 \ \ \ \ \ \  \text{for}\  i=2,3,\dots
\end{aligned}
\end{align}

Hence, 

\begin{align}
\lambda^t=(1^3)
\end{align}

The model then is defined completely by Table \ref{tab:SU3max}. We can then draw the Coulomb brane configuration and read the quiver. Let us do it carefully step by step.

\begin{table}[h]
	\centering
	\begin{tabular}{ l l l }
	\toprule
	 &$\bm{n}$ &\multicolumn{1}{c}{$\vec{\bm{l}}$}\\ 
	\midrule 
	D5 & $3$ & $(2,2,2)$	\\
	NS5 & $3$ & $(1,1,1)$  \\
	\bottomrule
	\end{tabular}
	\caption{This table shows the data for the theory with Higgs branch as the closure of the maximal nilpotent orbit of $\mathfrak{sl}_3$.}
	\label{tab:SU3max}
\end{table}

\begin{figure}[h]
	\centering
	\begin{subfigure}[t]{.49\textwidth}
    \centering
	\begin{tikzpicture}
		\draw (1.5,0)--(1.5,2);
		\draw (2,0)--(2,2);
		\draw (4,0)--(4,2);
		\draw (2,.4)--(4,.4)
				(2,.6)--(4,.6)
				(1.5,.5)--(2,.5);
		\draw (2.5,1) node[cross] {};
		\draw (3,1) node[cross] {};
		\draw (3.5,1) node[cross] {};
	\end{tikzpicture}

        \caption{}
    \end{subfigure}
    \hfill
	\begin{subfigure}[t]{.49\textwidth}
    \centering
	\begin{tikzpicture}
	\tikzstyle{gauge} = [circle, draw];
	\tikzstyle{flavour} = [regular polygon,regular polygon sides=4,draw];
	\node (g1) [gauge,label=below:{$1$}] {};
	\node (g2) [gauge,right of=g1,label=below:{$2$}] {};
	\node (f2) [flavour,above of=g2,label=above:{$3$}] {};
	\draw (g2)--(f2)
			(g1)--(g2)
		;
	\end{tikzpicture}
        \caption{}
    \end{subfigure}
    \hfill
 	\caption{Model with $n_s=n_d=3$, $\vec{l}_s=(1,1,1)$ and $\vec{l}_d=(2,2,2)$. (a) is the Coulomb brane configuration. (b) is the quiver.}
	\label{fig:SU3maximal}
\end{figure}
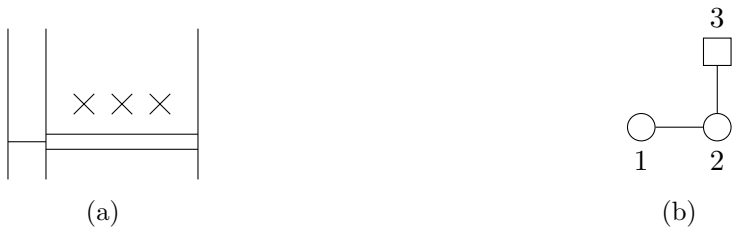

In order to obtain the Coulomb brane configuration and read the quiver of the model we want the D3-branes to stretch only between NS5-branes. This means that the $\vec{l}_d$ only counts number of NS5-branes to the left of each D5-brane. Since $\vec{l}_d=(2,2,2)$ all D5-branes are placed in the interval between the second and the third NS5-branes, starting from the left, see fig. \ref{fig:SU3maximal} (a). In the general case, since $\vec{l}_d=(N-1,N-1,\dots, N-1)$, all D5-branes are placed between the two rightmost NS5-branes. Now starting from the left we add D3-branes between neighboring NS5-branes to ensure that $\vec{l}_s$ is realized.\\

In this example, we start with the first leftmost NS fivebrane, since its linking number is $1$, a unique D3-brane is required to stretch between the NS5-brane and its neighbor to the right. The second NS5-brane, the one in the middle, needs $2$ D3-branes to be added to its right, to obtain linking number 1. We can check that the last NS5-brane already has linking number $3-2=1$, so the Coulomb brane configurationlooks like fig. \ref{fig:SU3maximal} (a). The quiver can be read from it as usual and is depicted in fig. \ref{fig:SU3maximal} (b).\\

\subsubsection*{Higgs Brane Configuration}

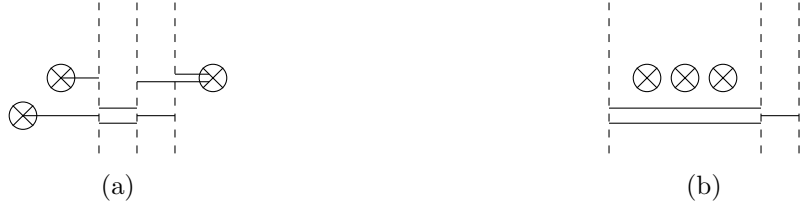
\begin{figure}[ht]
	\centering
	\begin{subfigure}[t]{.49\textwidth}
		\centering
		\begin{tikzpicture}
		\draw[dashed] (2.5,0)--(2.5,2)
			(3,0)--(3,2)
			(3.5,0)--(3.5,2);
		\draw (1.5,.5) node[cross] {};
		\draw (2,1) node[cross] {};
		\draw (4,1) node[cross] {};
		\draw (1.5,.5) node[circle,draw] {};
		\draw (2,1) node[circle,draw] {};
		\draw (4,1) node[circle,draw] {};
		\draw 	(1.5,.5)--(2.5,.5)
				(2,1)--(2.5,1)
				(3.95,1.05)--(3.5,1.05)
				(3.95,.95)--(3,.95);
		\draw 	(2.5,.6)--(3,.6)
				(2.5,.4)--(3,.4)
				(3,.5)--(3.5,.5);
	\end{tikzpicture}
        \caption{}
    \end{subfigure}
	\hfill
	\begin{subfigure}[t]{.49\textwidth}
		\centering
		\begin{tikzpicture}
		\draw[dashed] (1.5,0)--(1.5,2)
			(3.5,0)--(3.5,2)
			(4,0)--(4,2);
		\draw (2,1) node[cross] {};
		\draw (2.5,1) node[cross] {};
		\draw (3,1) node[cross] {};
		\draw (2,1) node[circle,draw] {};
		\draw (2.5,1) node[circle,draw] {};
		\draw (3,1) node[circle,draw] {};
		\draw 	(1.5,.6)--(3.5,.6)
				(1.5,.4)--(3.5,.4)
				(3.5,.5)--(4,.5);
	\end{tikzpicture}
        \caption{}
    \end{subfigure}
 	\caption{Model with $n_s=n_d=3$, $\vec{l}_s=(1,1,1)$ and $\vec{l}_d=(2,2,2)$. (a) represents the Higgs branch as obtained by aligning the D3-branes in the Coulomb branch with the D5-branes and then proceding to a maximal splitting. (b) is the same model after a phase transtition where all fixed D3-branes have been annihilated. In this phase, the self-duality: $\mathcal{M}_C=\mathcal{M}_H$ becomes manifest.}
	\label{fig:SU3maxHiggs}
\end{figure}

A phase transition to the Higgs brane configuration can be performed. As before, we align the D3-branes with the D5-branes, and then do a maximal splitting of the D3-branes (we split them in the most general way). All the resulting split D3-branes should be either fixed between a NS5-brane and a D5-brane or freely moving along their $\vec{y}_i$ directions. Finally we could perform some Hanany-Witten transitions to get rid of the fixed threebranes. The result right after the splitting is given in  fig. \ref{fig:SU3maxHiggs} (a).  In fig. \ref{fig:SU3maxHiggs} (b) we have annihilated the fixed threebranes via Hanany-Witten transitions.\\

We can see from fig. \ref{fig:SU3maxHiggs} that the Higgs branch is a variety with three quaternionic dimensions, i.e. $3 \times 4= 12$ real dimensions, where $3$ is the number of threebranes that generate the moduli. This corresponds to the dimensions of the closure of the maximal orbit of $\mathfrak{sl}_3$. In fact the Hilbert series of the Higgs branch for this quiver has been computed recently by \cite{Hanany2016a}. As we explained before, this variety is:

\begin{align}
	\bar{\mathcal{O}}_{(3)}=\mathcal{O}_{(3)} \cup \mathcal{O}_{(2,1)} \cup \mathcal{O}_{(1^3)}
\end{align}
 
Where $\mathcal{O}_\lambda=PSL(3)\cdot X_\lambda$ are the orbits generated by all the Jordan normal matrices:

\begin{align}
X_{(3)}&=\left(\begin{array}{ccc}
		0&1&0\\
		0&0&1\\
		0&0&0\\
		\end{array}\right) &
X_{(2,1)}&=\left(\begin{array}{ccc}
		0&1&0\\
		0&0&0\\
		0&0&0\\
		\end{array}\right) &
X_{(1^3)}&=\left(\begin{array}{ccc}
		0&0&0\\
		0&0&0\\
		0&0&0\\
		\end{array}\right)
\end{align}

Therefore the Higgs branch $\mathcal{M}_H=\bar{\mathcal{O}}_{(3)}$ can be thought as the set of all possible $3\times 3$ Jordan matrices with complex entries. This is also the set of all possible $3 \times 3$ matrices with zero eigenvalues, or the set of all possible $3 \times 3$ matrices, belonging to the adjoint representation of $\mathfrak{sl}_3$, with zero Casimir invariants:

\begin{align}
		\mathcal{M}_H = \{M_{3\times 3}| \text{ all Casimir invariants }\kappa_i=0\}
\end{align}

In the general case of the closure of the maximal orbit of $\mathfrak{sl}_N$, the quiver in fig. \ref{fig:SU3maximal} (b) generalizes to a quiver  with gauge group:

\begin{align}
G=U(1)\times U(2)\times\dots\times U(N-1)
\end{align}

and a single $SU(N)$ flavor node  connected to the $U(N-1)$ gauge node. These quivers and their Higgs branches where already known by \cite{Kraft1979}.  In recent years, physicists have recovered them under the name of $T(SU(N))$ theories \cite{Gaiotto2008}, thanks to the work of \cite{kronheimer1990}.

\subsubsection*{Mirror model}

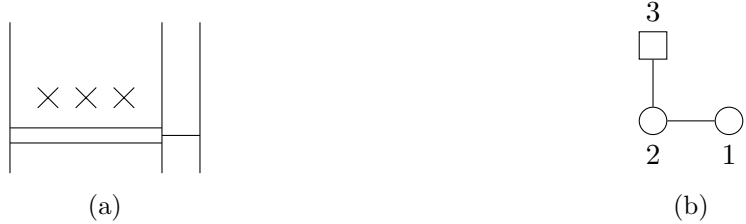
\begin{figure}[ht]
	\centering
	\begin{subfigure}[t]{.49\textwidth}
		\centering
		\begin{tikzpicture}
		\draw (1.5,0)--(1.5,2)
			(3.5,0)--(3.5,2)
			(4,0)--(4,2);
		\draw (2,1) node[cross] {};
		\draw (2.5,1) node[cross] {};
		\draw (3,1) node[cross] {};
		\draw 	(1.5,.6)--(3.5,.6)
				(1.5,.4)--(3.5,.4)
				(3.5,.5)--(4,.5);
	\end{tikzpicture}
        \caption{}
    \end{subfigure}
	\hfill
	\begin{subfigure}[t]{.49\textwidth}
    \centering
	\begin{tikzpicture}
	\tikzstyle{gauge} = [circle, draw];
	\tikzstyle{flavour} = [regular polygon,regular polygon sides=4,draw];
	\node (g1) [gauge,label=below:{$1$}] {};
	\node (g2) [gauge,left of=g1,label=below:{$2$}] {};
	\node (f2) [flavour,above of=g2,label=above:{$3$}] {};
	\draw (g2)--(f2)
			(g1)--(g2)
		;
	\end{tikzpicture}
        \caption{}
    \end{subfigure}
 	\caption{Model with $n_s=n_d=3$, $\vec{l}_s=(2,2,2)$ and $\vec{l}_d=(1,1,1)$. (a) is the Coulomb brane configuration, obtained via mirror duality from the Higgs branch of the dual model. (b) is the quiver as read from (a). }
	\label{fig:SU3maxMirror}
\end{figure}

Let us compute the mirror duality. After performing the S-duality in the Higgs brane configuration we obtain fig. \ref{fig:SU3maxMirror} (a). This gives the quiver in fig. \ref{fig:SU3maxMirror} (b). Now we see that this is exactly the same model we started with. Therefore we say that the model  is \emph{self-mirror} and the Coulomb branch and Higgs branch have the same geometry:

\begin{align}
		\mathcal{M}_C = \mathcal{M}_H = \bar{\mathcal{O}}_{(3)}
\end{align}

This is once again a different generalization of the model with $\mathcal{M}_C = \mathcal{M}_H = \mathbb{C}^2/\mathbb{Z}_2.$ This was the $A_1$ variety and also the $a_1$, but it is also corresponds to the closure of the maximal nilpotent orbit of $\mathfrak{sl}_2$. In general, the family of models whose Higgs branch corresponds to the closure of the maximal nilpotent orbit of $\mathfrak{sl}_N$ is self-dual, and their Coulomb branch is the same variety.\\

\section{The Kraft-Procesi Transition}\label{sec:K}

We are ready to introduce the main novelty of this paper: the \emph{Kraft-Procesi transitions}. This physical process can be understood as a transition between different models. This gives a structure for families of quiver gauge theories associated to different closures of nilpotent orbits of the same algebra, via their moduli spaces. In this section we will discuss the brane dynamics that characterize such structure and such transitions.

\subsection{Example: $\mathfrak{sl}_3$ Transitions, Maximal to Minimal}

Let us start directly by showing an example of the Higgs mechanism that produces a Kraft-Procesi transition. We start by considering the model introduced in Section \ref{sec:SU3max}. This is a self-dual model with $\mathcal{M}_H=\mathcal{M}_C=\bar{\mathcal{O}}_{(3)}$. It is defined by the linking numbers $\vec{l}_d=(2,2,2)$ and $\vec{l}_s=(1,1,1)$. The Higgs brane configuration of the model is depicted in fig. \ref{fig:SU3maxHiggs} (b).\\

Let us start the transition to a new model by \emph{Higgsing away minimal singularities} (these are the minimal singularities found in  \cite{Brieskorn1970}). To find the minimal singularity we focus on only one of the D3-branes that can be \emph{Higgsed away} and study what is the moduli generated by it: this will be the minimal singularity, a singular subvariety of the Higgs branch.
By a D3-brane that can be \emph{Higgsed away} we mean a threebrane that can align at least with two NS5-branes, generating a massless vectorplet that admits nonzero vacuum expectation values $(\vec{x},a)$. In fig. \ref{fig:SU3maxHiggs} (b), the two leftmost D3-branes fulfill this condition.\\

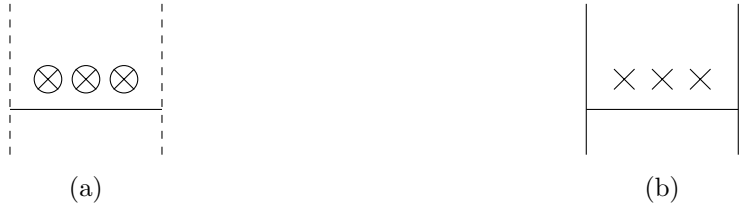
\begin{figure}[ht]
	\centering
	\begin{subfigure}[t]{.49\textwidth}
		\centering
		\begin{tikzpicture}
		\draw[dashed] (1.5,0)--(1.5,2)
			(3.5,0)--(3.5,2);
		\draw (2,1) node[cross] {};
		\draw (2.5,1) node[cross] {};
		\draw (3,1) node[cross] {};
		\draw (2,1) node[circle,draw] {};
		\draw (2.5,1) node[circle,draw] {};
		\draw (3,1) node[circle,draw] {};
		\draw 	(1.5,.6)--(3.5,.6);
	\end{tikzpicture}
        \caption{}
    \end{subfigure}
    \hfill
	\begin{subfigure}[t]{.49\textwidth}
		\centering
		\begin{tikzpicture}
		\draw (1.5,0)--(1.5,2)
			(3.5,0)--(3.5,2);
		\draw (2,1) node[cross] {};
		\draw (2.5,1) node[cross] {};
		\draw (3,1) node[cross] {};
		\draw 	(1.5,.6)--(3.5,.6);
	\end{tikzpicture}
        \caption{}
    \end{subfigure}
    \hfill
 	\caption{(a) Moduli generated by one of the leftmost D3-branes in the Higgs brane configuration depicted in fig. \ref{fig:SU3maxHiggs} (b). (b) is the S-dual moduli. It corresponds to the Coulomb brane configuration for the quiver with one $U(1)$ gauge node and one $SU(3)$ flavour node. We have already mentioned that the Coulomb branch for this quiver is the singularity $A_2$.}
	\label{fig:SU3singA2}
\end{figure}

Hence, we focus on one of the leftmost D3-branes in the Higgs brane configuration and \emph{freeze} the other two, they are now \emph{spectators}. The spectator threebranes can be anywhere and still observe the same transition. Mathematically, the single D3-brane is called a \emph{transverse slice}, as its motion does not affect the behavior of the remaining D3-branes. The moduli space generated by the single D3-brane can easily be found, it is depicted in fig. \ref{fig:SU3singA2} (a). We can see that it is the same variety as the Coulomb branch of a model with gauge group $G=U(1)$ and three flavors, therefore is 
\begin{align}
A_2:=\mathbb{C}^2/\mathbb{Z}_3
\end{align}

This constitutes an explicit construction that shows how $A_2\subset \bar{\mathcal{O}}_{(3)}$ is a subvariety of the closure of the maximal nilpotent orbit of $\mathfrak{sl}_3$. This is part of the more general result by Brieskorn \cite{Brieskorn1970}, in which there is a minimal singular subvariety $A_{N-1}\subset \bar{\mathcal{O}}_{(N)}$ for the algebra $\mathfrak{sl}_N$.\\

Let us see the Higgs mechanism that removes the minimal singularity $A_2$ from the Higgs branch. As usual we go to the singular point, where the $\vec{y}$ position of the D3-brane coincides with the positions of the 3 NS5-branes:

\begin{align}
\vec{y}=\vec{w}_1=\vec{w}_2=\vec{w}_3
\end{align}

Then we split the D3-brane. A maximal splitting splits the brane in 4 segments: the leftmost and rightmost segments are fixed, since they have one end in a NS5-brane and the other end in a D5-brane. The two intermediate segments can now move freely along their $\vec{x}_i$ directions, giving nonzero vacuum expectation value to two different massless vector multiplets: $(\vec{x}_1,a_1)$ and $(\vec{x}_2,a_2)$. So far, this is just a phase transition to a mixed phase of the model with some operators in the Higgs branch and some operators in the Coulomb branch.\\

\begin{figure}[h]
	\centering
	\begin{subfigure}[t]{.3\textwidth}
    \centering
	\begin{tikzpicture}
		\draw[dashed] (-1.5,0)--(-1.5,2);
		\draw[dashed] (-2,0)--(-2,2);
		\draw[dashed] (-4,0)--(-4,2);
		\draw (-2,.4)--(-4,.4)
				(-2,.6)--(-4,.6)
				(-1.5,.5)--(-2,.5);
		\draw (-2.5,1) node[cross] {};
		\draw (-3,1) node[cross] {};
		\draw (-3.5,1) node[cross] {};
		\draw (-2.5,1) node[circle,draw] {};
		\draw (-3,1) node[circle,draw] {};
		\draw (-3.5,1) node[circle,draw] {};
	\end{tikzpicture}
	\caption{}
 	\end{subfigure}
 	\hfill
 	\begin{subfigure}[t]{.3\textwidth}
    \centering
	\begin{tikzpicture}
		\draw[dashed] (-1.5,0)--(-1.5,2);
		\draw[dashed] (-2,0)--(-2,2);
		\draw[dashed] (-4,0)--(-4,2);
		\draw (-2,.4)--(-4,.4)
				(-2,1)--(-4,1)
				(-1.5,.5)--(-2,.5);
		\draw (-2.5,1) node[cross] {};
		\draw (-3,1) node[cross] {};
		\draw (-3.5,1) node[cross] {};
		\draw (-2.5,1) node[circle,draw] {};
		\draw (-3,1) node[circle,draw] {};
		\draw (-3.5,1) node[circle,draw] {};
	\end{tikzpicture}
	\caption{}
 	\end{subfigure}
 	\hfill
 	\begin{subfigure}[t]{.3\textwidth}
    \centering
	\begin{tikzpicture}
		\draw[dashed] (-1.5,0)--(-1.5,2);
		\draw[dashed] (-2,0)--(-2,2);
		\draw[dashed] (-4,0)--(-4,2);
		\draw (-2,.4)--(-4,.4)
				(-2,1)--(-2.5,1)
				(-3.5,1)--(-4,1)
				(-1.5,.5)--(-2,.5);
		\draw (-2.5,1) node[cross] {};
		\draw (-3,1) node[cross] {};
		\draw (-3.5,1) node[cross] {};
		\draw (-2.5,1) node[circle,draw] {};
		\draw (-3,1) node[circle,draw] {};
		\draw (-3.5,1) node[circle,draw] {};
	\end{tikzpicture}
	\caption{}
 	\end{subfigure}
 	\caption{$A_2$ Kraft-Procesi transition. (a) represents the Higgs branch of the model with $n_s=n_d=3$, $\vec{l}_s=(1,1,1)$ and $\vec{l}_d=(2,2,2)$. (b) represents the singular point when $\vec{y}=\vec{w}_1=\vec{w}_2=\vec{w}_3$. (c) represents the new model after taking $\vec{x}_1$ and $\vec{x}_2$ to infinity. The new linking numbers are $\vec{l}_s=(0,1,2)$, $\vec{l}_d=(2,2,2)$. This defines a new model which fulfils the prescription to have the Higgs branch as the closure of the nilpotent orbit with $\lambda^t=(2,1)$, hence $\lambda=(2,1)$. This is the minimal orbit of $\mathfrak{sl}_3$.}
	\label{fig:SU3KPA3}
\end{figure}
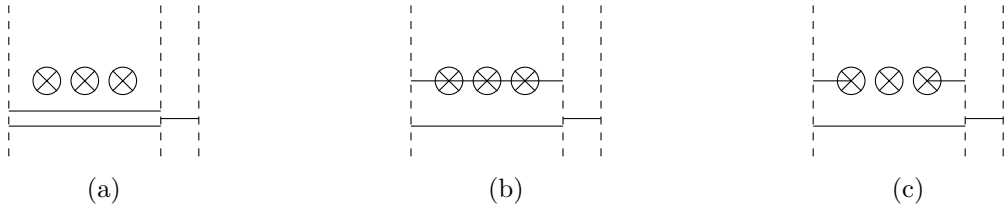

We can consider the remaining threebranes that are still in the Higgs branch and ask what is the variety that they generate. We can transition to the model to which this is just a pure Higgs branch, by completely removing the degrees of freedom in the mixed Coulomb branch. By this we mean that we take the segments of the split D3-branes that propagate along the $\vec{x}$ direction to infinity. Physically, this is equivalent to fixing a scale for the scalar fields, such that all their nonzero VEVs consist of a dimensionless number multiplying this scale. Then, we can flow the number of the VEVs in the mixed Coulomb branch along the RG flow to the infrared, taking their values to infinity, while keeping all other numbers for the VEVs in the mixed Higgs branch at order one. We consider that these D3-brane segments that were in the mixed Coulomb branch do not belong to the model any more, and we study what is the model whose Higgs brane configuration is the result of this process. After removing the two D3-brane segments, the linking numbers of the fivebranes have changed, and therefore we have a different model from the one we started with. The whole process is described in fig. \ref{fig:SU3KPA3}.\\

	
The resulting model has a Higgs brane configuration as in fig. \ref{fig:SU3KPA3} (c). We can see that the linking numbers for the D5-branes are still the same, $\vec{l}_d=(2,2,2)$ but the ones for the NS5-branes have changed. The first NS5-brane from the left now has linking number of 0, since there is one D5-brane to its left, and there is a total of 1 D3-brane ending from its left: $1-1=0$. The second NS5-brane from the left has linking number 1, since there is a D5-brane to its left and no D3-branes ending on it. The third NS5-brane has linking number of 2, since there is one D5-brane to its left and 1 D3-brane ending on it from the right: $1+1=2$. Hence we write 
\begin{align}
\vec{l}_s=(0,1,2)
\end{align}

We see that this still falls into the family of models presented in the previous section for which the Higgs branch is the closure of a nilpotent orbit $\mathcal{O}_\lambda \subset \mathfrak{sl}_3$. In this case $\lambda^t=\vec{l}_s$, this means that each part $\lambda^t_i$ of $\lambda^t$, will correspond to a linking number in $\vec{l}_s$, since the parts equal to 0 can be neglected we have $\lambda^t=(2,1)$. Therefore $\lambda=(2,1)$. This is the minimal orbit of $\mathfrak{sl}_3$.

\subsubsection*{Minimal Orbit of $\mathfrak{sl}_3$: $3d\ \mathcal{N}=4$ SQED with 3 Flavours}

Let us examine the result of our first Kraft-Procesi transition. The linking numbers of the models are:

\begin{align}
\vec{l}_d&=(2,2,2) \\
\vec{l}_s&=(0,1,2)
\end{align}

The Higgs branch is the closure of the minimal nilpotent orbit of $\mathfrak{sl}_3$:

\begin{align}
\mathcal{M}_H=a_2
\end{align}

We know that the model with gauge group $G=U(1)$ and $3$ flavors has exactly this Higgs branch and also 

\begin{align}
\mathcal{M}_C=A_2
\end{align}

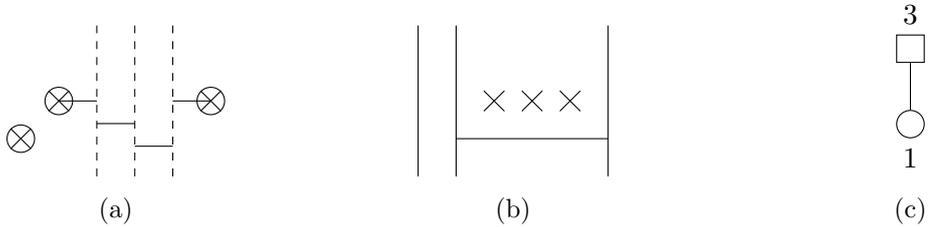
\begin{figure}[h]
	\centering
    \begin{subfigure}[t]{.3\textwidth}
    \centering
	\begin{tikzpicture}
		\draw[dashed] (2.5,0)--(2.5,2)
			(3,0)--(3,2)
			(3.5,0)--(3.5,2);
		\draw (1.5,.5) node[cross] {};
		\draw (2,1) node[cross] {};
		\draw (4,1) node[cross] {};
		\draw (1.5,.5) node[circle,draw] {};
		\draw (2,1) node[circle,draw] {};
		\draw (4,1) node[circle,draw] {};
		\draw 	(2,1)--(2.5,1)
				(4,1)--(3.5,1);
		\draw 	(2.5,.7)--(3,.7)
				(3,.4)--(3.5,.4);
	\end{tikzpicture}
        \caption{}
    \end{subfigure}
    \hfill
	\begin{subfigure}[t]{.3\textwidth}
    \centering
	\begin{tikzpicture}
		\draw (1.5,0)--(1.5,2);
		\draw (2,0)--(2,2);
		\draw (4,0)--(4,2);
		\draw (2,.5)--(4,.5);
		\draw (2.5,1) node[cross] {};
		\draw (3,1) node[cross] {};
		\draw (3.5,1) node[cross] {};
	\end{tikzpicture}

        \caption{}
    \end{subfigure}
    \hfill
	\begin{subfigure}[t]{.3\textwidth}
    \centering
	\begin{tikzpicture}
	\tikzstyle{gauge} = [circle, draw];
	\tikzstyle{flavour} = [regular polygon,regular polygon sides=4,draw];
	\node (g2) [gauge,right of=g1,label=below:{$1$}] {};
	\node (f2) [flavour,above of=g2,label=above:{$3$}] {};
	\draw (g2)--(f2)
		;
	\end{tikzpicture}
        \caption{}
    \end{subfigure}
    \hfill
 	\caption{Model with $n_s=n_d=3$, $\vec{l}_s=(0,1,2)$ and $\vec{l}_d=(2,2,2)$. (a) is the Higgs brane configuration. (b) represents the Coulomb brane configuration. (c) represents the quiver.}
	\label{fig:SU3minimal}
\end{figure}

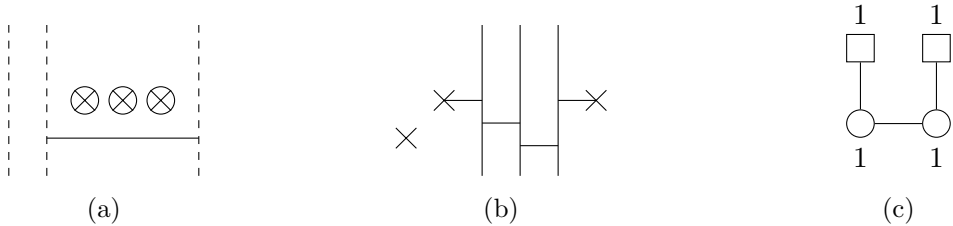
\begin{figure}[ht]
	\centering
	\begin{subfigure}[t]{.3\textwidth}
    \centering
	\begin{tikzpicture}
		\draw[dashed] (1.5,0)--(1.5,2)
					(2,0)--(2,2)
					(4,0)--(4,2);
		\draw (2,.5)--(4,.5);
		\draw (2.5,1) node[cross] {};
		\draw (3,1) node[cross] {};
		\draw (3.5,1) node[cross] {};
		\draw (2.5,1) node[circle,draw] {};
		\draw (3,1) node[circle,draw] {};
		\draw (3.5,1) node[circle,draw] {};
	\end{tikzpicture}
        \caption{}
    \end{subfigure}
    \hfill
    \begin{subfigure}[t]{.3\textwidth}
    \centering
	\begin{tikzpicture}
		\draw (2.5,0)--(2.5,2)
			(3,0)--(3,2)
			(3.5,0)--(3.5,2);
		\draw (1.5,.5) node[cross] {};
		\draw (2,1) node[cross] {};
		\draw (4,1) node[cross] {};
		\draw 	(2,1)--(2.5,1)
				(4,1)--(3.5,1);
		\draw 	(2.5,.7)--(3,.7)
				(3,.4)--(3.5,.4);
	\end{tikzpicture}
        \caption{}
    \end{subfigure}
    \hfill
	\begin{subfigure}[t]{.3\textwidth}
    \centering
	\begin{tikzpicture}
	\tikzstyle{gauge} = [circle, draw];
	\tikzstyle{flavour} = [regular polygon,regular polygon sides=4, draw];
	\node (g1) [gauge, label=below:{$1$}] {};
	\node (g2) [gauge,right of=g1, label=below:{$1$}] {};				
	\node (f1) [flavour,above of=g1, label=above:{$1$}] {};
	\node (f2) [flavour,above of=g2, label=above:{$1$}] {};
	\draw (f1)--(g1)--(g2)--(f2)
		;
	\end{tikzpicture}
        \caption{}
    \end{subfigure}
    \hfill
 	\caption{Model with $n_s=n_d=3$,  $\vec{l}_s=(2,2,2)$ and $\vec{l}_d=(0,1,2)$. (a) is the Higgs brane configuration. (b) represents the Coulomb brane configuration. (c) represents the quiver.}
	\label{fig:SU3minimalMirror}
\end{figure}

Let us show that this is in fact the model that we found. To show this we just need to take the Higgs brane configuration of fig. \ref{fig:SU3KPA3} (c), perform a phase transition to the Coulomb brane configuration and then read the quiver. The phase transition is performed as usual, to make it more explicit let us perform some Hanany-Witten transitions still in the Higgs brane configuration to obtain fig. \ref{fig:SU3minimal} (a). After alignment and maximal splitting we obtain the Coulomb brane configuration as in fig. \ref{fig:SU3minimal} (b). The quiver can then be read as fig. \ref{fig:SU3minimal} (c). The mirror model can be computed as usual and is presented in fig. \ref{fig:SU3minimalMirror}.

\subsection{Example: $\mathfrak{sl}_3$ Transitions, Minimal to Trivial}

Let us explore one more example, but this time consider the model with $\mathcal{M}_H=a_2$ and $\mathcal{M}_C=A_2$ as the starting point. We can look at the Higgs branch, that is a closure of a nilpotent orbit, and perform a new Kraft-Procesi transition on it.\\

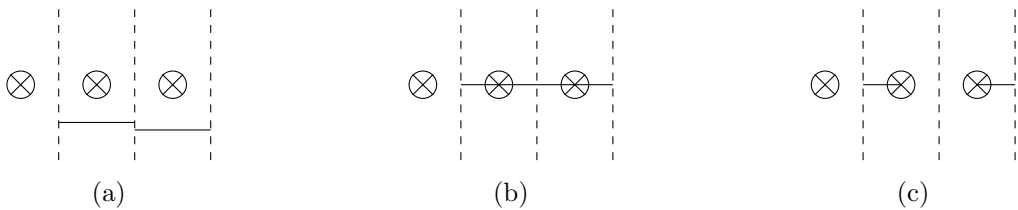
\begin{figure}[h]
	\centering
	\begin{subfigure}[t]{.3\textwidth}
    \centering
	\begin{tikzpicture}
		\draw[dashed] (-1.5,0)--(-1.5,2);
		\draw[dashed] (-2.5,0)--(-2.5,2);
		\draw[dashed] (-3.5,0)--(-3.5,2);
		\draw (-1.5,.4)--(-2.5,.4)
				(-2.5,.5)--(-3.5,.5);
		\draw (-2,1) node[cross] {};
		\draw (-3,1) node[cross] {};
		\draw (-4,1) node[cross] {};
		\draw (-2,1) node[circle,draw] {};
		\draw (-3,1) node[circle,draw] {};
		\draw (-4,1) node[circle,draw] {};
	\end{tikzpicture}
	\caption{}
 	\end{subfigure}
 	\hfill
 	\begin{subfigure}[t]{.3\textwidth}
    \centering
	\begin{tikzpicture}
		\draw[dashed] (-1.5,0)--(-1.5,2);
		\draw[dashed] (-2.5,0)--(-2.5,2);
		\draw[dashed] (-3.5,0)--(-3.5,2);
		\draw (-1.5,1)--(-3.5,1);
		\draw (-2,1) node[cross] {};
		\draw (-3,1) node[cross] {};
		\draw (-4,1) node[cross] {};
		\draw (-2,1) node[circle,draw] {};
		\draw (-3,1) node[circle,draw] {};
		\draw (-4,1) node[circle,draw] {};
	\end{tikzpicture}
	\caption{}
 	\end{subfigure}
 	\hfill
 	\begin{subfigure}[t]{.3\textwidth}
    \centering
	\begin{tikzpicture}
		\draw[dashed] (-1.5,0)--(-1.5,2);
		\draw[dashed] (-2.5,0)--(-2.5,2);
		\draw[dashed] (-3.5,0)--(-3.5,2);
		\draw (-1.5,1)--(-2,1)
		(-3,1)--(-3.5,1);
		\draw (-2,1) node[cross] {};
		\draw (-3,1) node[cross] {};
		\draw (-4,1) node[cross] {};
		\draw (-2,1) node[circle,draw] {};
		\draw (-3,1) node[circle,draw] {};
		\draw (-4,1) node[circle,draw] {};
	\end{tikzpicture}
	\caption{}
 	\end{subfigure}
 	\caption{$a_2$ Kraft-Procesi transition. (a) represents the Higgs branch of the Model with $n_s=n_d=3$, $\vec{l}_s=(0,1,2)$ and $\vec{l}_d=(2,2,2)$. (b) represents the singular point when $\vec{y}_1=\vec{y}_2=\vec{w}_2=\vec{w}_3$. (c) represents the new model after taking $\vec{x}$ to infinity. The new linking numbers are $\vec{l}_s=(0,0,3)$, $l_d=(2,2,2)$. This defines a new model which fulfils the prescription to have the Higgs branch as the closure of the nilpotent orbit with $\lambda^t=(3)$ and $\lambda=(1^3)$. This is the trivial orbit of $\mathfrak{sl}_3$.}
	\label{fig:SU3KPa3}
\end{figure}

Let us first perform a phase transition on the Higgs brane configuration of fig. \ref{fig:SU3minimal} (a) to a brane configuration in which there are no fixed D3-branes. The result is fig. \ref{fig:SU3KPa3} (a). Once again we would like to find a minimal singularity. There are no minimal singularities of quaternionic dimension 1, since there is no single D3-brane that can be aligned with at least 2 NS5-branes (remember that to be able to perform any Higgsing we need to align and then split D3-branes to get at least one segment stretching between two D5-branes). There are two D5-branes that could have a segment of D3-brane stretching between them: the second and the third ones from the left in fig. \ref{fig:SU3KPa3} (a). This can be achieved if we align \emph{both} of the D3-branes with the two aforementioned D5-branes. This means that we would be focusing in \emph{all} the D3-branes to find the minimal singularity in the Higgs branch. The moduli generated by these branes is the Higgs branch itself, and this is therefore the physical realization of the fact that the minimal singularity in the variety $a_2$ is $a_2$ itself. The singularity can be removed by applying the Kraft-Procesi transition as before, taking the position $\vec{x}$ of the D3-brane segment that stretches between the two D5-branes after the alignment and the splitting to infinity. This process is depicted in fig. \ref{fig:SU3KPa3}.\\

The resulting model has no D3-branes propagating in the Higgs brane configuration, this means that its Higgs branch is the trivial variety (a single point). This actually corresponds to the closure of the trivial nilpotent orbit of $\mathfrak{sl}_3$. The new model has linking numbers $\vec{l}_d=(2,2,2)$ and $\vec{l}_s=(0,0,3)$. We obtain the partition $\lambda^t=(3)$ and therefore $\lambda=(1^3)$.

\subsubsection*{Hasse Diagram}

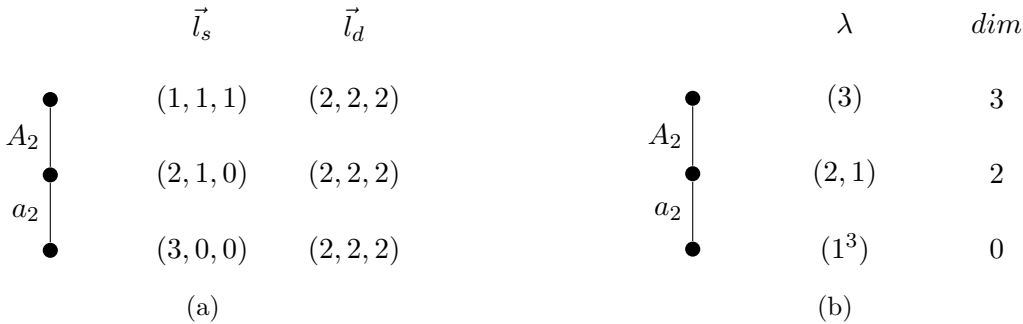
\begin{figure}[h]
	\centering
\begin{subfigure}[t]{.45\textwidth}
    \centering
	\begin{tikzpicture}
		\tikzstyle{hasse} = [circle, fill,inner sep=2pt];
		\node [hasse] (r) [] {};
		\node [hasse] (m) [below of=r] {};
		\node [hasse] (t) [below of=m] {};
		\draw (r) edge [] node[label=left:$A_2$] {} (m)
			(m) edge [] node[label=left:$a_2$] {} (t);
		\node (er) [right of=r] {};
		\node (dr) [right of=er] {$(1,1,1)$};
		\node (cr) [right of=dr] {};
		\node (br) [right of=cr] {$(2,2,2)$};
		\node (em) [right of=m] {};
		\node (dm) [right of=em] {$(2,1,0)$};
		\node (cm) [right of=dm] {};
		\node (bm) [right of=cm] {$(2,2,2)$};
		\node (et) [right of=t] {};
		\node (dt) [right of=et] {$(3,0,0)$};
		\node (ct) [right of=dt] {};
		\node (bt) [right of=ct] {$(2,2,2)$};
		\node (d) [above of=dr] {$\vec{l}_s$};
		\node (b) [above of=br] {$\vec{l}_d$};
	\end{tikzpicture}
	\caption{}
 	\end{subfigure}
 	\hfill
 	\begin{subfigure}[t]{.45\textwidth}
    \centering
	\begin{tikzpicture}
		\tikzstyle{hasse} = [circle, fill,inner sep=2pt];
		\node [hasse] (r) [] {};
		\node [hasse] (m) [below of=r] {};
		\node [hasse] (t) [below of=m] {};
		\draw (r) edge [] node[label=left:$A_2$] {} (m)
			(m) edge [] node[label=left:$a_2$] {} (t);		\node (er) [right of=r] {};
		\node (dr) [right of=er] {$(3)$};
		\node (cr) [right of=dr] {};
		\node (br) [right of=cr] {3};
		\node (em) [right of=m] {};
		\node (dm) [right of=em] {$(2,1)$};
		\node (cm) [right of=dm] {};
		\node (bm) [right of=cm] {2};
		\node (et) [right of=t] {};
		\node (dt) [right of=et] {$(1^3)$};
		\node (ct) [right of=dt] {};
		\node (bt) [right of=ct] {0};
		\node (d) [above of=dr] {$\lambda$};
		\node (b) [above of=br] {$dim$};
	\end{tikzpicture}
	\caption{}
 	\end{subfigure}
	\caption{Hasse diagram for the models with Higgs branch being the closure of a nilpotent orbit of $\mathfrak{sl}_3$. fig. (a) represents the brane configurations, where the linking numbers $\vec{l}_s$ and $\vec{l}_d$ are provided for each orbit. fig. (b) depicts the information of the orbits. With respect to the brane configurations, $\lambda$ is the transpose partition of $\vec{l}_s$, and $dim$ is the number of D3-branes that generate the Higgs branch in each model. This second diagram corresponds to the one developed in  \cite{Kraft1982}.}
	\label{fig:HasseSU3}
	
\end{figure}

We can summarize in fig. \ref{fig:HasseSU3} (a) the transition between different models in a Hasse diagram with the information: linking numbers, minimal transitions. The linking numbers determine the partition and dimension of the nilpotent orbits of $\mathfrak{sl}_3$ and vice-versa. If these are included in the diagram instead of the linking numbers the resulting Hasse diagrams correspond to the ones developed by Kraft and Procesi \cite{Kraft1982}, fig. \ref{fig:HasseSU3} (b). Note that \cite{Kraft1982} addresses all classical algebras, for the case of $\mathfrak{sl}_N$ algebras discussed here these diagrams appeared a year earlier in \cite{Kraft1981}.

\subsection{General Definition}

We can implement the Kraft-Procesi transition between closures of nilpotent orbits, as a transition between models in Type IIB superstring theory, and by induction a transition between their corresponding effective worldvolume quiver gauge theories. So far we have focused on transitions performed in the Higgs brane configurations but a mirror duality could see the transitions in Coulomb brane configurations in the exactly same fashion.\\

The general steps to perform this transition in the Higgs brane configuration of a model whose Higgs branch is the closure of a nilpotent orbit of $\mathfrak{sl}_n$ are:

\begin{enumerate}
	\item Find all minimal singularities that are subvarieties of the Higgs branch $\mathcal{M}_H$. Each of them corresponds to a different Kraft-Procesi transition. Choose one of them, let us denote it with $\mathcal{V}\subset \mathcal{M}_H$. The transition then inherits the name of the minimal singularity, it will be called a \emph{$\mathcal{V}$ KP transition}. If $\mathcal{M}_H$ is the closure of a nilpotent orbit of $\mathfrak{sl}_n$ the variety $\mathcal{V}$ has to be a singularity of either $A_k$ or $a_k$ type, where $k < n$.
	\item Remove the singularity $\mathcal{V}$ from the Higgs branch $\mathcal{M}_H$ via the Higgs mechanism. Consider the resulting variety $\mathcal{M}'_H$ as the Higgs branch of a new model. $\mathcal{M}'_H$ will be a closure of a nilpotent orbit of $\mathfrak{sl}_n$ with dimension:
	\begin{align}
		dim(\mathcal{M}'_H)=dim(\mathcal{M}_H)-dim(\mathcal{V})
	\end{align}
\end{enumerate}

Let us go through both steps in more depth.

\subsubsection{Finding all Minimal Singularities}

As mentioned above, there are only two possible ways of doing a minimal Higgsing in a Higgs brane configuration. One is to remove only one D3-brane in the Higgs branch, and with it the $A_k$ singularity that the threebrane generated ($k+1$ is the number of NS5-branes that coincide with the D3-brane in the singular point). The other way is to remove a set of D3-branes that align together to create a single D3-brane in the Coulomb branch, the moduli they generate in the Higgs brane configuration is always an $a_k$ variety, where $k$ is the number of D3-branes that are initially involved.\\

To find all such minimal singularities the first step is to perform a phase transition that takes the model to a Higgs brane configuration where all fixed D3-branes have been annihilated and only D3-branes with both ends in D5-branes remain. Once in this brane configuration every NS5-brane will be in an interval between two D5-branes, with no D3-branes ending on it. If we order such intervals from left to right with integer labels $1,2,\dots$ the linking numbers of the NS5-branes will be equal to the label of the interval they occupy.\\

To find all $A_n$ subvarieties we look at all intervals with two or more NS5-branes in them. For each of these intervals, an $A_k$ will arise, where $k+1$ will be the number of NS5-branes in the interval.\\

To find all $a_n$ varieties we look at the intervals with exactly one NS5-brane in them, if there are two of these intervals adjacent, they correspond to an $a_2$ singularity. If there are two of these intervals with $k-2$ intervals with no NS5-branes in between, they all correspond to an $a_k$ singularity.\\

\subsubsection{Removing the Minimal Singularity}

Each of the subvarieties $\mathcal{V}_i\in \M_H$ found in the previous section can give rise to their own $\mathcal{V}_i$ KP transition. Let us choose one of them and denote it $\mathcal{V}$. To remove it we just perform a Higgsing mechanism that sends the D3-branes involved (one if it is an $A_k$ singularity and $k$ of them if it is of the $a_k$ type) to the Coulomb branch. The difference is that this time, instead of just sending them to the Coulomb branch, we will send the vacuum expectation values of the scalar fields in the new massless vectormultiplets to infinity, thus removing the D3-branes from the brane configuration, and producing a new pure Higgs branch $\mathcal{M}'_H$ that will correspond to a new quiver gauge theory.

\subsection{Example: $\mathfrak{sl}_4$ KP Transitions}

Let us study one more example that will further illustrate the KP transitions. This time we show how starting with the self-dual model corresponding to the closure of the maximal orbit of $\mathfrak{sl}_4$ we can produce \emph{all} other models that correspond to all other closures of nilpotent orbits of the algebra.\\

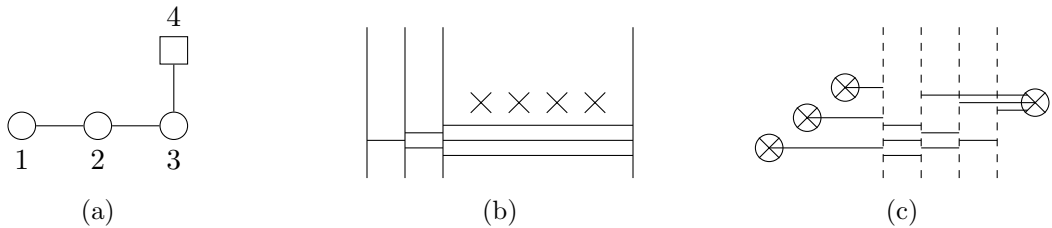
\begin{figure}[h]
	\centering
	\begin{subfigure}[t]{.3\textwidth}
    \centering
	\begin{tikzpicture}
	\tikzstyle{gauge} = [circle, draw];
	\tikzstyle{flavour} = [regular polygon,regular polygon sides=4,draw];
	\node (g1) [gauge, label=below:{$1$}] {};
	\node (g2) [gauge,right of=g1, label=below:{$2$}] {};
	\node (g3) [gauge,right of=g2, label=below:{$3$}] {};
	\node (f3) [flavour,above of=g3, label=above:{$4$}] {};
	\draw (g1)--(g2)--(g3)--(f3)
		;
	\end{tikzpicture}
        \caption{}
    \end{subfigure}
    \hfill
	\begin{subfigure}[t]{.3\textwidth}
    \centering
	\begin{tikzpicture}
		\draw (1,0)--(1,2);
		\draw (1.5,0)--(1.5,2);
		\draw (2,0)--(2,2);
		\draw (4.5,0)--(4.5,2);
		\draw 	(1,.5)--(1.5,.5)
				(1.5,.4)--(2,.4)
				(1.5,.6)--(2,.6)
				(2,.3)--(4.5,.3)
				(2,.5)--(4.5,.5)
				(2,.7)--(4.5,.7);
		\draw (2.5,1) node[cross] {};
		\draw (3,1) node[cross] {};
		\draw (3.5,1) node[cross] {};
		\draw (4,1) node[cross] {};
	\end{tikzpicture}

        \caption{}
    \end{subfigure}
    \hfill
    \begin{subfigure}[t]{.3\textwidth}
    \centering
	\begin{tikzpicture}
		\draw[dashed] 	(2.5,0)--(2.5,2)
						(3,0)--(3,2)
						(3.5,0)--(3.5,2)
						(4,0)--(4,2);
		\draw (1,.4) node[cross] {};
		\draw (1.5,.8) node[cross] {};
		\draw (2,1.2) node[cross] {};
		\draw (4.5,1) node[cross] {};
		\draw (1,.4) node[circle,draw] {};
		\draw (1.5,.8) node[circle,draw] {};
		\draw (2,1.2) node[circle,draw] {};
		\draw (4.5,1) node[circle,draw] {};
		\draw 	(1,.4)--(2.5,.4)
				(1.5,.8)--(2.5,.8)
				(2,1.2)--(2.5,1.2);
		\draw	(4.4,1.1)--(3,1.1)
				(4.5,1)--(3.5,1)
				(4.4,.9)--(4,.9);
		\draw 	(4,.5)--(3.5,.5)
				(3.5,.4)--(3,.4)
				(3.5,.6)--(3,.6)
				(3,.7)--(2.5,.7)
				(3,.5)--(2.5,.5)
				(3,.3)--(2.5,.3)
				;
	\end{tikzpicture}
        \caption{}
    \end{subfigure}
 	\caption{Model with $n_s=n_d=4$, $\vec{l}_s=(1,1,1,1)$ and $\vec{l}_d=(3,3,3,3)$. (a) is the quiver. (b) represents the Coulomb branch. (c) represents the Higgs branch.}
	\label{fig:SU4maximal}
\end{figure}

The initial model, and all other models we  reach via KP transitions, belong to the $T_{\lambda^t}(SU(4))$ family of theories. The model with the closure of the maximal nilpotent orbit as the Higgs branch is: 

\begin{align}
T_{\lambda^t}(SU(4))=T_{(1^4)}(SU(4))
\end{align}

since $\lambda=(4)$ is the corresponding partition. Using Table \ref{tab:Tsigma} we obtain the data for the model: $\vec{l}_d=(3,3,3,3)$ and $\vec{l}_s=(1,1,1,1)$. With these linking numbers we can recover the Coulomb brane configuration and read the quiver from it; we also get the Higgs brane configuration. They are presented in fig. \ref{fig:SU4maximal}.\\

\begin{figure}[h]
	\centering
	\begin{subfigure}[t]{.3\textwidth}
    \centering
	\begin{tikzpicture}
		\draw[dashed] 	(-1,0)--(-1,2)
					(-1.5,0)--(-1.5,2)
					(-2,0)--(-2,2)
					(-4.5,0)--(-4.5,2);
		\draw 	(-1,.5)--(-1.5,.5)
				(-1.5,.4)--(-2,.4)
				(-1.5,.6)--(-2,.6)
				(-2,.3)--(-4.5,.3)
				(-2,.5)--(-4.5,.5)
				(-2,.7)--(-4.5,.7);
		\draw (-2.5,1) node[cross] {};
		\draw (-3,1) node[cross] {};
		\draw (-3.5,1) node[cross] {};
		\draw (-4,1) node[cross] {};
		\draw (-2.5,1) node[circle,draw] {};
		\draw (-3,1) node[circle,draw] {};
		\draw (-3.5,1) node[circle,draw] {};
		\draw (-4,1) node[circle,draw] {};
	\end{tikzpicture}
        \caption{}
    \end{subfigure}
    \hfill
	\begin{subfigure}[t]{.3\textwidth}
    \centering
	\begin{tikzpicture}
		\draw[dashed] 	(-1,0)--(-1,2)
					(-1.5,0)--(-1.5,2)
					(-2,0)--(-2,2)
					(-4.5,0)--(-4.5,2);
		\draw 	(-1,.5)--(-1.5,.5)
				(-1.5,.4)--(-2,.4)
				(-1.5,.6)--(-2,.6)
				(-2,.3)--(-4.5,.3)
				(-2,.5)--(-4.5,.5)
				(-2,1)--(-4.5,1);
		\draw (-2.5,1) node[cross] {};
		\draw (-3,1) node[cross] {};
		\draw (-3.5,1) node[cross] {};
		\draw (-4,1) node[cross] {};
		\draw (-2.5,1) node[circle,draw] {};
		\draw (-3,1) node[circle,draw] {};
		\draw (-3.5,1) node[circle,draw] {};
		\draw (-4,1) node[circle,draw] {};
	\end{tikzpicture}

        \caption{}
    \end{subfigure}
    \hfill
    \begin{subfigure}[t]{.3\textwidth}
    \centering
	\begin{tikzpicture}
		\draw[dashed] 	(-1,0)--(-1,2)
					(-1.5,0)--(-1.5,2)
					(-2,0)--(-2,2)
					(-4.5,0)--(-4.5,2);
		\draw 	(-1,.5)--(-1.5,.5)
				(-1.5,.4)--(-2,.4)
				(-1.5,.6)--(-2,.6)
				(-2,.3)--(-4.5,.3)
				(-2,.5)--(-4.5,.5)
				(-2,1)--(-2.5,1)
				(-4,1)--(-4.5,1);
		\draw (-2.5,1) node[cross] {};
		\draw (-3,1) node[cross] {};
		\draw (-3.5,1) node[cross] {};
		\draw (-4,1) node[cross] {};
		\draw (-2.5,1) node[circle,draw] {};
		\draw (-3,1) node[circle,draw] {};
		\draw (-3.5,1) node[circle,draw] {};
		\draw (-4,1) node[circle,draw] {};
	\end{tikzpicture}
        \caption{}
    \end{subfigure}
 	\caption{$A_3$ Kraft-Procesi transition. (a) represents the Higgs branch of the Model with $n_s=n_d=4$, $\vec{l}_s=(1,1,1,1)$ and $\vec{l}_d=(3,3,3,3)$. (b) represents the singular point when $\vec{y}=\vec{w}_1=\vec{w}_2=\vec{w}_3=\vec{w}_4$. (c) represents the new model after taking $\vec{x}_1$, $\vec{x}_2$ and $\vec{x}_3$ to infinity. The new linking numbers are $\vec{l}_s=(0,1,1,2)$, $\vec{l}_d=(3,3,3,3)$. This defines a new model which fulfils the prescription to have the Higgs branch as the closure of the nilpotent orbit with $\lambda^t=(2,1^2)$ and hence $\lambda=(3,1)$. This is the \emph{subregular} orbit of $\mathfrak{sl}_4$.}
	\label{fig:SU4transA3}
\end{figure}
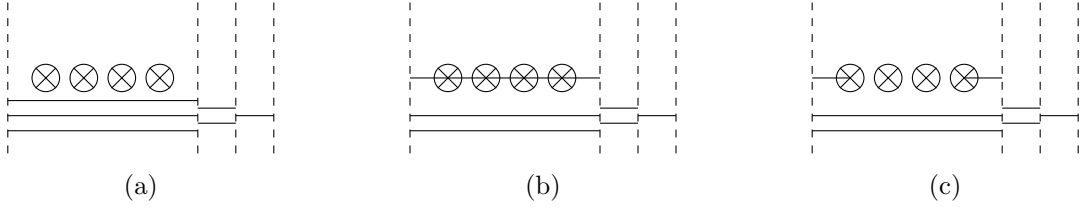

In the model of fig. \ref{fig:SU4maximal}, $\mathcal{M}_H=\mathcal{M}_C=\bar{\mathcal{O}}_{(4)}$.  We choose the Higgs branch and decide to perform the KP transitions on the Higgs brane configuration (if we choose the Coulomb branch, then we obtain the mirror models of our results, which in the following discussion are obtained by performing S-duality after each KP transition). The starting point, according to the general prescription we perform Hanany-Witten transitions to annihilate all fixed threebranes in fig. \ref{fig:SU4maximal} (c), the result is fig. \ref{fig:SU4transA3} (a).\\

Now we look for minimal singularities that are subvarieties of $\mathcal{M}_H$. As explained in the previous section we look for intervals between D5-branes with one or more NS5-branes in between. Since all the NS5-branes are in the first interval only this interval will host minimal singularities. In this case, there are $4$ NS5-branes, since this is bigger than one, this corresponds to an $A_n$ singularity, in particular $A_3=\mathbb{C}^2/\mathbb{Z}_4$. This is a singularity of real dimension $4$ that is generated by a single D3-brane moving freely in this first interval. Since this is the only minimal singularity in the Higgs brane configuration we say that there is only one possible KP transition from the closure of the orbit related to $\lambda=(4)$, and this is an $A_3$ KP transition, reproducing the result of Brieskorn \cite{Brieskorn1970}.\\

We can perform the $A_3$ KP transition on the Higgs brane configuration, the process is depicted in fig. \ref{fig:SU4transA3}. The procedure is always the same, we select the singular point of $A_3$, this is when the D3-brane aligns with the four NS5-branes:

\begin{align}
	\vec{y}=\vec{w}_1=\vec{w}_2=\vec{w}_3=\vec{w}_4
\end{align}

This is shown in fig. \ref{fig:SU4transA3} (b). In this singular point there are three new vectorplets that became massless and  admit nonzero vacuum expectation values $(\vec{x}_i,a_i)$ when the previously massless hyper multiplet $(\vec{y},b)$ becomes massive, abandoning the Higgs branch. This corresponds to split the D3-brane into different segments that end in the NS5-branes. The first and the last segments are fixed, since they will end on a D5-brane and a NS5-brane. The three segments in the middle that end in two NS5-branes will correspond to the new three massless vector multiplets. In order to obtain a new model, instead of just a different phase of the model we started with, we need to remove completely those three segments. To do this we take their $\vec{x}_i$ positions to infinity. The result is shown in fig. \ref{fig:SU4transA3} (c). This results in a new model with linking numbers for the NS5-branes: $\vec{l}_s=(0,1,1,2)$.\\

Let us now study the resulting model. It has linking numbers $\vec{l}_s=(0,1,1,2)$ and $\vec{l}_d=(3,3,3,3)$. It corresponds to the model with Higgs branch being the closure of the orbit of $\mathfrak{sl}_4$ determined by $\lambda=(3,1)$, since $\vec{l}_s$ is identified with the transpose partition $\lambda^t=(2,1^2)$, and therefore $\lambda=(3,1)$. We write:

\begin{align}
\mathcal{M}_H=\bar{\mathcal{O}}_{(3,1)}
\end{align}

\begin{figure}[h]
	\centering
	\begin{subfigure}[t]{.3\textwidth}
    \centering
	\begin{tikzpicture}
	\tikzstyle{gauge} = [circle,draw];
	\tikzstyle{flavour} = [regular polygon,regular polygon sides=4,draw];
	\node (g2) [gauge,label=below:{$1$}] {};
	\node (g3) [gauge,right of=g2,label=below:{$2$}] {};
	\node (f3) [flavour,above of=g3,label=above:{$4$}] {};
	\draw (g2)--(g3)--(f3)
		;
	\end{tikzpicture}
        \caption{}
    \end{subfigure}
    \hfill
	\begin{subfigure}[t]{.3\textwidth}
    \centering
	\begin{tikzpicture}
		\draw (1,0)--(1,2);
		\draw (1.5,0)--(1.5,2);
		\draw (2,0)--(2,2);
		\draw (4.5,0)--(4.5,2);
		\draw 	(1.5,.5)--(2,.5)
				(2,.4)--(4.5,.4)
				(2,.6)--(4.5,.6);
		\draw (2.5,1) node[cross] {};
		\draw (3,1) node[cross] {};
		\draw (3.5,1) node[cross] {};
		\draw (4,1) node[cross] {};
	\end{tikzpicture}

        \caption{}
    \end{subfigure}
    \hfill
    \begin{subfigure}[t]{.3\textwidth}
    \centering
	\begin{tikzpicture}
		\draw[dashed] 	(2.5,0)--(2.5,2)
						(3,0)--(3,2)
						(3.5,0)--(3.5,2)
						(4,0)--(4,2);
		\draw (1,.4) node[cross] {};
		\draw (1.5,.8) node[cross] {};
		\draw (2,1.2) node[cross] {};
		\draw (4.5,1) node[cross] {};
		\draw (1,.4) node[circle,draw] {};
		\draw (1.5,.8) node[circle,draw] {};
		\draw (2,1.2) node[circle,draw] {};
		\draw (4.5,1) node[circle,draw] {};
		\draw 	(1.5,.8)--(2.5,.8)
				(2,1.2)--(2.5,1.2);
		\draw	(4.45,1.05)--(3.5,1.05)
				(4.45,.95)--(4,.95);
		\draw 	(4,.5)--(3.5,.5)
				(3.5,.4)--(3,.4)
				(3.5,.6)--(3,.6)
				(3,.7)--(2.5,.7)
				(3,.5)--(2.5,.5)
				;
	\end{tikzpicture}
        \caption{}
    \end{subfigure}
 	\caption{Model with $n_s=n_d=4$, $\vec{l}_s=(0,1,1,2)$ and $\vec{l}_d=(3,3,3,3)$. (a) is the quiver. (b) represents the Coulomb branch. (c) represents the Higgs branch.}
	\label{fig:SU4subreg}
\end{figure}
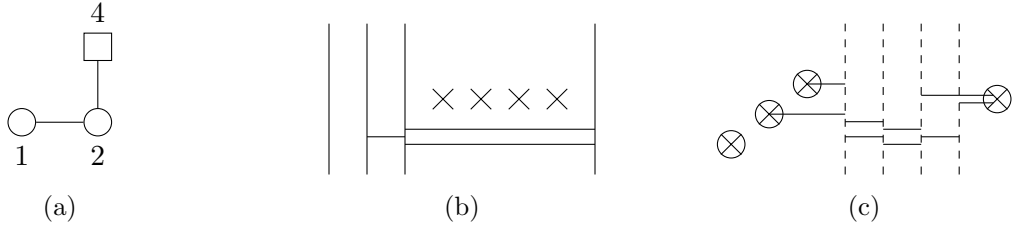

Fig. \ref{fig:SU4transA3} (c) is its Higgs brane configuration, performing a phase transition we obtain its Coulomb brane configuration and read the corresponding quiver from it. The result of this is displayed in fig. \ref{fig:SU4subreg}.\\

\begin{figure}[h]
	\centering
	\begin{subfigure}[t]{.3\textwidth}
    \centering
	\begin{tikzpicture}
	\tikzstyle{gauge} = [circle, draw];
	\tikzstyle{flavour} = [regular polygon,regular polygon sides=4,draw];
	\node (g1) [gauge,label=below:{$2$}] {};
	\node (g2) [gauge,right of=g1,label=below:{$2$}] {};
	\node (g3) [gauge,right of=g2,label=below:{$1$}] {};
	\node (f1) [flavour,above of=g1,label=above:{$2$}] {};
	\node (f2) [flavour,above of=g2,label=above:{$1$}] {};
	\draw (f1)--(g1)--(g2)--(g3)
			(f2)--(g2)
		;
	\end{tikzpicture}
        \caption{}
    \end{subfigure}
      \hfill
    \begin{subfigure}[t]{.3\textwidth}
    \centering
	\begin{tikzpicture}
		\draw 	(2.5,0)--(2.5,2)
						(3,0)--(3,2)
						(3.5,0)--(3.5,2)
						(4,0)--(4,2);
		\draw (1,.4) node[cross] {};
		\draw (1.5,.8) node[cross] {};
		\draw (2,1.2) node[cross] {};
		\draw (4.5,1) node[cross] {};
		\draw 	(1.5,.8)--(2.5,.8)
				(2,1.2)--(2.5,1.2);
		\draw	(4.45,1.05)--(3.5,1.05)
				(4.45,.95)--(4,.95);
		\draw 	(4,.5)--(3.5,.5)
				(3.5,.4)--(3,.4)
				(3.5,.6)--(3,.6)
				(3,.7)--(2.5,.7)
				(3,.5)--(2.5,.5)
				;
	\end{tikzpicture}
        \caption{}
    \end{subfigure}
    \hfill
	\begin{subfigure}[t]{.3\textwidth}
    \centering
	\begin{tikzpicture}
		\draw[dashed] (1,0)--(1,2)
				(1.5,0)--(1.5,2)
				(2,0)--(2,2)
				(4.5,0)--(4.5,2);
		\draw 	(1.5,.5)--(2,.5)
				(2,.4)--(4.5,.4)
				(2,.6)--(4.5,.6);
		\draw (2.5,1) node[cross] {};
		\draw (3,1) node[cross] {};
		\draw (3.5,1) node[cross] {};
		\draw (4,1) node[cross] {};
		\draw (2.5,1) node[circle,draw] {};
		\draw (3,1) node[circle,draw] {};
		\draw (3.5,1) node[circle,draw] {};
		\draw (4,1) node[circle,draw] {};
	\end{tikzpicture}

        \caption{}
    \end{subfigure}
 	\caption{Model with $n_s=n_d=4$, $\vec{l}_s=(3,3,3,3)$ and $\vec{l}_d=(0,1,1,2)$. (a) is the quiver. (b) represents the Coulomb branch. (c) represents the Higgs branch.}
	\label{fig:SU4subregMirror}
\end{figure}
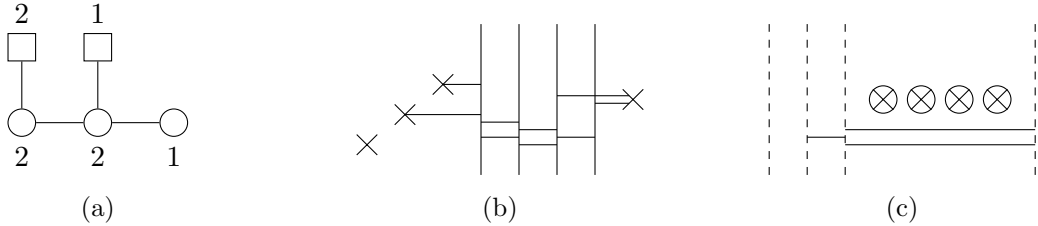

We can perform a mirror duality on this model to obtain the theory with $\mathcal{M}_C=\bar{\mathcal{O}}_{(3,1)}$. This can be performed in any of the ways explained before, for example, performing an \mbox{S-duality} on the Higgs brane configuration to obtain the mirror Coulomb brane configuration and read the quiver from it. The mirror model has $\vec{l}_d=(0,1,1,2)$ and $\vec{l}_s=(3,3,3,3)$. Its quiver and brane configurations are depicted in fig. \ref{fig:SU4subregMirror}.\\

\begin{figure}[h]
	\centering
	\begin{subfigure}[t]{.3\textwidth}
    \centering
	\begin{tikzpicture}
		\draw[dashed] 	(-1,0)--(-1,2)
					(-1.5,0)--(-1.5,2)
					(-2.5,0)--(-2.5,2)
					(-4,0)--(-4,2);
		\draw 	(-1,.5)--(-1.5,.5)
				(-1.5,.4)--(-2.5,.4)
				(-1.5,.6)--(-2.5,.6)
				(-2.5,.3)--(-4,.3)
				(-2.5,.5)--(-4,.5);
		\draw (-2,1) node[cross] {};
		\draw (-3,1) node[cross] {};
		\draw (-3.5,1) node[cross] {};
		\draw (-4.5,1) node[cross] {};
		\draw (-2,1) node[circle,draw] {};
		\draw (-3,1) node[circle,draw] {};
		\draw (-3.5,1) node[circle,draw] {};
		\draw (-4.5,1) node[circle,draw] {};
	\end{tikzpicture}
        \caption{}
    \end{subfigure}
    \hfill
	\begin{subfigure}[t]{.3\textwidth}
    \centering
	\begin{tikzpicture}
		\draw[dashed] 	(-1,0)--(-1,2)
					(-1.5,0)--(-1.5,2)
					(-2.5,0)--(-2.5,2)
					(-4,0)--(-4,2);
		\draw 	(-1,.5)--(-1.5,.5)
				(-1.5,.4)--(-2.5,.4)
				(-1.5,.6)--(-2.5,.6)
				(-2.5,.3)--(-4,.3)
				(-2.5,1)--(-4,1);
		\draw (-2,1) node[cross] {};
		\draw (-3,1) node[cross] {};
		\draw (-3.5,1) node[cross] {};
		\draw (-4.5,1) node[cross] {};
		\draw (-2,1) node[circle,draw] {};
		\draw (-3,1) node[circle,draw] {};
		\draw (-3.5,1) node[circle,draw] {};
		\draw (-4.5,1) node[circle,draw] {};
	\end{tikzpicture}

        \caption{}
    \end{subfigure}
    \hfill
    \begin{subfigure}[t]{.3\textwidth}
    \centering
	\begin{tikzpicture}
		\draw[dashed] 	(-1,0)--(-1,2)
					(-1.5,0)--(-1.5,2)
					(-2.5,0)--(-2.5,2)
					(-4,0)--(-4,2);
		\draw 	(-1,.5)--(-1.5,.5)
				(-1.5,.4)--(-2.5,.4)
				(-1.5,.6)--(-2.5,.6)
				(-2.5,.3)--(-4,.3)
				(-2.5,1)--(-3,1)
				(-3.5,1)--(-4,1);
		\draw (-2,1) node[cross] {};
		\draw (-3,1) node[cross] {};
		\draw (-3.5,1) node[cross] {};
		\draw (-4.5,1) node[cross] {};
		\draw (-2,1) node[circle,draw] {};
		\draw (-3,1) node[circle,draw] {};
		\draw (-3.5,1) node[circle,draw] {};
		\draw (-4.5,1) node[circle,draw] {};
	\end{tikzpicture}
        \caption{}
    \end{subfigure}
 	\caption{$A_1$ Kraft-Procesi transition. (a) represents the Higgs branch of the Model with $n_s=n_d=4$, $\vec{l}_s=(0,1,1,2)$ and $\vec{l}_d=(3,3,3,3)$. (b) represents the singular point when $\vec{y}=\vec{w}_1=\vec{w}_2$. (c) represents the new model after taking $\vec{x}$ to infinity. The new linking numbers are $\vec{l}_s=(0,0,2,2)$, $\vec{l}_d=(3,3,3,3)$. This defines a new model which fulfils the prescription to have the Higgs branch as the closure of the nilpotent orbit with $\lambda^t=(2^2)$, hence $\lambda=(2^2)$. This is the \emph{next to minimal} orbit of $\mathfrak{sl}_4$.}
	\label{fig:SU4transA1}
\end{figure}
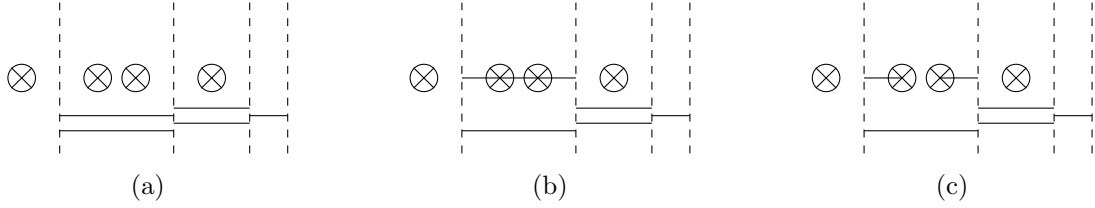

So far we have found two models and their mirrors, corresponding to the closure of two nilpotent orbits of $\mathfrak{sl}_4$: the maximal orbit $\lambda=(4)$, and the known as \emph{subregular} $\lambda=(3,1)$. We have also established the existence of an $A_3$ KP transition that takes the model of the maximal to the model of the subregular. Let us continue, exploring now what are the KP transitions that can be performed in the model correspondent to $\lambda=(3,1)$. Let us once more choose the model with $\mathcal{M}_H=\bar{\mathcal{O}}_{(3,1)}$, its Higgs brane configuration should be taken to the phase where all fixed threebranes have been annihilated. This can be achieved by performing two Hanany-Witten transitions in fig. \ref{fig:SU4transA3} (c). The result is fig. \ref{fig:SU4transA1} (a).\\

Let us find all possible minimal singularities $\mathcal{V}_i\subset\bar{\mathcal{O}}_{(3,1)}$. There are two NS5-branes in the first interval between D5-branes starting from the left. This determines the existence of a minimal singularity $A_1$ and a subsequent $A_1$ KP transition. There is one NS5-brane in the second interval between D5-branes. This could be one end of an $a_n$ minimal singularity, however there are no other cases of intervals containing exactly one NS5-brane. Therefore there is no such minimal singularity as a subvariety of the Higgs branch. The only minimal singularity that can be found is therefore $\mathcal{V}= A_1$.\\

Hence, we can perform an $A_1$ KP transition Higgsing away one D3-brane from the leftmost interval between D5-branes. The process (see fig. \ref{fig:SU4transA1}) is by now familiar, we align the D3-brane that generates $\mathcal{V}$ with the 2 NS5-branes present in the interval:

\begin{align}
	\vec{y}=\vec{w}_1=\vec{w}_2
\end{align}

We split the D3-brane into three segments, the middle one now stretches between two NS5-branes, and its position $\vec{x}$ is part of a vectorplet that has become massless and admits nonzero VEVs $(\vec{x},a)$ when the hyper containing the fields  $(\vec{y},b)$  becomes massive. By taking $\vec{x}$ to infinity we fully remove this D3-brane segment from the model and reach a new model with new linking numbers for the NS5-branes $\vec{l}_s=(0,0,2,2)$.\\

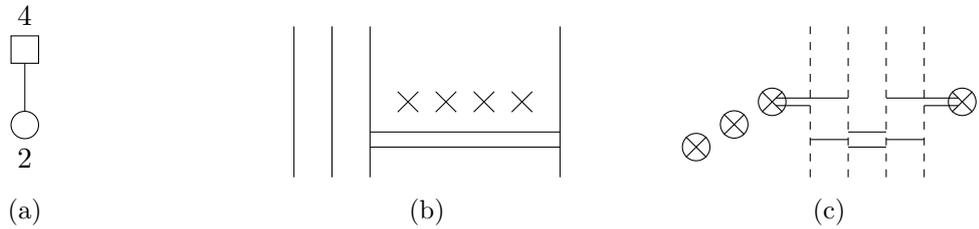
\begin{figure}[h]
	\centering
	\begin{subfigure}[t]{.3\textwidth}
    \centering
	\begin{tikzpicture}
	\tikzstyle{gauge} = [circle, draw];
	\tikzstyle{flavour} = [regular polygon,regular polygon sides=4,draw];
	\node (g3) [gauge,right of=g2,label=below:{$2$}] {};
	\node (f3) [flavour,above of=g3,label=above:{$4$}] {};
	\draw (g3)--(f3)
		;
	\end{tikzpicture}
        \caption{}
    \end{subfigure}
    \hfill
	\begin{subfigure}[t]{.3\textwidth}
    \centering
	\begin{tikzpicture}
		\draw (1,0)--(1,2);
		\draw (1.5,0)--(1.5,2);
		\draw (2,0)--(2,2);
		\draw (4.5,0)--(4.5,2);
		\draw 	(2,.4)--(4.5,.4)
				(2,.6)--(4.5,.6);
		\draw (2.5,1) node[cross] {};
		\draw (3,1) node[cross] {};
		\draw (3.5,1) node[cross] {};
		\draw (4,1) node[cross] {};
	\end{tikzpicture}

        \caption{}
    \end{subfigure}
    \hfill
    \begin{subfigure}[t]{.3\textwidth}
    \centering
	\begin{tikzpicture}
		\draw[dashed] 	(2.5,0)--(2.5,2)
						(3,0)--(3,2)
						(3.5,0)--(3.5,2)
						(4,0)--(4,2);
		\draw (1,.4) node[cross] {};
		\draw (1.5,.7) node[cross] {};
		\draw (2,1) node[cross] {};
		\draw (4.5,1) node[cross] {};
		\draw (1,.4) node[circle,draw] {};
		\draw (1.5,.7) node[circle,draw] {};
		\draw (2,1) node[circle,draw] {};
		\draw (4.5,1) node[circle,draw] {};
		\draw 	(2.05,1.05)--(3,1.05)
				(2.05,.95)--(2.5,.95);
		\draw	(4.45,1.05)--(3.5,1.05)
				(4.45,.95)--(4,.95);
		\draw 	(4,.5)--(3.5,.5)
				(3.5,.4)--(3,.4)
				(3.5,.6)--(3,.6)
				(3,.5)--(2.5,.5)
				;
	\end{tikzpicture}
        \caption{}
    \end{subfigure}
 	\caption{Model with $n_s=n_d=4$, $\vec{l}_s=(0,0,2,2)$ and $\vec{l}_d=(3,3,3,3)$. (a) is the quiver. (b) represents the Coulomb branch. (c) represents the Higgs branch.}
	\label{fig:SU4ntm}
\end{figure}

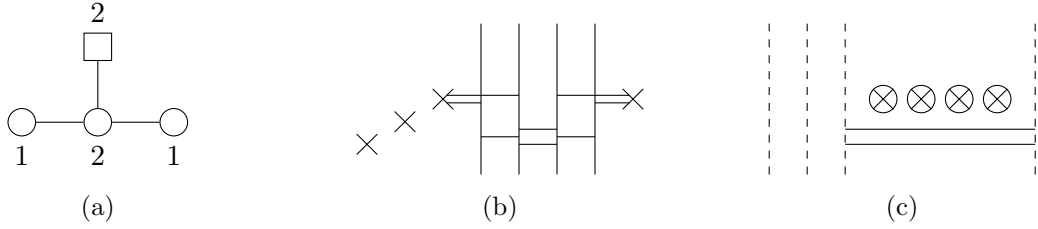
\begin{figure}[h]
	\centering
	\begin{subfigure}[t]{.3\textwidth}
    \centering
	\begin{tikzpicture}
	\tikzstyle{gauge} = [circle,draw];
	\tikzstyle{flavour} = [regular polygon,regular polygon sides=4,draw];
	\node (g1) [gauge,label=below:{$1$}] {};
	\node (g2) [gauge,right of=g1,label=below:{$2$}] {};
	\node (g3) [gauge,right of=g2,label=below:{$1$}] {};
	\node (f2) [flavour,above of=g2,label=above:{$2$}] {};
	\draw (g1)--(g2)--(g3)
			(f2)--(g2)
		;
	\end{tikzpicture}
        \caption{}
    \end{subfigure}
      \hfill
    \begin{subfigure}[t]{.3\textwidth}
    \centering
	\begin{tikzpicture}
		\draw 	(2.5,0)--(2.5,2)
						(3,0)--(3,2)
						(3.5,0)--(3.5,2)
						(4,0)--(4,2);
		\draw (1,.4) node[cross] {};
		\draw (1.5,.7) node[cross] {};
		\draw (2,1) node[cross] {};
		\draw (4.5,1) node[cross] {};
		\draw 	(2.05,1.05)--(3,1.05)
				(2.05,.95)--(2.5,.95);
		\draw	(4.45,1.05)--(3.5,1.05)
				(4.45,.95)--(4,.95);
		\draw 	(4,.5)--(3.5,.5)
				(3.5,.4)--(3,.4)
				(3.5,.6)--(3,.6)
				(3,.5)--(2.5,.5)
				;
	\end{tikzpicture}
        \caption{}
    \end{subfigure}
    \hfill
	\begin{subfigure}[t]{.3\textwidth}
    \centering
	\begin{tikzpicture}
		\draw[dashed] (1,0)--(1,2)
				(1.5,0)--(1.5,2)
				(2,0)--(2,2)
				(4.5,0)--(4.5,2);
		\draw 	(2,.4)--(4.5,.4)
				(2,.6)--(4.5,.6);
		\draw (2.5,1) node[cross] {};
		\draw (3,1) node[cross] {};
		\draw (3.5,1) node[cross] {};
		\draw (4,1) node[cross] {};
		\draw (2.5,1) node[circle,draw] {};
		\draw (3,1) node[circle,draw] {};
		\draw (3.5,1) node[circle,draw] {};
		\draw (4,1) node[circle,draw] {};
	\end{tikzpicture}

        \caption{}
    \end{subfigure}
 	\caption{Model with $n_s=n_d=4$, $\vec{l}_s=(3,3,3,3)$ and $\vec{l}_d=(0,0,2,2)$. (a) is the quiver. (b) represents the Coulomb branch. (c) represents the Higgs branch.}
	\label{fig:SU4ntmMirror}
\end{figure}

The new model fulfills the conditions to have $\mathcal{M}_H=\bar{\mathcal{O}}_{(2^2)}$. We have reached the closure of a new nilpotent orbit of $\mathfrak{sl}_4$. The Coulomb brane configuration and the quiver can be computed as usual and are displayed in fig. \ref{fig:SU4ntm}. The mirror model with $\vec{l}_d=(0,0,2,2)$ and $\vec{l}_s=(3,3,3,3)$ and $\mathcal{M}_C=\bar{\mathcal{O}}_{(2^2)}$ can also be calculated and are shown in fig. \ref{fig:SU4ntmMirror}.\\

\begin{figure}[h]
	\centering
	\begin{subfigure}[t]{.3\textwidth}
    \centering
	\begin{tikzpicture}
		\draw[dashed] 	(-1,0)--(-1,2)
					(-1.5,0)--(-1.5,2)
					(-3,0)--(-3,2)
					(-3.5,0)--(-3.5,2);
		\draw 	(-1,.5)--(-1.5,.5)
				(-1.5,.4)--(-3,.4)
				(-1.5,.6)--(-3,.6)
				(-3,.5)--(-3.5,.5);
		\draw (-2,1) node[cross] {};
		\draw (-2.5,1) node[cross] {};
		\draw (-4,1) node[cross] {};
		\draw (-4.5,1) node[cross] {};
		\draw (-2,1) node[circle,draw] {};
		\draw (-2.5,1) node[circle,draw] {};
		\draw (-4,1) node[circle,draw] {};
		\draw (-4.5,1) node[circle,draw] {};
	\end{tikzpicture}
        \caption{}
    \end{subfigure}
    \hfill
	\begin{subfigure}[t]{.3\textwidth}
    \centering
	\begin{tikzpicture}
		\draw[dashed] 	(-1,0)--(-1,2)
					(-1.5,0)--(-1.5,2)
					(-3,0)--(-3,2)
					(-3.5,0)--(-3.5,2);
		\draw 	(-1,.5)--(-1.5,.5)
				(-1.5,.4)--(-3,.4)
				(-1.5,1)--(-3,1)
				(-3,.5)--(-3.5,.5);
		\draw (-2,1) node[cross] {};
		\draw (-2.5,1) node[cross] {};
		\draw (-4,1) node[cross] {};
		\draw (-4.5,1) node[cross] {};
		\draw (-2,1) node[circle,draw] {};
		\draw (-2.5,1) node[circle,draw] {};
		\draw (-4,1) node[circle,draw] {};
		\draw (-4.5,1) node[circle,draw] {};
	\end{tikzpicture}

        \caption{}
    \end{subfigure}
    \hfill
    \begin{subfigure}[t]{.3\textwidth}
    \centering
	\begin{tikzpicture}
		\draw[dashed] 	(-1,0)--(-1,2)
					(-1.5,0)--(-1.5,2)
					(-3,0)--(-3,2)
					(-3.5,0)--(-3.5,2);
		\draw 	(-1,.5)--(-1.5,.5)
				(-1.5,.4)--(-3,.4)
				(-1.5,1)--(-2,1)
				(-2.5,1)--(-3,1)
				(-3,.5)--(-3.5,.5);
		\draw (-2,1) node[cross] {};
		\draw (-2.5,1) node[cross] {};
		\draw (-4,1) node[cross] {};
		\draw (-4.5,1) node[cross] {};
		\draw (-2,1) node[circle,draw] {};
		\draw (-2.5,1) node[circle,draw] {};
		\draw (-4,1) node[circle,draw] {};
		\draw (-4.5,1) node[circle,draw] {};
	\end{tikzpicture}
        \caption{}
    \end{subfigure}
 	\caption{$A_1$ Kraft-Procesi transition. (a) represents the Higgs branch of the Model with $n_s=n_d=4$, $\vec{l}_s=(0,0,2,2)$ and $\vec{l}_d=(3,3,3,3)$. (b) represents the singular point when $\vec{y}=\vec{w}_1=\vec{w}_2$. (c) represents the new model after taking $\vec{x}$ to infinity. The new linking numbers are $\vec{l}_s=(0,0,1,3)$, $\vec{l}_d=(3,3,3,3)$. This defines a new model which fulfils the prescription to have the Higgs branch as the closure of the nilpotent orbit with $\lambda^t=(3,1)$ and hence $\lambda=(2,1^2)$. This is the minimal orbit of $\mathfrak{sl}_4$.}
	\label{fig:SU4transa1}
\end{figure}
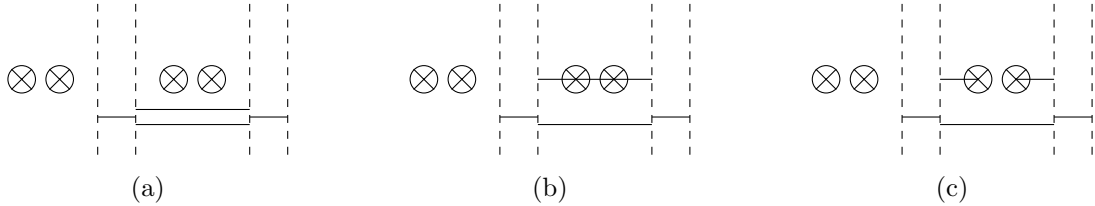

Now we have found the models for closures of orbits $\lambda=(4)$, $\lambda=(3,1)$, $\lambda=(2^2)$ and their mirror duals. We have also found an $A_3$ KP transition from $\lambda=(4)$ to $\lambda=(3,1)$ and an $A_1$ KP transition from $\lambda=(3,1)$ to $\lambda=(2^2)$. Let us explore next what KP transitions can be performed on the model corresponding to $\lambda=(2^2)$. We perform a phase transition on the Higgs brane configuration on the model with $\mathcal{M}_H=\bar{\mathcal{O}}_{(2^2)}$, fig. \ref{fig:SU4transA1} (c), to annihilate all fixed threebranes. The result is \ref{fig:SU4transa1} (a).\\

In this Higgs brane configuration, the second interval between D5-branes starting from the left is the only one containing NS5-branes. It contains two of them, so it corresponds to a $A_1$ minimal singularity of $\bar{\mathcal{O}}_{(2^2)}$. Therefore there is only one possible $A_1$ KP transition that can be performed.\\

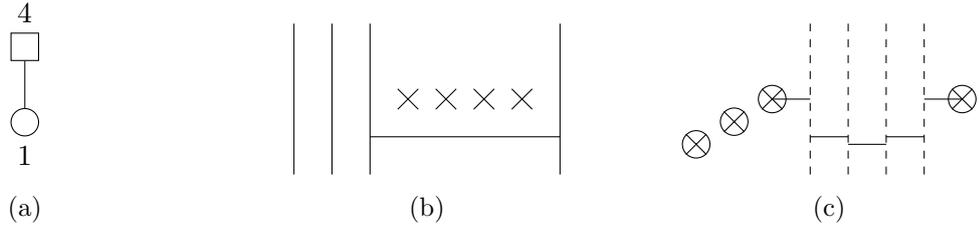
\begin{figure}[h]
	\centering
	\begin{subfigure}[t]{.3\textwidth}
    \centering
	\begin{tikzpicture}
	\tikzstyle{gauge} = [circle, draw];
	\tikzstyle{flavour} = [regular polygon,regular polygon sides=4,draw];
	\node (g3) [gauge,right of=g2,label=below:{$1$}] {};
	\node (f3) [flavour,above of=g3,label=above:{$4$}] {};
	\draw (g3)--(f3)
		;
	\end{tikzpicture}
        \caption{}
    \end{subfigure}
    \hfill
	\begin{subfigure}[t]{.3\textwidth}
    \centering
	\begin{tikzpicture}
		\draw (1,0)--(1,2);
		\draw (1.5,0)--(1.5,2);
		\draw (2,0)--(2,2);
		\draw (4.5,0)--(4.5,2);
		\draw 	(2,.5)--(4.5,.5);
		\draw (2.5,1) node[cross] {};
		\draw (3,1) node[cross] {};
		\draw (3.5,1) node[cross] {};
		\draw (4,1) node[cross] {};
	\end{tikzpicture}

        \caption{}
    \end{subfigure}
    \hfill
    \begin{subfigure}[t]{.3\textwidth}
    \centering
	\begin{tikzpicture}
		\draw[dashed] 	(2.5,0)--(2.5,2)
						(3,0)--(3,2)
						(3.5,0)--(3.5,2)
						(4,0)--(4,2);
		\draw (1,.4) node[cross] {};
		\draw (1.5,.7) node[cross] {};
		\draw (2,1) node[cross] {};
		\draw (4.5,1) node[cross] {};
		\draw (1,.4) node[circle,draw] {};
		\draw (1.5,.7) node[circle,draw] {};
		\draw (2,1) node[circle,draw] {};
		\draw (4.5,1) node[circle,draw] {};
		\draw 	(2,1)--(2.5,1);
		\draw	(4.5,1)--(4,1);
		\draw 	(4,.5)--(3.5,.5)
				(3.5,.4)--(3,.4)
				(3,.5)--(2.5,.5)
				;
	\end{tikzpicture}
        \caption{}
    \end{subfigure}
 	\caption{Model with $n_s=n_d=4$, $\vec{l}_s=(0,0,1,3)$ and $\vec{l}_d=(3,3,3,3)$. (a) is the quiver. (b) represents the Coulomb branch. (c) represents the Higgs branch.}
	\label{fig:SU4minimal}
\end{figure}

\begin{figure}[h]
	\centering
	\begin{subfigure}[t]{.3\textwidth}
    \centering
	\begin{tikzpicture}
	\tikzstyle{gauge} = [circle, draw];
	\tikzstyle{flavour} = [regular polygon,regular polygon sides=4,draw];
	\node (g1) [gauge,label=below:{$1$}] {};
	\node (g2) [gauge,right of=g1,label=below:{$1$}] {};
	\node (g3) [gauge,right of=g2,label=below:{$1$}] {};
	\node (f1) [flavour,above of=g1,label=above:{$1$}] {};
	\node (f3) [flavour,above of=g3,label=above:{$1$}] {};
	\draw (g1)--(g2)--(g3)
			(f1)--(g1)
			(f3)--(g3)
		;
	\end{tikzpicture}
        \caption{}
    \end{subfigure}
      \hfill
    \begin{subfigure}[t]{.3\textwidth}
    \centering
	\begin{tikzpicture}
		\draw	(2.5,0)--(2.5,2)
						(3,0)--(3,2)
						(3.5,0)--(3.5,2)
						(4,0)--(4,2);
		\draw (1,.4) node[cross] {};
		\draw (1.5,.7) node[cross] {};
		\draw (2,1) node[cross] {};
		\draw (4.5,1) node[cross] {};
		\draw 	(2,1)--(2.5,1);
		\draw	(4.5,1)--(4,1);
		\draw 	(4,.5)--(3.5,.5)
				(3.5,.4)--(3,.4)
				(3,.5)--(2.5,.5)
				;
	\end{tikzpicture}
        \caption{}
    \end{subfigure}
    \hfill
	\begin{subfigure}[t]{.3\textwidth}
    \centering
	\begin{tikzpicture}
		\draw[dashed] (1,0)--(1,2)
				(1.5,0)--(1.5,2)
				(2,0)--(2,2)
				(4.5,0)--(4.5,2);
		\draw 	(2,.5)--(4.5,.5);
		\draw (2.5,1) node[cross] {};
		\draw (3,1) node[cross] {};
		\draw (3.5,1) node[cross] {};
		\draw (4,1) node[cross] {};
		\draw (2.5,1) node[circle,draw] {};
		\draw (3,1) node[circle,draw] {};
		\draw (3.5,1) node[circle,draw] {};
		\draw (4,1) node[circle,draw] {};
	\end{tikzpicture}

        \caption{}
    \end{subfigure}
 	\caption{Model with $n_s=n_d=4$, $\vec{l}_s=(3,3,3,3)$ and $\vec{l}_d=(0,0,1,3)$. (a) is the quiver. (b) represents the Coulomb branch. (c) represents the Higgs branch.}
	\label{fig:SU4minimalMirror}
\end{figure}
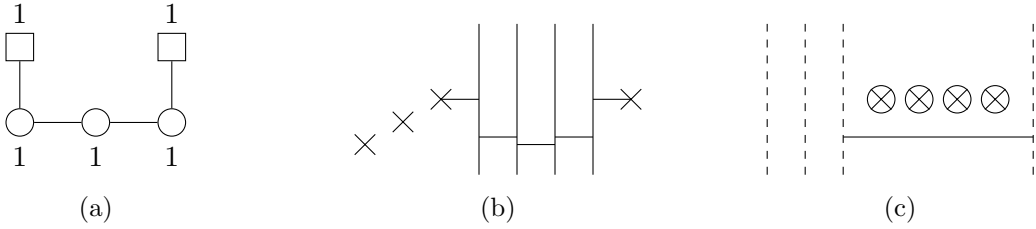

The $A_1$ KP transition is depicted in fig. \ref{fig:SU4transa1}. The resulting model has linking numbers $\vec{l}_s=(0,0,1,3)$ and $\vec{l}_d=(3,3,3,3)$. Its Coulomb brane configuration and quiver can be computed as usual, they are shown in fig. \ref{fig:SU4minimal}. This is the model with $\mathcal{M}_H=\bar{\mathcal{O}}_{(2,1^2)}=a_3$ and $\mathcal{M}_C=A_3$. The Higgs branch is therefore the closure of the minimal nilpotent orbit of $\mathfrak{sl}_4$. The mirror model is depicted in fig. \ref{fig:SU4minimalMirror}.\\

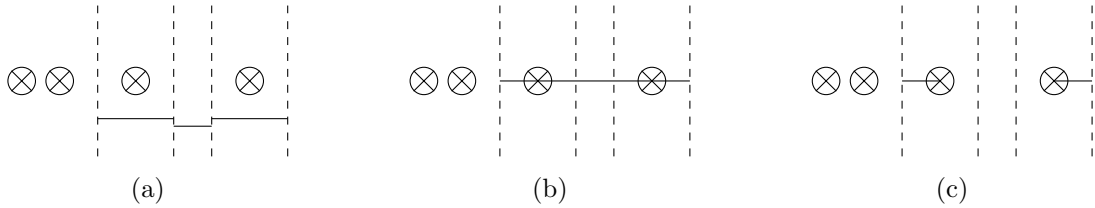
\begin{figure}[h]
	\centering
	\begin{subfigure}[t]{.3\textwidth}
    \centering
	\begin{tikzpicture}
		\draw[dashed] 	(-1,0)--(-1,2)
					(-2,0)--(-2,2)
					(-2.5,0)--(-2.5,2)
					(-3.5,0)--(-3.5,2);
		\draw 	(-1,.5)--(-2,.5)
				(-2,.4)--(-2.5,.4)
				(-2.5,.5)--(-3.5,.5);
		\draw (-1.5,1) node[cross] {};
		\draw (-3,1) node[cross] {};
		\draw (-4,1) node[cross] {};
		\draw (-4.5,1) node[cross] {};
		\draw (-1.5,1) node[circle,draw] {};
		\draw (-3,1) node[circle,draw] {};
		\draw (-4,1) node[circle,draw] {};
		\draw (-4.5,1) node[circle,draw] {};
	\end{tikzpicture}
        \caption{}
    \end{subfigure}
    \hfill
	\begin{subfigure}[t]{.3\textwidth}
    \centering
	\begin{tikzpicture}
		\draw[dashed] 	(-1,0)--(-1,2)
					(-2,0)--(-2,2)
					(-2.5,0)--(-2.5,2)
					(-3.5,0)--(-3.5,2);
		\draw 	(-1,1)--(-3.5,1);
		\draw (-1.5,1) node[cross] {};
		\draw (-3,1) node[cross] {};
		\draw (-4,1) node[cross] {};
		\draw (-4.5,1) node[cross] {};
		\draw (-1.5,1) node[circle,draw] {};
		\draw (-3,1) node[circle,draw] {};
		\draw (-4,1) node[circle,draw] {};
		\draw (-4.5,1) node[circle,draw] {};
	\end{tikzpicture}

        \caption{}
    \end{subfigure}
    \hfill
    \begin{subfigure}[t]{.3\textwidth}
    \centering
	\begin{tikzpicture}
		\draw[dashed] 	(-1,0)--(-1,2)
					(-2,0)--(-2,2)
					(-2.5,0)--(-2.5,2)
					(-3.5,0)--(-3.5,2);
		\draw 	(-1,1)--(-1.5,1)
				(-3,1)--(-3.5,1);
		\draw (-1.5,1) node[cross] {};
		\draw (-3,1) node[cross] {};
		\draw (-4,1) node[cross] {};
		\draw (-4.5,1) node[cross] {};
		\draw (-1.5,1) node[circle,draw] {};
		\draw (-3,1) node[circle,draw] {};
		\draw (-4,1) node[circle,draw] {};
		\draw (-4.5,1) node[circle,draw] {};
	\end{tikzpicture}
        \caption{}
    \end{subfigure}
 	\caption{$a_3$ Kraft-Procesi transition. (a) represents the Higgs branch of the Model with $n_s=n_d=4$, $\vec{l}_s=(0,0,1,3)$ and $\vec{l}_d=(3,3,3,3)$. (b) represents the singular point when $\vec{y}_1=\vec{y}_2=\vec{y}_3=\vec{w}_1=\vec{w}_2$. (c) represents the new model after taking $\vec{x}$ to infinity. The new linking numbers are $\vec{l}_s=(0,0,0,4)$, $\vec{l}_d=(3,3,3,3)$. This defines a new model which fulfils the prescription to have the Higgs branch as the closure of the nilpotent orbit with $\lambda=(1^4)$. This is the trivial orbit of $\mathfrak{sl}_4$.}
	\label{fig:SU4transa3}
\end{figure}

So far we have found models corresponding with the closures of the nilpotent orbits $\lambda=(4)$, $\lambda=(3,1)$, $\lambda=(2^2)$, $\lambda=(2,1^2)$ and their mirror models. We also found KP transitions $A_3$, $A_1$ and $A_1$ between each of the orbits. Let us study the KP that can be performed in the model corresponding to the closure for the minimal nilpotent orbit $\lambda=(2,1^2)$. Once more we find the Higgs brane configuration in which there are no fixed threebranes for the model with $\M_H=\Or_{(2,1^2)}$. This is depicted in fig. \ref{fig:SU4transa3} (a).\\

Let us find the minimal singularity. There is one NS5-brane in the first interval between D5-branes starting from the left, together with the NS5-brane in the last interval and the empty intervals in between give rise to an $a_n$ singularity, where $n$ is the number of intervals. In this case it is an $a_3$ minimal singularity. This is the only possible minimal singularity and in this case it also corresponds with the Higgs branch. Therefore there is an $a_3$ KP transition that can be performed in the closure of the orbit $a_3$. This transition is depicted in fig. \ref{fig:SU4transa3}. The result is a model with no D3-branes, in the Higgs branch. Therefore a Higgs branch with dimension zero, just one point, the trivial variety. This corresponds with the closure of the orbit with $\lambda=(1^4)$.\\

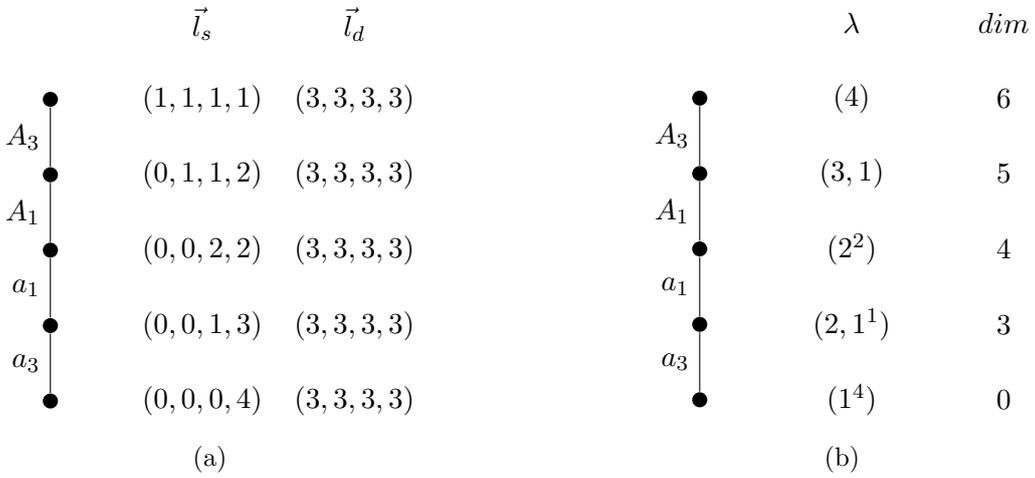
\begin{figure}[h]
	\centering
\begin{subfigure}[t]{.45\textwidth}
    \centering
	\begin{tikzpicture}
		\tikzstyle{hasse} = [circle, fill,inner sep=2pt];
		\node [hasse] (1) [] {};
		\node [hasse] (2) [below of=1] {};
		\node [hasse] (3) [below of=2] {};
		\node [hasse] (4) [below of=3] {};
		\node [hasse] (5) [below of=4] {};
		\draw (1) edge [] node[label=left:$A_3$] {} (2)
			(2) edge [] node[label=left:$A_1$] {} (3)
			(3) edge [] node[label=left:$a_1$] {} (4)
			(4) edge [] node[label=left:$a_3$] {} (5);
		\node (e1) [right of=1] {};
		\node (d1) [right of=e1] {$(1,1,1,1)$};
		\node (c1) [right of=d1] {};
		\node (b1) [right of=c1] {$(3,3,3,3)$};
		\node (e2) [right of=2] {};
		\node (d2) [right of=e2] {$(0,1,1,2)$};
		\node (c2) [right of=d2] {};
		\node (b2) [right of=c2] {$(3,3,3,3)$};
		\node (e3) [right of=3] {};
		\node (d3) [right of=e3] {$(0,0,2,2)$};
		\node (c3) [right of=d3] {};
		\node (b3) [right of=c3] {$(3,3,3,3)$};
		\node (e4) [right of=4] {};
		\node (d4) [right of=e4] {$(0,0,1,3)$};
		\node (c4) [right of=d4] {};
		\node (b4) [right of=c4] {$(3,3,3,3)$};
		\node (e5) [right of=5] {};
		\node (d5) [right of=e5] {$(0,0,0,4)$};
		\node (c5) [right of=d5] {};
		\node (b5) [right of=c5] {$(3,3,3,3)$};
		\node (d) [above of=d1] {$\vec{l}_s$};
		\node (b) [above of=b1] {$\vec{l}_d$};
	\end{tikzpicture}
	\caption{}
 	\end{subfigure}
 	\hfill
 	\begin{subfigure}[t]{.45\textwidth}
    \centering
	\begin{tikzpicture}
		\tikzstyle{hasse} = [circle, fill,inner sep=2pt];
		\node [hasse] (1) [] {};
		\node [hasse] (2) [below of=1] {};
		\node [hasse] (3) [below of=2] {};
		\node [hasse] (4) [below of=3] {};
		\node [hasse] (5) [below of=4] {};
		\draw (1) edge [] node[label=left:$A_3$] {} (2)
			(2) edge [] node[label=left:$A_1$] {} (3)
			(3) edge [] node[label=left:$a_1$] {} (4)
			(4) edge [] node[label=left:$a_3$] {} (5);
		\node (e1) [right of=1] {};
		\node (d1) [right of=e1] {$(4)$};
		\node (c1) [right of=d1] {};
		\node (b1) [right of=c1] {$6$};
		\node (e2) [right of=2] {};
		\node (d2) [right of=e2] {$(3,1)$};
		\node (c2) [right of=d2] {};
		\node (b2) [right of=c2] {$5$};
		\node (e3) [right of=3] {};
		\node (d3) [right of=e3] {$(2^2)$};
		\node (c3) [right of=d3] {};
		\node (b3) [right of=c3] {$4$};
		\node (e4) [right of=4] {};
		\node (d4) [right of=e4] {$(2,1^1)$};
		\node (c4) [right of=d4] {};
		\node (b4) [right of=c4] {$3$};
		\node (e5) [right of=5] {};
		\node (d5) [right of=e5] {$(1^4)$};
		\node (c5) [right of=d5] {};
		\node (b5) [right of=c5] {$0$};
		\node (d) [above of=d1] {$\lambda$};
		\node (b) [above of=b1] {$dim$};
	\end{tikzpicture}
	\caption{}
 	\end{subfigure}
	\caption{Hasse diagram for the models whose Higgs branch is the closure of a nilpotent orbit of $\mathfrak{sl}_4$. (a) represents the brane configurations, where the linking numbers $\vec{l}_s$ and $\vec{l}_d$ are provided for each orbit. (b) depicts the information of the orbits. With respect to the brane configurations $\lambda$ is the transpose partition of $\vec{l}_s$, and $dim$ is the number of D3-branes that generate the Higgs branch in each model. This second diagram also corresponds to the one found in \cite{Kraft1982}.}
	\label{fig:HasseSU4}
	
\end{figure}

This was the last remaining partition of $N=4$. Therefore we have found models corresponding to the closures of all nilpotent orbits of $\mathfrak{sl}_4$, their mirror duals and all the KP transitions. All of them were found just by starting with the self-dual model corresponding to the closure of the maximal partition, performing \emph{all} possible KP transitions on it, then performing all possible KP transitions in the resulting models, etc., iterating until the trivial orbit was reached. This procedure can always be implemented, starting with the model corresponding to the closure of the  maximal nilpotent orbit of any algebra of the form $\mathfrak{sl}_N$. We can summarize once more all models that we found in this case and all KP transitions between them in a Hasse diagram, fig. \ref{fig:HasseSU4}.

\section{The Matrix Formalism}\label{sec:F}

It was shown in the previous section that KP transitions can be used to find all models such that either its Higgs branch or its Coulomb branch is the closure of a nilpotent orbit of $\mathfrak{sl}_N$. This can be done by starting with the self-dual model corresponding to the closure of the maximal nilpotent orbit $\M_C=\M_H=\Or_{(N)}$. In this section we present a way to perform this KP transitions in an efficient way, encoding the data of the Higgs brane configurations into $2\times (N+1)$ matrices $M$ with integer elements.\\

\subsection{The Formalism}

The first step to turn the Higgs brane configuration into a matrix is the same as the first step in a KP transition: take the Higgs brane configuration to a phase where all fixed threebranes have been annihilated.\\

Once this is done a $2\times (N+1)$ matrix $M$ can be written\footnote{One should not mistake these matrices with the matrices defined in Section \ref{sec:HBC}. They are entirely different objects.}, such that $M_{1j}$ is the number of NS5-branes contained in the $j$th interval between D5-branes, starting from the left, and assigning $j=1$ to the section to the left of the leftmost D5-brane and $j=N+1$ to the section to the right of the rightmost D5-brane. Similarly, $M_{2j}$ is the number of D3-branes in the $j$th interval between D5-branes.\\

For example, the Higgs brane configuration for the self-dual model corresponding to the closure of the maximal orbit of $\mathfrak{sl}_3$, with partition $\lambda=(3)$, depicted in fig. \ref{fig:SU3KPA3} (a),  has matrix:

\begin{align}
	M(\lambda)=\left(\begin{array}{cccc}
		0&3&0&0\\
		0&2&1&0\\
		\end{array}\right)
\end{align}

More generally, for the self-dual model corresponding to the closure of the orbit of the maximal nilpotent orbit of $\mathfrak{sl}_N$ with $\lambda=(N)$ we have:

\begin{align}
	M(\lambda)=\left(\begin{array}{clllccc}
		0&N&0&\dots&0&0&0\\
		0&N-1&N-2&\dots &2&1&0\\
		\end{array}\right)
\end{align}

\subsection{$A_n$ KP Transition}

We only need to consider the elements in the first row that are different from zero in order to find the minimal singularities. This is because there would be NS5-branes in that interval that we can use to perform the Higgsing.\\

The $A_n$ transition is always performed on a single interval, let us consider the minimal singularity on its own, its matrix is:\\ 

\begin{align}
M({A_n})=\left(\begin{array}{ccc}
		0&(n+1)&0\\
		0&1&0\\
		\end{array}\right)
\end{align}

The $A_n$ KP transition removes the D3-brane, and two of the NS5-branes can move to the right and to the left of the interval, passing through the right and left D5-branes and annihilating fixed threebranes via Hanany-Witten transitions. The matrix of the resulting model is:
\begin{align}
M=\left(\begin{array}{ccc}
		1&(n-1)&1\\
		0&0&0\\
		\end{array}\right)
\end{align}

Therefore, whenever we find this as part of the nilpotent orbit matrix, this transition can be performed, removing one D3-brane, and moving away two NS5-branes to annihilate the fixed D3-segments and obtain a new phase that can be encoded in a matrix. For a generic nilpotent orbit $\lambda$ of the algebra $\mathfrak{sl}_N$, any element of the matrix $M(\lambda)$ of the form $M_{1j}>1$ such that $j$ is neither $1$ or $N+1$ gives rise to an  $A_{n}$ transition where $n=M_{1j}-1$.\\

For example in the maximal orbit of $\mathfrak{sl}_3$, with partition $\lambda=(3)$ and matrix:

\begin{align}
	M(\lambda)=\left(\begin{array}{cccc}
		0&3&0&0\\
		0&2&1&0\\
		\end{array}\right)
\end{align}

we find that there is an $A_2$ singularity of the form: \begin{align}
	M(A_2)=\left(\begin{array}{ccc}
		0&3&0\\
		0&1&0\\
		\end{array}\right)
\end{align}

embedded on the second column of $M(\lambda)$. If we remove it, by performing an $A_2$ KP transition, we find:

\begin{align}
	M(\lambda')=\left(\begin{array}{cccc}
		1&1&1&0\\
		0&1&1&0\\
		\end{array}\right)
\end{align}

The linking numbers of the NS fivebranes have changed to $\vec{l}_s=(0,1,2)$, corresponding to $\lambda'^t=(2,1)$. Therefore the new matrix corresponds to partition $\lambda'=(2,1)$. We can easily check that this corresponds with the brane configuration in fig. \ref{fig:SU3KPa3} (a). To summarize, this matrix manipulation corresponds to the $A_2$ KP transition depicted in fig. \ref{fig:SU3KPA3}.

\subsection{$a_n$ KP Transition}

This type of transition always involve only two NS5-branes. If they are in the same interval we say it is $a_1=A_1$, if they are in adjacent intervals we say it is $a_2$, etc.\\

 Since $a_1$ is already accounted for, we only need to be concerned about $a_n$ KP transitions with $n>1$. The minimal singularity has a matrix of the form:
\begin{align}
	M(a_n)=\left(\begin{array}{ccccccc}
		0&1&0&\dots&0&1&0\\
		0&1&1&\dots &1&1&0\\
		\end{array}\right)
\end{align}
		where there are $n-2$ columns with 0s in the first row and 1s in the second, between the two columns with $1$s in the first row. The transition is to:
\begin{align}
M=\left(\begin{array}{ccccccc}
		1&0&0&\dots&0&0&1\\
		0&0&0&\dots &0&0&0\\
		\end{array}\right)
\end{align}
where the $n$ D3-branes have been removed. For $a_n$ with $n>1$ to be a minimal KP singularity the NS5-branes have to be alone in their intervals, i.e. we cannot use a NS5-brane from an interval with $M_{1j}>1$ to perform an $a_n$ different from $a_1$ if we want to restrict the transitions to minimal KP singularities.\\

For example, there is an $a_2$ singularity in the brane system for $\lambda=(2,1)$:

\begin{align}
		M(\lambda)=\left(\begin{array}{cccc}
		1&1&1&0\\
		0&1&1&0\\
		\end{array}\right)
\end{align}

The result after performing an $a_2$ KP transition is:
\begin{align}
		M(\lambda')=\left(\begin{array}{cccc}
		2&0&0&1\\
		0&0&0&0\\
		\end{array}\right)
\end{align}		
The model represented by the matrix $M(\lambda')$ has linking numbers $\vec{l}_s=(0,0,3)$, corresponding to $\lambda'^t=(3)$. Therefore the new matrix corresponds to partition $\lambda'=(1^3)$. This is the trivial orbit. This corresponds to the KP transition described in fig. \ref{fig:SU3KPa3}.

\subsection{Example: KP Transitions for all Nilpotent Orbits of $\mathfrak{sl}_4$}

Let us now illustrate once more the matrix formalism by showing the computation for the already familiar $T_{\lambda^t}(SU(4))$ theories. Remember that if we wanted to obtain the quivers we only need to recover the Higgs brane configuration from the matrix of each orbit.\\

The starting matrix corresponds to $\lambda=(4)$:
\begin{align} 
		M(\lambda)=\left(\begin{array}{ccccc}
		0&4&0&0&0\\
		0&3&2&1&0\\
		\end{array}\right)
\end{align}

All the elements in the first row are zero, except for $M_{12}=4$. Since it is bigger than $1$ it corresponds to an $A_n$ transition, with $n=M_{12}-1$, i.e. to $A_3$. So there is only one KP transition, of type $A_3$, that can be performed on the maximal orbit. After performing it we obtain:\\

\begin{align}
M(\lambda)=\left(\begin{array}{ccccc}
		1&2&1&0&0\\
		0&2&2&1&0\\
		\end{array}\right)
\end{align}		
		
		This matrix corresponds to linking numbers $\vec{l}_s=(0,1,1,2)$, giving $\lambda^t=(2,1^2)$ hence $\lambda=(3,1)$.\\
		
Now we repeat the same process, starting by looking at all the elements in the first row of the matrix that are different from $0$. The first and the last columns correspond to NS5-branes outside of any interval between D5-branes. In the second column, corresponding to the first and leftmost interval, there are 2 NS5-branes. They form an $A_1$ minimal singularity with one of the D3-branes in the interval. Hence, an $A_1$ KP partition can be performed. In the third column there is 1 NS5-brane, indicated by $M_{13}=1$, this could give rise to an $a_n$, but it does not, since there is no other column with only 1 NS5-brane that it could pair with.\\

Therefore, there is only one $A_1$ KP transition available form the matrix of $\lambda=(3,1)$. After performing it the result is:

\begin{align}
		M(\lambda)=\left(\begin{array}{ccccc}
		2&0&2&0&0\\
		0&1&2&1&0\\
		\end{array}\right)
\end{align}
		
		with linking numbers $\vec{l}_s=(0,0,2,2)$, giving $\lambda^t=(2^2)$, hence $\lambda=(2^2)$.\\
		
We once again look for KP minimal singularities: only the third column could be a candidate. There are 2 NS5-branes, together with one of the D3-branes make up for a $A_1$ singularity. Therefore an $A_1$ KP transition can be performed, the result is

\begin{align} 
		M(\lambda)=\left(\begin{array}{ccccc}
		2&1&0&1&0\\
		0&1&1&1&0\\
		\end{array}\right)
\end{align}

		with linking numbers $\vec{l}_s=(0,0,1,3)$, giving $\lambda^t=(3,1)$, hence $\lambda=(2,1^2)$.\\

Looking for singularities we find $M_{12}=1$, this is a candidate for an $a_n$ singularity with $n>1$. In this case we can pair it up with the NS5-brane in interval 4, i.e. $M_{14}=1$. Hence, both NS5-branes, from intervals 2 and 4, and the three D3-branes, from intervals 2, 3 and 4, make up an $a_3$ singularity. Removing the singularity via an $a_3$ KP transition we obtain 

\begin{align}
M=\left(\begin{array}{ccccc}
		3&0&0&0&1\\
		0&0&0&0&0\\
		\end{array}\right)
\end{align}

with linking numbers $l_s=(0,0,0,4)$, giving $\lambda^t=(4)$, $\lambda=(1^4)$. This is the minimal orbit and it marks the end of the iteration.\\
		
If we make a diagram where the nodes are the orbits, and there are edges connecting them where we found a KP transition we recover the KP Hasse diagram from fig. \ref{fig:HasseSU4} (b). Note that in this formalism the quaternionic dimension is just:

\begin{align}
	dim:=\sum_j M_{2j}
\end{align}

\section{Results}\label{sec:R}

With the matrix formalism we can write a computer algorithm that is able to calculate all matrices (i.e. all brane configurations and all quivers) and KP transitions for all nilpotent orbits of any $\mathfrak{sl}_N$ algebra, starting from the matrix of the maximal nilpotent orbit.\\

\subsection{Tables with Results from the Matrix Formalism}

In this section we present all the results that have been produced with this algorithm. For each value of $N$ we include a table that contains all matrices for all models of the form $T_{\lambda^t}(SU(N))$. The corresponding partition and quaternionic dimension can be read from the matrix and are also included. The quivers for both $T_{\lambda^t}(SU(N))$ and $T^{\lambda^t}(SU(N))$ can easily be recovered from the matrices, as was shown in the example in the next section. The algorithm can also provide the nature of the KP transition that is required in each step. These have been added to the matrix data in the form of Hasse diagrams\footnote{Note that there is an equivalence $a_1=A_1$}.

\begin{table}[h]
	\centering
	\begin{subfigure}[]{0.1\textwidth}
	    \centering
	\begin{tikzpicture}[node distance=40pt]
		\tikzstyle{hasse} = [circle, fill,inner sep=2pt];
		\node at (0,0)[]{\large{$\mathfrak{sl}_2$}};
		\node at (0,-0.5) [hasse] (1) [] {};
		\node [hasse] (2) [below of=1] {};
		\draw (1) edge [] node[label=left:$A_1$] {} (2);
	\end{tikzpicture}
	\end{subfigure}
	\begin{subfigure}[]{0.5\textwidth}
	\begin{tabular}{ c c c }
	\toprule
	\textbf{Matrix} & \textbf{Partition} & $dim$ \\ 
	\midrule \addlinespace[1.5ex]
	$ \left(\begin{array}{ccc}
 0 & 2 & 0 \\
 0 & 1 & 0 \\
\end{array}\right)$ & 2 & 1 \\ 
	\addlinespace[1.5ex]
$ \left(\begin{array}{ccc}
 1 & 0 & 1 \\
 0 & 0 & 0 \\
\end{array}\right)$ & 1,1 & $0$ \\  \addlinespace[1.5ex]
	\bottomrule
	\end{tabular}
	\end{subfigure}
	\caption{Results obtained applying the matrix formalism to $\mathfrak{sl}_2$.}
\end{table}

%
%
%
%

\begin{table}[h]
	\centering
	\begin{subfigure}[]{0.1\textwidth}
	    \centering
	\begin{tikzpicture}[node distance=40pt]
		\tikzstyle{hasse} = [circle, fill,inner sep=2pt];
		\node at (0,0)[]{\large{$\mathfrak{sl}_3$}};
		\node at (0,-0.5) [hasse] (1) [] {};
		\node [hasse] (2) [below of=1] {};
		\node [hasse] (3) [below of=2] {};
		\draw (1) edge [] node[label=left:$A_2$] {} (2)
			(2) edge [] node[label=left:$a_2$] {} (3);
	\end{tikzpicture}
	\end{subfigure}
	\begin{subfigure}[]{0.5\textwidth}
	\begin{tabular}{ c c c }
	\toprule
	\textbf{Matrix} & \textbf{Partition} & $dim$ \\ 
	\midrule \addlinespace[1.5ex]
	$\left(
\begin{array}{cccc}
 0 & 3 & 0 & 0 \\
 0 & 2 & 1 & 0 \\
\end{array}
\right)$ & 3 & $3$ \\ 
	\addlinespace[1.5ex]
	 $\left(
\begin{array}{cccc}
 1 & 1 & 1 & 0 \\
 0 & 1 & 1 & 0 \\
\end{array}
\right)$ & 2,1 & 2 \\ 
	\addlinespace[1.5ex]
 	$\left(
\begin{array}{cccc}
 2 & 0 & 0 & 1 \\
 0 & 0 & 0 & 0 \\
\end{array}
\right)$ & 1,1,1 & 0 \\ \addlinespace[1.5ex]
	\bottomrule
	\end{tabular}
	\end{subfigure}
		\caption{Results obtained applying the matrix formalism to $\mathfrak{sl}_3$.}
\end{table}

%
%
%
%
%


\begin{table}[h]
	\centering
	\begin{subfigure}[]{0.1\textwidth}
	    \centering
	\begin{tikzpicture}[node distance=40pt]
		\tikzstyle{hasse} = [circle, fill,inner sep=2pt];
		\node at (0,0)[]{\large{$\mathfrak{sl}_4$}};
		\node at (0,-0.5) [hasse] (1) [] {};
		\node [hasse] (2) [below of=1] {};
		\node [hasse] (3) [below of=2] {};
		\node [hasse] (4) [below of=3] {};
		\node [hasse] (5) [below of=4] {};
		\draw (1) edge [] node[label=left:$A_3$] {} (2)
			(2) edge [] node[label=left:$A_1$] {} (3)
			(3) edge [] node[label=left:$a_1$] {} (4)
			(4) edge [] node[label=left:$a_3$] {} (5);
	\end{tikzpicture}
	\end{subfigure}
	\begin{subfigure}[]{0.5\textwidth}
	\centering
	\begin{tabular}{ c l c }
	\toprule
	\textbf{Matrices} & $\lambda$ & $dim$ \\ 
	\midrule	 \addlinespace[1.5ex]
$\left(
\begin{array}{ccccc}
 0 & 4 & 0 & 0 & 0 \\
 0 & 3 & 2 & 1 & 0 \\
\end{array}
\right)$ & 4 & 6 \\	\addlinespace[1.5ex]
 $\left(
\begin{array}{ccccc}
 1 & 2 & 1 & 0 & 0 \\
 0 & 2 & 2 & 1 & 0 \\
\end{array}
\right)$ & 3,1 & 5 \\	\addlinespace[1.5ex]
 $\left(
\begin{array}{ccccc}
 2 & 0 & 2 & 0 & 0 \\
 0 & 1 & 2 & 1 & 0 \\
\end{array}
\right)$ & 2,2 & 4 \\	\addlinespace[1.5ex]
 $\left(
\begin{array}{ccccc}
 2 & 1 & 0 & 1 & 0 \\
 0 & 1 & 1 & 1 & 0 \\
\end{array}
\right)$ & 2,1,1 & 3 \\	\addlinespace[1.5ex]
 $\left(
\begin{array}{ccccc}
 3 & 0 & 0 & 0 & 1 \\
 0 & 0 & 0 & 0 & 0 \\
\end{array}
\right)$ & 1,1,1,1 & 0 \\	\addlinespace[1.5ex]
\bottomrule
	\end{tabular}
	\end{subfigure}
	\caption{Results obtained applying the matrix formalism to $\mathfrak{sl}_4$.}
\end{table}

\pagebreak
\begin{table}[h]
	\centering
	\begin{subfigure}[]{0.1\textwidth}
	    \centering
	\begin{tikzpicture}[node distance=40pt]
	\tikzstyle{hasse} = [circle, fill,inner sep=2pt];
		\node at (0,0)[]{\large{$\mathfrak{sl}_5$}};
		\node at (0,-0.5) [hasse] (1) [] {};
		\node [hasse] (2) [below of=1] {};
		\node [hasse] (3) [below of=2] {};
		\node [hasse] (4) [below of=3] {};
		\node [hasse] (5) [below of=4] {};
		\node [hasse] (6) [below of=5] {};
		\node [hasse] (7) [below of=6] {};
		\draw (1) edge [] node[label=left:$A_4$] {} (2)
			(2) edge [] node[label=left:$A_2$] {} (3)
			(3) edge [] node[label=left:$a_1$] {} (4)
			(4) edge [] node[label=left:$A_1$] {} (5)
			(5) edge [] node[label=left:$a_2$] {} (6)
			(6) edge [] node[label=left:$a_4$] {} (7);
	\end{tikzpicture}
	\end{subfigure}
	\begin{subfigure}[]{0.5\textwidth}
	\centering
	\begin{tabular}{ c l c }
	\toprule
	\textbf{Matrices} & $\lambda$ & $dim$ \\ 
	\midrule	 \addlinespace[1.5ex]
 $\left(
\begin{array}{cccccc}
 0 & 5 & 0 & 0 & 0 & 0 \\
 0 & 4 & 3 & 2 & 1 & 0 \\
\end{array}
\right)$ & 5 & 10 \\	\addlinespace[1.5ex]
 $\left(
\begin{array}{cccccc}
 1 & 3 & 1 & 0 & 0 & 0 \\
 0 & 3 & 3 & 2 & 1 & 0 \\
\end{array}
\right)$ & 4,1 & 9 \\	\addlinespace[1.5ex]
 $\left(
\begin{array}{cccccc}
 2 & 1 & 2 & 0 & 0 & 0 \\
 0 & 2 & 3 & 2 & 1 & 0 \\
\end{array}
\right)$ & 3,2 & 8 \\	\addlinespace[1.5ex]
 $\left(
\begin{array}{cccccc}
 2 & 2 & 0 & 1 & 0 & 0 \\
 0 & 2 & 2 & 2 & 1 & 0 \\
\end{array}
\right)$ & 3,1,1 & 7 \\	\addlinespace[1.5ex]
 $\left(
\begin{array}{cccccc}
 3 & 0 & 1 & 1 & 0 & 0 \\
 0 & 1 & 2 & 2 & 1 & 0 \\
\end{array}
\right)$ & 2,2,1 & 6 \\	\addlinespace[1.5ex]
 $\left(
\begin{array}{cccccc}
 3 & 1 & 0 & 0 & 1 & 0 \\
 0 & 1 & 1 & 1 & 1 & 0 \\
\end{array}
\right)$ & 2,1,1,1 & 4 \\	\addlinespace[1.5ex]
 $\left(
\begin{array}{cccccc}
 4 & 0 & 0 & 0 & 0 & 1 \\
 0 & 0 & 0 & 0 & 0 & 0 \\
\end{array}
\right)$ & 1,1,1,1,1 & 0 \\	\addlinespace[1.5ex]
	\bottomrule
	\end{tabular}
	\end{subfigure}
	\caption{Results obtained applying the matrix formalism to $\mathfrak{sl}_5$.}
\end{table}

\begin{table}[h]
	\centering
	\begin{subfigure}[]{0.2\textwidth}
	    \centering
	\begin{tikzpicture}[node distance=40pt]
	\tikzstyle{hasse} = [circle, fill,inner sep=2pt];
		\node at (0,0)[]{\large{$\mathfrak{sl}_6$}};
		\node at (-0.7,-0.5) [] (1a) [] {};
		\node at (0,-0.5) [hasse] (1b) [] {};
		\node at (0.7,-0.5) [] (1c) [] {};
		\node [] (2a) [below of=1a] {};
		\node [hasse] (2b) [below of=1b] {};
		\node [] (2c) [below of=1c] {};
		\node  (3a) [below of=2a] {};
		\node [hasse] (3b) [below of=2b] {};
		\node  (3c) [below of=2c] {};
		\node [hasse] (4a) [below of=3a] {};
		\node (4b) [below of=3b] {};
		\node (4c) [below of=3c] {};
		\node (5a) [below of=4a] {};
		\node  (5b) [below of=4b] {};
		\node [hasse] (5c) [below of=4c] {};
		\node (6a) [below of=5a] {};
		\node [hasse] (6b) [below of=5b] {};
		\node  (6c) [below of=5c] {};
		\node  (7a) [below of=6a] {};
		\node  (7b) [below of=6b] {};
		\node [hasse] (7c) [below of=6c] {};
		\node [hasse] (8a) [below of=7a] {};
		\node (8b) [below of=7b] {};
		\node (8c) [below of=7c] {};
		\node (9a) [below of=8a] {};
		\node [hasse] (9b) [below of=8b] {};
		\node (9c) [below of=8c] {};
		\node  (10a) [below of=9a] {};
		\node [hasse] (10b) [below of=9b] {};
		\node  (10c) [below of=9c] {};
		\node (11a) [below of=10a] {};
		\node [hasse] (11b) [below of=10b] {};
		\node (11c) [below of=10c] {};
		\draw (1b) edge [] node[label=left:$A_5$] {} (2b)
			(2b) edge [] node[label=left:$A_3$] {} (3b)
			(3b) edge [] node[label=left:$a_1$] {} (4a)
			(4a) edge [] node[label=left:$A_2$] {} (6b)
			(3b) edge [] node[label=right:$A_1$] {} (5c)
			(5c) edge [] node[label=right:$A_2$] {} (6b)
			(6b) edge [] node[label=right:$a_2$] {} (7c)
			(7c) edge [] node[label=right:$a_1$] {} (9b)
			(6b) edge [] node[label=left:$a_2$] {} (8a)
			(8a) edge [] node[label=left:$A_1$] {} (9b)
			(9b) edge [] node[label=left:$a_3$] {} (10b)
			(10b) edge [] node[label=left:$a_5$] {} (11b);
	\end{tikzpicture}
	\end{subfigure}
	\begin{subfigure}[]{0.5\textwidth}
	\centering
	\begin{tabular}{ c l c }
	\toprule
	\textbf{Matrices} & $\lambda$ & $dim$ \\ 
	\midrule	 \addlinespace[1.5ex]
	 $\left(
\begin{array}{ccccccc}
 0 & 6 & 0 & 0 & 0 & 0 & 0 \\
 0 & 5 & 4 & 3 & 2 & 1 & 0 \\
\end{array}
\right)$ & 6 & 15 \\   \addlinespace[1.5ex]
 $\left(
\begin{array}{ccccccc}
 1 & 4 & 1 & 0 & 0 & 0 & 0 \\
 0 & 4 & 4 & 3 & 2 & 1 & 0 \\
\end{array}
\right)$ & 5,1 & 14 \\  \addlinespace[1.5ex]
 $\left(
\begin{array}{ccccccc}
 2 & 2 & 2 & 0 & 0 & 0 & 0 \\
 0 & 3 & 4 & 3 & 2 & 1 & 0 \\
\end{array}
\right)$ & 4,2 & 13 \\  \addlinespace[1.5ex]
 $\left(
\begin{array}{ccccccc}
 2 & 3 & 0 & 1 & 0 & 0 & 0 \\
 0 & 3 & 3 & 3 & 2 & 1 & 0 \\
\end{array}
\right)$ & 4,1,1 & 12 \\  \addlinespace[1.5ex]
 $\left(
\begin{array}{ccccccc}
 3 & 0 & 3 & 0 & 0 & 0 & 0 \\
 0 & 2 & 4 & 3 & 2 & 1 & 0 \\
\end{array}
\right)$ & 3,3 & 12 \\  \addlinespace[1.5ex]
 $\left(
\begin{array}{ccccccc}
 3 & 1 & 1 & 1 & 0 & 0 & 0 \\
 0 & 2 & 3 & 3 & 2 & 1 & 0 \\
\end{array}
\right)$ & 3,2,1 & 11 \\  \addlinespace[1.5ex]
 $\left(
\begin{array}{ccccccc}
 3 & 2 & 0 & 0 & 1 & 0 & 0 \\
 0 & 2 & 2 & 2 & 2 & 1 & 0 \\
\end{array}
\right)$ & 3,1,1,1 & 9 \\  \addlinespace[1.5ex]
 $\left(
\begin{array}{ccccccc}
 4 & 0 & 0 & 2 & 0 & 0 & 0 \\
 0 & 1 & 2 & 3 & 2 & 1 & 0 \\
\end{array}
\right)$ & 2,2,2 & 9 \\  \addlinespace[1.5ex]
 $\left(
\begin{array}{ccccccc}
 4 & 0 & 1 & 0 & 1 & 0 & 0 \\
 0 & 1 & 2 & 2 & 2 & 1 & 0 \\
\end{array}
\right)$ & 2,2,1,1 & 8 \\  \addlinespace[1.5ex]
 $\left(
\begin{array}{ccccccc}
 4 & 1 & 0 & 0 & 0 & 1 & 0 \\
 0 & 1 & 1 & 1 & 1 & 1 & 0 \\
\end{array}
\right)$ & 2,1,1,1,1 & 5 \\  \addlinespace[1.5ex]
 $\left(
\begin{array}{ccccccc}
 5 & 0 & 0 & 0 & 0 & 0 & 1 \\
 0 & 0 & 0 & 0 & 0 & 0 & 0 \\
\end{array}
\right)$ & 1,1,1,1,1,1 & 0 \\  \addlinespace[1.5ex]
	\bottomrule
	\end{tabular}
	\end{subfigure}
	\caption{Results obtained applying the matrix formalism to $\mathfrak{sl}_6$.}
\end{table}

\begin{table}[h]
	\centering
	\begin{subfigure}[]{0.2\textwidth}
	    \centering
	\begin{tikzpicture}[node distance=40pt]
	\tikzstyle{hasse} = [circle, fill,inner sep=2pt];
		\node at (0,0)[]{\large{$\mathfrak{sl}_7$}};
		\node at (-0.7,-0.5) [] (1a) [] {};
		\node at (0,-0.5) [hasse] (1b) [] {};
		\node at (0.7,-0.5) [] (1c) [] {};
		\node [] (2a) [below of=1a] {};
		\node [hasse] (2b) [below of=1b] {};
		\node [] (2c) [below of=1c] {};
		\node  (3a) [below of=2a] {};
		\node [hasse] (3b) [below of=2b] {};
		\node  (3c) [below of=2c] {};
		\node [hasse] (4a) [below of=3a] {};
		\node (4b) [below of=3b] {};
		\node (4c) [below of=3c] {};
		\node  (5a) [below of=4a] {};
		\node (5b) [below of=4b] {};
		\node [hasse] (5c) [below of=4c] {};
		\node (6a) [below of=5a] {};
		\node [hasse] (6b) [below of=5b] {};
		\node (6c) [below of=5c] {};
		\node  (7a) [below of=6a] {};
		\node  (7b) [below of=6b] {};
		\node [hasse] (7c) [below of=6c] {};
		\node [hasse] (8a) [below of=7a] {};
		\node  (8b) [below of=7b] {};
		\node (8c) [below of=7c] {};
		\node (9a) [below of=8a] {};
		\node  (9b) [below of=8b] {};
		\node [hasse] (9c) [below of=8c] {};
		\node  (10a) [below of=9a] {};
		\node [hasse] (10b) [below of=9b] {};
		\node  (10c) [below of=9c] {};
		\node (11a) [below of=10a] {};
		\node  (11b) [below of=10b] {};
		\node [hasse] (11c) [below of=10c] {};
		\node [hasse] (12a) [below of=11a] {};
		\node (12b) [below of=11b] {};
		\node  (12c) [below of=11c] {};
		\node (13a) [below of=12a] {};
		\node [hasse] (13b) [below of=12b] {};
		\node (13c) [below of=12c] {};
		\node  (14a) [below of=13a] {};
		\node [hasse] (14b) [below of=13b] {};
		\node  (14c) [below of=13c] {};
		\node (15a) [below of=14a] {};
		\node [hasse] (15b) [below of=14b] {};
		\node (15c) [below of=14c] {};
		\draw (1b) edge [] node[label=left:$A_6$] {} (2b)
			(2b) edge [] node[label=left:$A_4$] {} (3b)
			(3b) edge [] node[label=left:$a_1$] {} (4a)
			(4a) edge [] node[label=left:$A_3$] {} (6b)
			(3b) edge [] node[label=right:$A_2$] {} (5c)
			(5c) edge [] node[label=right:$A_2$] {} (6b)
			(6b) edge [] node[label=right:$A_1$] {} (7c)
			(7c) edge [] node[label=right:$A_1$] {} (9c)
			(6b) edge [] node[label=left:$a_2$] {} (8a)
			(8a) edge [] node[label=left:$A_2$] {} (10b)
			(9c) edge [] node[label=right:$A_1$] {} (10b)
			(10b) edge [] node[label=left:$a_3$] {} (12a)
			(12a) edge [] node[label=left:$A_1$] {} (13b)
			(10b) edge [] node[label=right:$a_2$] {} (11c)
			(11c) edge [] node[label=right:$a_2$] {} (13b)
			(13b) edge [] node[label=left:$a_4$] {} (14b)
			(14b) edge [] node[label=left:$a_6$] {} (15b);
	\end{tikzpicture}
	\end{subfigure}
	\begin{subfigure}[]{0.5\textwidth}
	\centering
	\begin{tabular}{ c l c }
	\toprule
	\textbf{Matrices} & $\lambda$ & $dim$ \\ 
	\midrule	 \addlinespace[1.5ex]
	$\left(
\begin{array}{cccccccc}
 0 & 7 & 0 & 0 & 0 & 0 & 0 & 0 \\
 0 & 6 & 5 & 4 & 3 & 2 & 1 & 0 \\
\end{array}
\right)$ & 7 & 21 \\  \addlinespace[1.5ex]
 $\left(
\begin{array}{cccccccc}
 1 & 5 & 1 & 0 & 0 & 0 & 0 & 0 \\
 0 & 5 & 5 & 4 & 3 & 2 & 1 & 0 \\
\end{array}
\right)$ & 6,1 & 20 \\  \addlinespace[1.5ex]
 $\left(
\begin{array}{cccccccc}
 2 & 3 & 2 & 0 & 0 & 0 & 0 & 0 \\
 0 & 4 & 5 & 4 & 3 & 2 & 1 & 0 \\
\end{array}
\right)$ & 5,2 & 19 \\  \addlinespace[1.5ex]
 $\left(
\begin{array}{cccccccc}
 2 & 4 & 0 & 1 & 0 & 0 & 0 & 0 \\
 0 & 4 & 4 & 4 & 3 & 2 & 1 & 0 \\
\end{array}
\right)$ & 5,1,1 & 18 \\  \addlinespace[1.5ex]
 $\left(
\begin{array}{cccccccc}
 3 & 1 & 3 & 0 & 0 & 0 & 0 & 0 \\
 0 & 3 & 5 & 4 & 3 & 2 & 1 & 0 \\
\end{array}
\right)$ & 4,3 & 18 \\  \addlinespace[1.5ex]
 $\left(
\begin{array}{cccccccc}
 3 & 2 & 1 & 1 & 0 & 0 & 0 & 0 \\
 0 & 3 & 4 & 4 & 3 & 2 & 1 & 0 \\
\end{array}
\right)$ & 4,2,1 & 17 \\  \addlinespace[1.5ex]
 $\left(
\begin{array}{cccccccc}
 4 & 0 & 2 & 1 & 0 & 0 & 0 & 0 \\
 0 & 2 & 4 & 4 & 3 & 2 & 1 & 0 \\
\end{array}
\right)$ & 3,3,1 & 16 \\  \addlinespace[1.5ex]
 $\left(
\begin{array}{cccccccc}
 3 & 3 & 0 & 0 & 1 & 0 & 0 & 0 \\
 0 & 3 & 3 & 3 & 3 & 2 & 1 & 0 \\
\end{array}
\right)$ & 4,1,1,1 & 15 \\  \addlinespace[1.5ex]
 $\left(
\begin{array}{cccccccc}
 4 & 1 & 0 & 2 & 0 & 0 & 0 & 0 \\
 0 & 2 & 3 & 4 & 3 & 2 & 1 & 0 \\
\end{array}
\right)$ & 3,2,2 & 15 \\  \addlinespace[1.5ex]
 $\left(
\begin{array}{cccccccc}
 4 & 1 & 1 & 0 & 1 & 0 & 0 & 0 \\
 0 & 2 & 3 & 3 & 3 & 2 & 1 & 0 \\
\end{array}
\right)$ & 3,2,1,1 & 14 \\  \addlinespace[1.5ex]
 $\left(
\begin{array}{cccccccc}
 5 & 0 & 0 & 1 & 1 & 0 & 0 & 0 \\
 0 & 1 & 2 & 3 & 3 & 2 & 1 & 0 \\
\end{array}
\right)$ & 2,2,2,1 & 12 \\  \addlinespace[1.5ex]
 $\left(
\begin{array}{cccccccc}
 4 & 2 & 0 & 0 & 0 & 1 & 0 & 0 \\
 0 & 2 & 2 & 2 & 2 & 2 & 1 & 0 \\
\end{array}
\right)$ & 3,1,1,1,1 & 11 \\  \addlinespace[1.5ex]
 $\left(
\begin{array}{cccccccc}
 5 & 0 & 1 & 0 & 0 & 1 & 0 & 0 \\
 0 & 1 & 2 & 2 & 2 & 2 & 1 & 0 \\
\end{array}
\right)$ & 2,2,1,1,1 & 10 \\  \addlinespace[1.5ex]
 $\left(
\begin{array}{cccccccc}
 5 & 1 & 0 & 0 & 0 & 0 & 1 & 0 \\
 0 & 1 & 1 & 1 & 1 & 1 & 1 & 0 \\
\end{array}
\right)$ & 2,1,1,1,1,1 & 6 \\  \addlinespace[1.5ex]
 $\left(
\begin{array}{cccccccc}
 6 & 0 & 0 & 0 & 0 & 0 & 0 & 1 \\
 0 & 0 & 0 & 0 & 0 & 0 & 0 & 0 \\
\end{array}
\right)$ & 1,1,1,1,1,1,1 & 0 \\  \addlinespace[1.5ex]
	\bottomrule
	\end{tabular}
	\end{subfigure}
	\caption{Results obtained applying the matrix formalism to $\mathfrak{sl}_7$.}
\end{table}

\begin{table}[h]
	\centering
	\begin{subfigure}[]{0.2\textwidth}
	    \centering
	\begin{tikzpicture}[node distance=27.5pt]\tikzstyle{hasse} = [circle, fill,inner sep=2pt];
		\node at (0,0)[]{\large{$\mathfrak{sl}_8$}};
		\node at (-0.7,-0.5) [] (1a) [] {};
		\node at (0,-0.5) [hasse] (1b) [] {};
		\node at (0.7,-0.5) [] (1c) [] {};
		\node at (1.4,-0.5) [] (1d) [] {};
		\node [] (2a) [below of=1a] {};
		\node [hasse] (2b) [below of=1b] {};
		\node [] (2c) [below of=1c] {};
		\node [] (2d) [below of=1d] {};
		\node  (3a) [below of=2a] {};
		\node [hasse] (3b) [below of=2b] {};
		\node (3c) [below of=2c] {};
		\node (3d) [below of=2d] {};
		\node [hasse] (4a) [below of=3a] {};
		\node (4b) [below of=3b] {};
		\node (4c) [below of=3c] {};
		\node  (4d) [below of=3d] {};
		\node  (5a) [below of=4a] {};
		\node  (5b) [below of=4b] {};
		\node [hasse] (5c) [below of=4c] {};
		\node  (5d) [below of=4d] {};
		\node  (6a) [below of=5a] {};
		\node  (6b) [below of=5b] {};
		\node (6c) [below of=5c] {};
		\node [hasse] (6d) [below of=5d] {};
		\node (7a) [below of=6a] {};
		\node [hasse] (7b) [below of=6b] {};
		\node (7c) [below of=6c] {};
		\node (7d) [below of=6d] {};
		\node  (8a) [below of=7a] {};
		\node (8b) [below of=7b] {};
		\node [hasse] (8c) [below of=7c] {};
		\node (8d) [below of=7d] {};
		\node [hasse] (9a) [below of=8a] {};
		\node  (9b) [below of=8b] {};
		\node  (9c) [below of=8c] {};
		\node  (9d) [below of=8d] {};
		\node (10a) [below of=9a] {};
		\node (10b) [below of=9b] {};
		\node [hasse] (10c) [below of=9c] {};
		\node  (10d) [below of=9d] {};
		\node (11a) [below of=10a] {};
		\node [hasse] (11b) [below of=10b] {};
		\node (11c) [below of=10c] {};
		\node (11d) [below of=10d] {};
		\node (12a) [below of=11a] {};
		\node (12b) [below of=11b] {};
		\node (12c) [below of=11c] {};
		\node [hasse] (12d) [below of=11d] {};
		\node (13a) [below of=12a] {};
		\node (13b) [below of=12b] {};
		\node [hasse] (13c) [below of=12c] {};
		\node (13d) [below of=12d] {};
		\node [hasse] (14a) [below of=13a] {};
		\node  (14b) [below of=13b] {};
		\node  (14c) [below of=13c] {};
		\node  (14d) [below of=13d] {};
		\node  (15a) [below of=14a] {};
		\node  (15b) [below of=14b] {};
		\node [hasse] (15c) [below of=14c] {};
		\node  (15d) [below of=14d] {};
		\node  (16a) [below of=15a] {};
		\node [hasse] (16b) [below of=15b] {};
		\node (16c) [below of=15c] {};
		\node  (16d) [below of=15d] {};
		\node  (17a) [below of=16a] {};
		\node  (17b) [below of=16b] {};
		\node  (17c) [below of=16c] {};
		\node [hasse] (17d) [below of=16d] {};
		\node  (18a) [below of=17a] {};
		\node  (18b) [below of=17b] {};
		\node [hasse] (18c) [below of=17c] {};
		\node  (18d) [below of=17d] {};
		\node [hasse] (19a) [below of=18a] {};
		\node  (19b) [below of=18b] {};
		\node (19c) [below of=18c] {};
		\node (19d) [below of=18d] {};
		\node  (20a) [below of=19a] {};
		\node [hasse] (20b) [below of=19b] {};
		\node  (20c) [below of=19c] {};
		\node  (20d) [below of=19d] {};
		\node  (21a) [below of=20a] {};
		\node [hasse] (21b) [below of=20b] {};
		\node  (21c) [below of=20c] {};
		\node  (21d) [below of=20d] {};
		\node  (22a) [below of=21a] {};
		\node [hasse] (22b) [below of=21b] {};
		\node  (22c) [below of=21c] {};
		\node  (22d) [below of=21d] {};
		\draw (1b) edge [] node[label=left:$A_7$] {} (2b)
				(2b) edge [] node[label=left:$A_5$] {} (3b)
				(3b) edge [] node[label=left:$A_1$] {} (4a)
				(3b) edge [] node[label=right:$A_3$] {} (5c)
				(4a) edge [] node[label=left:$A_4$] {} (7b)
				(5c) edge [] node[pos=0.1, left] {$A_2$} (7b)
				(5c) edge [] node[label=right:$A_1$] {} (6d)
				(7b) edge [] node[label=left:$a_2$] {} (9a)
				(7b) edge [] node[pos=0.9, left] {$A_2$} (8c)
				(6d) edge [] node[label=right:$A_3$] {} (8c)
				(9a) edge [] node[label=left:$A_3$] {} (11b)
				(8c) edge [] node[label=right:$A_1$] {} (10c)
				(10c) edge [] node[pos=0, left] {$A_1$} (11b)
				(11b) edge [] node[pos=0.9, left] {$A_1$} (13c)
				(10c) edge [] node[label=right:$A_1$] {} (12d)
				(12d) edge [] node[label=right:$A_1$] {} (13c)
				(13c) edge [] node[label=right:$A_1$] {} (15c)
				(11b) edge [] node[label=left:$a_3$] {} (14a)
				(14a) edge [] node[label=left:$A_2$] {} (16b)
				(15c) edge [] node[pos=0.1, left] {$a_2$} (16b)
				(15c) edge [] node[label=right:$a_3$] {} (17d)
				(16b) edge [] node[pos=0.8, left] {$a_2$} (18c)
				(17d) edge [] node[label=right:$A_1$] {} (18c)
				(16b) edge [] node[label=left:$a_4$] {} (19a)
				(19a) edge [] node[label=left:$A_1$] {} (20b)
				(18c) edge [] node[label=right:$a_3$] {} (20b)
				(20b) edge [] node[label=left:$a_5$] {} (21b)
				(21b) edge [] node[label=left:$a_7$] {} (22b);
			\end{tikzpicture}
	\end{subfigure}
	\begin{subfigure}[]{0.5\textwidth}
	\centering
	\fontsize{9}{10.8}\selectfont
	\begin{tabular}{ c l c }
	\toprule
	\textbf{Matrices} & $\lambda$ & $dim$ \\ 
	\midrule	 \addlinespace[1ex]
	 $\left(
\begin{array}{ccccccccc}
 0 & 8 & 0 & 0 & 0 & 0 & 0 & 0 & 0 \\
 0 & 7 & 6 & 5 & 4 & 3 & 2 & 1 & 0 \\
\end{array}
\right)$ & 8 & 28 \\  \addlinespace[1ex]
 $\left(
\begin{array}{ccccccccc}
 1 & 6 & 1 & 0 & 0 & 0 & 0 & 0 & 0 \\
 0 & 6 & 6 & 5 & 4 & 3 & 2 & 1 & 0 \\
\end{array}
\right)$ & 7,1 & 27 \\  \addlinespace[1ex]
 $\left(
\begin{array}{ccccccccc}
 2 & 4 & 2 & 0 & 0 & 0 & 0 & 0 & 0 \\
 0 & 5 & 6 & 5 & 4 & 3 & 2 & 1 & 0 \\
\end{array}
\right)$ & 6,2 & 26 \\  \addlinespace[1ex]
 $\left(
\begin{array}{ccccccccc}
 2 & 5 & 0 & 1 & 0 & 0 & 0 & 0 & 0 \\
 0 & 5 & 5 & 5 & 4 & 3 & 2 & 1 & 0 \\
\end{array}
\right)$ & 6,1,1 & 25 \\  \addlinespace[1ex]
 $\left(
\begin{array}{ccccccccc}
 3 & 2 & 3 & 0 & 0 & 0 & 0 & 0 & 0 \\
 0 & 4 & 6 & 5 & 4 & 3 & 2 & 1 & 0 \\
\end{array}
\right)$ & 5,3 & 25 \\  \addlinespace[1ex]
 $\left(
\begin{array}{ccccccccc}
 4 & 0 & 4 & 0 & 0 & 0 & 0 & 0 & 0 \\
 0 & 3 & 6 & 5 & 4 & 3 & 2 & 1 & 0 \\
\end{array}
\right)$ & 4,4 & 24 \\  \addlinespace[1ex]
 $\left(
\begin{array}{ccccccccc}
 3 & 3 & 1 & 1 & 0 & 0 & 0 & 0 & 0 \\
 0 & 4 & 5 & 5 & 4 & 3 & 2 & 1 & 0 \\
\end{array}
\right)$ & 5,2,1 & 24 \\  \addlinespace[1ex]
 $\left(
\begin{array}{ccccccccc}
 4 & 1 & 2 & 1 & 0 & 0 & 0 & 0 & 0 \\
 0 & 3 & 5 & 5 & 4 & 3 & 2 & 1 & 0 \\
\end{array}
\right)$ & 4,3,1 & 23 \\  \addlinespace[1ex]
 $\left(
\begin{array}{ccccccccc}
 3 & 4 & 0 & 0 & 1 & 0 & 0 & 0 & 0 \\
 0 & 4 & 4 & 4 & 4 & 3 & 2 & 1 & 0 \\
\end{array}
\right)$ & 5,1,1,1 & 22 \\  \addlinespace[1ex]
 $\left(
\begin{array}{ccccccccc}
 4 & 2 & 0 & 2 & 0 & 0 & 0 & 0 & 0 \\
 0 & 3 & 4 & 5 & 4 & 3 & 2 & 1 & 0 \\
\end{array}
\right)$ & 4,2,2 & 22 \\  \addlinespace[1ex]
 $\left(
\begin{array}{ccccccccc}
 4 & 2 & 1 & 0 & 1 & 0 & 0 & 0 & 0 \\
 0 & 3 & 4 & 4 & 4 & 3 & 2 & 1 & 0 \\
\end{array}
\right)$ & 4,2,1,1 & 21 \\  \addlinespace[1ex]
 $\left(
\begin{array}{ccccccccc}
 5 & 0 & 1 & 2 & 0 & 0 & 0 & 0 & 0 \\
 0 & 2 & 4 & 5 & 4 & 3 & 2 & 1 & 0 \\
\end{array}
\right)$ & 3,3,2 & 21 \\  \addlinespace[1ex]
 $\left(
\begin{array}{ccccccccc}
 5 & 0 & 2 & 0 & 1 & 0 & 0 & 0 & 0 \\
 0 & 2 & 4 & 4 & 4 & 3 & 2 & 1 & 0 \\
\end{array}
\right)$ & 3,3,1,1 & 20 \\  \addlinespace[1ex]
 $\left(
\begin{array}{ccccccccc}
 4 & 3 & 0 & 0 & 0 & 1 & 0 & 0 & 0 \\
 0 & 3 & 3 & 3 & 3 & 3 & 2 & 1 & 0 \\
\end{array}
\right)$ & 4,1,1,1,1 & 18 \\  \addlinespace[1ex]
 $\left(
\begin{array}{ccccccccc}
 5 & 1 & 0 & 1 & 1 & 0 & 0 & 0 & 0 \\
 0 & 2 & 3 & 4 & 4 & 3 & 2 & 1 & 0 \\
\end{array}
\right)$ & 3,2,2,1 & 19 \\  \addlinespace[1ex]
 $\left(
\begin{array}{ccccccccc}
 5 & 1 & 1 & 0 & 0 & 1 & 0 & 0 & 0 \\
 0 & 2 & 3 & 3 & 3 & 3 & 2 & 1 & 0 \\
\end{array}
\right)$ & 3,2,1,1,1 & 17 \\  \addlinespace[1ex]
 $\left(
\begin{array}{ccccccccc}
 6 & 0 & 0 & 0 & 2 & 0 & 0 & 0 & 0 \\
 0 & 1 & 2 & 3 & 4 & 3 & 2 & 1 & 0 \\
\end{array}
\right)$ & 2,2,2,2 & 16 \\  \addlinespace[1ex]
 $\left(
\begin{array}{ccccccccc}
 6 & 0 & 0 & 1 & 0 & 1 & 0 & 0 & 0 \\
 0 & 1 & 2 & 3 & 3 & 3 & 2 & 1 & 0 \\
\end{array}
\right)$ & 2,2,2,1,1 & 15 \\  \addlinespace[1ex]
 $\left(
\begin{array}{ccccccccc}
 5 & 2 & 0 & 0 & 0 & 0 & 1 & 0 & 0 \\
 0 & 2 & 2 & 2 & 2 & 2 & 2 & 1 & 0 \\
\end{array}
\right)$ & 3,1,1,1,1,1 & 13 \\  \addlinespace[1ex]
 $\left(
\begin{array}{ccccccccc}
 6 & 0 & 1 & 0 & 0 & 0 & 1 & 0 & 0 \\
 0 & 1 & 2 & 2 & 2 & 2 & 2 & 1 & 0 \\
\end{array}
\right)$ & 2,2,1,1,1,1 & 12 \\  \addlinespace[1ex]
 $\left(
\begin{array}{ccccccccc}
 6 & 1 & 0 & 0 & 0 & 0 & 0 & 1 & 0 \\
 0 & 1 & 1 & 1 & 1 & 1 & 1 & 1 & 0 \\
\end{array}
\right)$ & 2,1,1,1,1,1,1 & 7 \\  \addlinespace[1ex]
 $\left(
\begin{array}{ccccccccc}
 7 & 0 & 0 & 0 & 0 & 0 & 0 & 0 & 1 \\
 0 & 0 & 0 & 0 & 0 & 0 & 0 & 0 & 0 \\
\end{array}
\right)$ & 1,1,1,1,1,1,1,1 & 0 \\  \addlinespace[1ex]
	\bottomrule
	\end{tabular}
	\end{subfigure}
	\caption{Results obtained applying the matrix formalism to $\mathfrak{sl}_8$.}
\end{table}

\begin{table}[h]
	\centering
	\begin{subfigure}[]{0.3\textwidth}
	    \centering
	\begin{tikzpicture}[node distance=20.4pt]
	\tikzstyle{hasse} = [circle, fill,inner sep=2pt];
		\node at (0.7,0)[]{\large{$\mathfrak{sl}_9$}};
		\node at (-0.7,-0.6) [] (1a) [] {};
		\node at (0,-0.6) [] (1b) [] {};
		\node at (0.7,-0.6) [hasse] (1c) [] {};
		\node at (1.4,-0.6) [] (1d) [] {};
		\node at (2.1,-0.6) [] (1e) [] {};
		\node [] (2a) [below of=1a] {};
		\node [] (2b) [below of=1b] {};
		\node [hasse] (2c) [below of=1c] {};
		\node [] (2d) [below of=1d] {};
		\node [] (2e) [below of=1e] {};
		\node  (3a) [below of=2a] {};
		\node (3b) [below of=2b] {};
		\node [hasse] (3c) [below of=2c] {};
		\node (3d) [below of=2d] {};
		\node (3e) [below of=2e] {};
		\node  (4a) [below of=3a] {};
		\node [hasse] (4b) [below of=3b] {};
		\node  (4c) [below of=3c] {};
		\node  (4d) [below of=3d] {};
		\node (4e) [below of=3e] {};
		\node  (5a) [below of=4a] {};
		\node  (5b) [below of=4b] {};
		\node  (5c) [below of=4c] {};
		\node  [hasse] (5d) [below of=4d] {};
		\node  (5e) [below of=4e] {};
		\node (6a) [below of=5a] {};
		\node (6b) [below of=5b] {};
		\node [hasse] (6c) [below of=5c] {};
		\node (6d) [below of=5d] {};
		\node  (6e) [below of=5e] {};
		\node  (7a) [below of=6a] {};
		\node  (7b) [below of=6b] {};
		\node (7c) [below of=6c] {};
		\node  (7d) [below of=6d] {};
		\node  [hasse] (7e) [below of=6e] {};
		\node (8a) [below of=7a] {};
		\node  (8b) [below of=7b] {};
		\node (8c) [below of=7c] {};
		\node [hasse] (8d) [below of=7d] {};
		\node  (8e) [below of=7e] {};
		\node  (9a) [below of=8a] {};
		\node [hasse] (9b) [below of=8b] {};
		\node  (9c) [below of=8c] {};
		\node  (9d) [below of=8d] {};
		\node  (9e) [below of=8e] {};
		\node (10a) [below of=9a] {};
		\node  (10b) [below of=9b] {};
		\node  (10c) [below of=9c] {};
		\node (10d) [below of=9d] {};
		\node  [hasse] (10e) [below of=9e] {};
		\node (11a) [below of=10a] {};
		\node  (11b) [below of=10b] {};
		\node   [hasse] (11c) [below of=10c] {};
		\node  (11d) [below of=10d] {};
		\node  (11e) [below of=10e] {};
		\node  (12a) [below of=11a] {};
		\node [hasse] (12b) [below of=11b] {};
		\node  (12c) [below of=11c] {};
		\node  (12d) [below of=11d] {};
		\node  (12e) [below of=11e] {};
		\node (13a) [below of=12a] {};
		\node  (13b) [below of=12b] {};
		\node  (13c) [below of=12c] {};
		\node [hasse] (13d) [below of=12d] {};
		\node  (13e) [below of=12e] {};
		\node  (14a) [below of=13a] {};
		\node  (14b) [below of=13b] {};
		\node [hasse] (14c) [below of=13c] {};
		\node  (14d) [below of=13d] {};
		\node  (14e) [below of=13e] {};
		\node  (15a) [below of=14a] {};
		\node  (15b) [below of=14b] {};
		\node  (15c) [below of=14c] {};
		\node  (15d) [below of=14d] {};
		\node [hasse] (15e) [below of=14e] {};
		\node [hasse] (16a) [below of=15a] {};
		\node  (16b) [below of=15b] {};
		\node  (16c) [below of=15c] {};
		\node  (16d) [below of=15d] {};
		\node  (16e) [below of=15e] {};
		\node (17a) [below of=16a] {};
		\node  (17b) [below of=16b] {};
		\node [hasse] (17c) [below of=16c] {};
		\node  (17d) [below of=16d] {};
		\node (17e) [below of=16e] {};
		\node  (18a) [below of=17a] {};
		\node  (18b) [below of=17b] {};
		\node  (18c) [below of=17c] {};
		\node [hasse] (18d) [below of=17d] {};
		\node (18e) [below of=17e] {};
		\node (19a) [below of=18a] {};
		\node [hasse] (19b) [below of=18b] {};
		\node  (19c) [below of=18c] {};
		\node  (19d) [below of=18d] {};
		\node  (19e) [below of=18e] {};
		\node  (20a) [below of=19a] {};
		\node  (20b) [below of=19b] {};
		\node  [hasse]  (20c) [below of=19c] {};
		\node  (20d) [below of=19d] {};
		\node (20e) [below of=19e] {};
		\node (21a) [below of=20a] {};
		\node  (21b) [below of=20b] {};
		\node (21c) [below of=20c] {};
		\node  (21d) [below of=20d] {};
		\node  [hasse] (21e) [below of=20e] {};
		\node (22a) [below of=21a] {};
		\node [hasse] (22b) [below of=21b] {};
		\node  (22c) [below of=21c] {};
		\node  (22d) [below of=21d] {};
		\node  (22e) [below of=21e] {};
		\node (23a) [below of=22a] {};
		\node  (23b) [below of=22b] {};
		\node (23c) [below of=22c] {};
		\node [hasse] (23d) [below of=22d] {};
		\node (23e) [below of=22e] {};
		\node (24a) [below of=23a] {};
		\node  (24b) [below of=23b] {};
		\node  (24c) [below of=23c] {};
		\node  (24d) [below of=23d] {};
		\node [hasse] (24e) [below of=23e] {};
		\node  (25a) [below of=24a] {};
		\node  (25b) [below of=24b] {};
		\node [hasse] (25c) [below of=24c] {};
		\node  (25d) [below of=24d] {};
		\node  (25e) [below of=24e] {};
		\node (26a) [below of=25a] {};
		\node  (26b) [below of=25b] {};
		\node  (26c) [below of=25c] {};
		\node [hasse] (26d) [below of=25d] {};
		\node  (26e) [below of=25e] {};
		\node (27a) [below of=26a] {};
		\node [hasse] (27b) [below of=26b] {};
		\node (27c) [below of=26c] {};
		\node (27d) [below of=26d] {};
		\node  (27e) [below of=26e] {};
		\node  (28a) [below of=27a] {};
		\node (28b) [below of=27b] {};
		\node [hasse] (28c) [below of=27c] {};
		\node  (28d) [below of=27d] {};
		\node  (28e) [below of=27e] {};
		\node (29a) [below of=28a] {};
		\node  (29b) [below of=28b] {};
		\node [hasse] (29c) [below of=28c] {};
		\node (29d) [below of=28d] {};
		\node  (29e) [below of=28e] {};
		\node  (30a) [below of=29a] {};
		\node  (30b) [below of=29b] {};
		\node [hasse] (30c) [below of=29c] {};
		\node  (30d) [below of=29d] {};
		\node (30e) [below of=29e] {};
		\draw (1c) edge [] node[label=left:$A_8$] {} (2c)
				(2c) edge [] node[label=left:$A_6$] {} (3c)
				(3c) edge [] node[label=left:$A_1$] {} (4b)
				(3c) edge [] node[label=right:$A_4$] {} (5d)
				(4b) edge [] node[label=left:$A_5$] {} (6c)
				(5d) edge [] node[label=right:$A_2$] {} (7e)
				(5d) edge [] node[pos=0.1,left]{$A_2$} (6c)
				(6c) edge [] node[label=left:$a_2$] {} (9b)
				(9b) edge [] node[label=left:$A_4$] {} (12b)
				(12b) edge [] node[pos=0.8,left]{$A_2$} (14c)
				(12b) edge [] node[label=left:$a_3$] {} (16a)
				(16a) edge [] node[label=left:$A_3$] {} (19b)
				(19b) edge [] node[label=left:$a_4$] {} (22b)
				(22b) edge [] node[label=left:$A_2$] {} (25c)
				(25c) edge [] node[label=left:$a_5$] {} (27b)
				(27b) edge [] node[label=left:$A_1$] {} (28c)
				(28c) edge [] node[label=left:$a_6$] {} (29c)
				(29c) edge [] node[label=left:$a_8$] {} (30c)
				(7e) edge [] node[label=right:$A_3$] {} (8d)
				(6c) edge [] node[pos=0.8,left]{$A_3$} (8d)
				(8d) edge [] node[pos=0.4, left]{$A_1$} (11c)
				(11c) edge [] node[pos=0.2,right]{$A_2$} (13d)
				(8d) edge [] node[label=right:$A_1$] {} (10e)
				(10e) edge [] node[label=right:$A_2$] {} (13d)
				(13d) edge [] node[pos=0,left]{$A_1$} (14c)
				(13d) edge [] node[label=right:$a_2$] {} (15e)
				(15e) edge [] node[label=right:$A_2$] {} (18d)
				(18d) edge [] node[label=right:$a_2$] {} (21e)
				(21e) edge [] node[label=right:$A_1$] {} (23d)
				(23d) edge [] node[label=right:$a_3$] {} (24e)
				(24e) edge [] node[label=right:$a_2$] {} (26d)
				(26d) edge [] node[label=right:$a_4$] {} (28c)
				(14c) edge [] node[left]{$A_1$} (17c)
				(17c) edge [] node[pos=1,left]{$A_1$} (18d)
				(17c) edge [] node[pos=0.2,left]{$a_2$} (19b)
				(18d) edge [] node[pos=0.7, right]{$a_2$}(20c)
				(20c) edge [] node[pos=0.6, left]{$A_1$} (23d)
				(23d) edge [] node[pos=0.2, left]{$a_3$} (25c)
				(25c) edge [] node[pos=0.3, below]{$a_2$} (26d)
				(11c) edge [] node[pos=0, left]{$A_1$} (12b)
				(19b) edge [] node[pos=1.2, left]{$A_1$} (20c);
	\end{tikzpicture}
	\end{subfigure}
	\begin{subfigure}[]{0.5\textwidth}
	\centering
	\fontsize{7.5}{9}\selectfont
	\begin{tabular}{ c l c }
	\toprule
	\textbf{Matrices} & $\lambda$ & $dim$ \\ 
	\midrule	 \addlinespace[1.2ex]
	$\left(
\begin{array}{cccccccccc}
 0 & 9 & 0 & 0 & 0 & 0 & 0 & 0 & 0 & 0 \\
 0 & 8 & 7 & 6 & 5 & 4 & 3 & 2 & 1 & 0 \\
\end{array}
\right)$ & 9 & 36 \\  \addlinespace[.2ex]
 $\left(
\begin{array}{cccccccccc}
 1 & 7 & 1 & 0 & 0 & 0 & 0 & 0 & 0 & 0 \\
 0 & 7 & 7 & 6 & 5 & 4 & 3 & 2 & 1 & 0 \\
\end{array}
\right)$ & 8,1 & 35 \\  \addlinespace[.2ex]
 $\left(
\begin{array}{cccccccccc}
 2 & 5 & 2 & 0 & 0 & 0 & 0 & 0 & 0 & 0 \\
 0 & 6 & 7 & 6 & 5 & 4 & 3 & 2 & 1 & 0 \\
\end{array}
\right)$ & 7,2 & 34 \\  \addlinespace[.2ex]
 $\left(
\begin{array}{cccccccccc}
 2 & 6 & 0 & 1 & 0 & 0 & 0 & 0 & 0 & 0 \\
 0 & 6 & 6 & 6 & 5 & 4 & 3 & 2 & 1 & 0 \\
\end{array}
\right)$ & 7,1,1 & 33 \\  \addlinespace[.2ex]
 $\left(
\begin{array}{cccccccccc}
 3 & 3 & 3 & 0 & 0 & 0 & 0 & 0 & 0 & 0 \\
 0 & 5 & 7 & 6 & 5 & 4 & 3 & 2 & 1 & 0 \\
\end{array}
\right)$ & 6,3 & 33 \\  \addlinespace[.2ex]
 $\left(
\begin{array}{cccccccccc}
 3 & 4 & 1 & 1 & 0 & 0 & 0 & 0 & 0 & 0 \\
 0 & 5 & 6 & 6 & 5 & 4 & 3 & 2 & 1 & 0 \\
\end{array}
\right)$ & 6,2,1 & 32 \\  \addlinespace[.2ex]
 $\left(
\begin{array}{cccccccccc}
 4 & 1 & 4 & 0 & 0 & 0 & 0 & 0 & 0 & 0 \\
 0 & 4 & 7 & 6 & 5 & 4 & 3 & 2 & 1 & 0 \\
\end{array}
\right)$ & 5,4 & 32 \\  \addlinespace[.2ex]
 $\left(
\begin{array}{cccccccccc}
 4 & 2 & 2 & 1 & 0 & 0 & 0 & 0 & 0 & 0 \\
 0 & 4 & 6 & 6 & 5 & 4 & 3 & 2 & 1 & 0 \\
\end{array}
\right)$ & 5,3,1 & 31 \\  \addlinespace[.2ex]
 $\left(
\begin{array}{cccccccccc}
 3 & 5 & 0 & 0 & 1 & 0 & 0 & 0 & 0 & 0 \\
 0 & 5 & 5 & 5 & 5 & 4 & 3 & 2 & 1 & 0 \\
\end{array}
\right)$ & 6,1,1,1 & 30 \\  \addlinespace[.2ex]
 $\left(
\begin{array}{cccccccccc}
 5 & 0 & 3 & 1 & 0 & 0 & 0 & 0 & 0 & 0 \\
 0 & 3 & 6 & 6 & 5 & 4 & 3 & 2 & 1 & 0 \\
\end{array}
\right)$ & 4,4,1 & 30 \\  \addlinespace[.2ex]
 $\left(
\begin{array}{cccccccccc}
 4 & 3 & 0 & 2 & 0 & 0 & 0 & 0 & 0 & 0 \\
 0 & 4 & 5 & 6 & 5 & 4 & 3 & 2 & 1 & 0 \\
\end{array}
\right)$ & 5,2,2 & 30 \\  \addlinespace[.2ex]
 $\left(
\begin{array}{cccccccccc}
 4 & 3 & 1 & 0 & 1 & 0 & 0 & 0 & 0 & 0 \\
 0 & 4 & 5 & 5 & 5 & 4 & 3 & 2 & 1 & 0 \\
\end{array}
\right)$ & 5,2,1,1 & 29 \\  \addlinespace[.2ex]
 $\left(
\begin{array}{cccccccccc}
 5 & 1 & 1 & 2 & 0 & 0 & 0 & 0 & 0 & 0 \\
 0 & 3 & 5 & 6 & 5 & 4 & 3 & 2 & 1 & 0 \\
\end{array}
\right)$ & 4,3,2 & 29 \\  \addlinespace[.2ex]
 $\left(
\begin{array}{cccccccccc}
 5 & 1 & 2 & 0 & 1 & 0 & 0 & 0 & 0 & 0 \\
 0 & 3 & 5 & 5 & 5 & 4 & 3 & 2 & 1 & 0 \\
\end{array}
\right)$ & 4,3,1,1 & 28 \\  \addlinespace[.2ex]
 $\left(
\begin{array}{cccccccccc}
 6 & 0 & 0 & 3 & 0 & 0 & 0 & 0 & 0 & 0 \\
 0 & 2 & 4 & 6 & 5 & 4 & 3 & 2 & 1 & 0 \\
\end{array}
\right)$ & 3,3,3 & 27 \\  \addlinespace[.2ex]
 $\left(
\begin{array}{cccccccccc}
 4 & 4 & 0 & 0 & 0 & 1 & 0 & 0 & 0 & 0 \\
 0 & 4 & 4 & 4 & 4 & 4 & 3 & 2 & 1 & 0 \\
\end{array}
\right)$ & 5,1,1,1,1 & 26 \\  \addlinespace[.2ex]
 $\left(
\begin{array}{cccccccccc}
 5 & 2 & 0 & 1 & 1 & 0 & 0 & 0 & 0 & 0 \\
 0 & 3 & 4 & 5 & 5 & 4 & 3 & 2 & 1 & 0 \\
\end{array}
\right)$ & 4,2,2,1 & 27 \\  \addlinespace[.2ex]
 $\left(
\begin{array}{cccccccccc}
 6 & 0 & 1 & 1 & 1 & 0 & 0 & 0 & 0 & 0 \\
 0 & 2 & 4 & 5 & 5 & 4 & 3 & 2 & 1 & 0 \\
\end{array}
\right)$ & 3,3,2,1 & 26 \\  \addlinespace[.2ex]
 $\left(
\begin{array}{cccccccccc}
 5 & 2 & 1 & 0 & 0 & 1 & 0 & 0 & 0 & 0 \\
 0 & 3 & 4 & 4 & 4 & 4 & 3 & 2 & 1 & 0 \\
\end{array}
\right)$ & 4,2,1,1,1 & 25 \\  \addlinespace[.2ex]
 $\left(
\begin{array}{cccccccccc}
 6 & 0 & 2 & 0 & 0 & 1 & 0 & 0 & 0 & 0 \\
 0 & 2 & 4 & 4 & 4 & 4 & 3 & 2 & 1 & 0 \\
\end{array}
\right)$ & 3,3,1,1,1 & 24 \\  \addlinespace[.2ex]
 $\left(
\begin{array}{cccccccccc}
 6 & 1 & 0 & 0 & 2 & 0 & 0 & 0 & 0 & 0 \\
 0 & 2 & 3 & 4 & 5 & 4 & 3 & 2 & 1 & 0 \\
\end{array}
\right)$ & 3,2,2,2 & 24 \\  \addlinespace[.2ex]
 $\left(
\begin{array}{cccccccccc}
 5 & 3 & 0 & 0 & 0 & 0 & 1 & 0 & 0 & 0 \\
 0 & 3 & 3 & 3 & 3 & 3 & 3 & 2 & 1 & 0 \\
\end{array}
\right)$ & 4,1,1,1,1,1 & 21 \\  \addlinespace[.2ex]
 $\left(
\begin{array}{cccccccccc}
 6 & 1 & 0 & 1 & 0 & 1 & 0 & 0 & 0 & 0 \\
 0 & 2 & 3 & 4 & 4 & 4 & 3 & 2 & 1 & 0 \\
\end{array}
\right)$ & 3,2,2,1,1 & 23 \\  \addlinespace[.2ex]
 $\left(
\begin{array}{cccccccccc}
 7 & 0 & 0 & 0 & 1 & 1 & 0 & 0 & 0 & 0 \\
 0 & 1 & 2 & 3 & 4 & 4 & 3 & 2 & 1 & 0 \\
\end{array}
\right)$ & 2,2,2,2,1 & 20 \\  \addlinespace[.2ex]
 $\left(
\begin{array}{cccccccccc}
 6 & 1 & 1 & 0 & 0 & 0 & 1 & 0 & 0 & 0 \\
 0 & 2 & 3 & 3 & 3 & 3 & 3 & 2 & 1 & 0 \\
\end{array}
\right)$ & 3,2,1,1,1,1 & 20 \\  \addlinespace[.2ex]
 $\left(
\begin{array}{cccccccccc}
 7 & 0 & 0 & 1 & 0 & 0 & 1 & 0 & 0 & 0 \\
 0 & 1 & 2 & 3 & 3 & 3 & 3 & 2 & 1 & 0 \\
\end{array}
\right)$ & 2,2,2,1,1,1 & 18 \\  \addlinespace[.2ex]
 $\left(
\begin{array}{cccccccccc}
 6 & 2 & 0 & 0 & 0 & 0 & 0 & 1 & 0 & 0 \\
 0 & 2 & 2 & 2 & 2 & 2 & 2 & 2 & 1 & 0 \\
\end{array}
\right)$ & 3,1,1,1,1,1,1 & 15 \\  \addlinespace[.2ex]
 $\left(
\begin{array}{cccccccccc}
 7 & 0 & 1 & 0 & 0 & 0 & 0 & 1 & 0 & 0 \\
 0 & 1 & 2 & 2 & 2 & 2 & 2 & 2 & 1 & 0 \\
\end{array}
\right)$ & 2,2,1,1,1,1,1 & 14 \\  \addlinespace[.2ex]
 $\left(
\begin{array}{cccccccccc}
 7 & 1 & 0 & 0 & 0 & 0 & 0 & 0 & 1 & 0 \\
 0 & 1 & 1 & 1 & 1 & 1 & 1 & 1 & 1 & 0 \\
\end{array}
\right)$ & 2,1,1,1,1,1,1,1 & 8 \\  \addlinespace[.2ex]
 $\left(
\begin{array}{cccccccccc}
 8 & 0 & 0 & 0 & 0 & 0 & 0 & 0 & 0 & 1 \\
 0 & 0 & 0 & 0 & 0 & 0 & 0 & 0 & 0 & 0 \\
\end{array}
\right)$ & 1,1,1,1,1,1,1,1,1 & 0 \\  \addlinespace[.2ex]
	\bottomrule
	\end{tabular}
	\end{subfigure}
	\caption{Results obtained applying the matrix formalism to $\mathfrak{sl}_9$.}
\end{table}

\clearpage

\pagebreak

\subsection{Quivers Obtained from the Matrices}

From the matrices $M(\lambda)$ that we just found, the Higgs brane configuration can be recovered. We can then obtain a quiver for a model with $\M_H=\Or_{(\lambda)}$ by performing a phase transition to the Coulomb brane configuration and reading the quiver. We can obtain the mirror quiver with $\M_C=\Or_{(\lambda)}$ by swapping the D5-branes with NS5-branes and vice-versa in the Higgs brane configuration obtained form $M(\lambda)$ and reading the quiver.\\

Alternatively, one can say that the matrix $M(\lambda)$ obtained in each step of the matrix formalism fixes the linking numbers $\vec{l}_s$ and $\vec{l}_d$ of a new model. This fully determines the quiver of the gauge theory. The mirror model is obtained by swapping $\vec{l}_s$ and $\vec{l}_d$. Consequently, each matrix $M(\lambda)$ fully characterizes two different quivers, one corresponding to a model with $\M_H=\Or_{(\lambda)}$, and the mirror, with $\M_C=\Or_{(\lambda)}$.\\

In the following tables we explicitly show the Higgs brane configuration corresponding to some of the matrices and the respective quivers.\\


\begin{table}[h]
	\centering
	\includegraphics{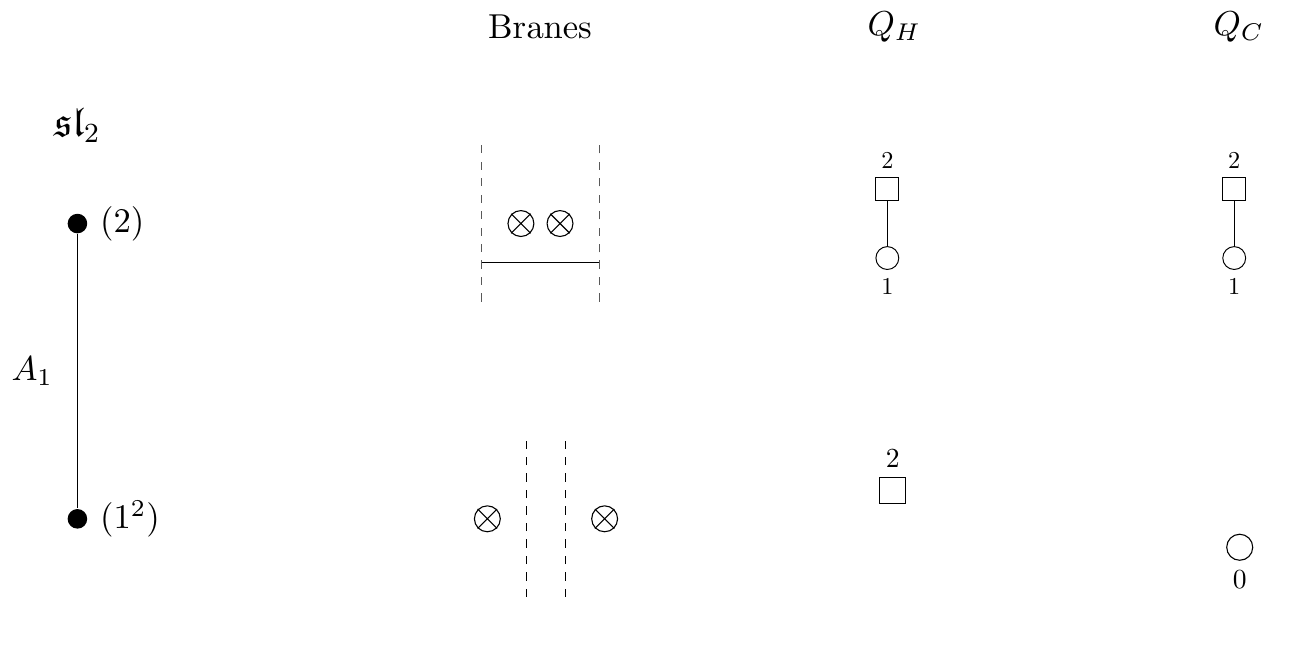}
	\caption{Hasse diagram with the partial order of all closures of nilpotent orbits of the algebra $\mathfrak{sl}_2$. The brane configurations can be obtained from the matrices that result from the matrix formalism computations. From each brane configuration we can obtain the quiver of the corresponding theory, labeled $Q_H$, for which the Higgs branch is the closure of the corresponding nilpotent orbit, and the quiver for the mirror model, denoted $Q_C$. For the mirrror model, the Coulomb branch is the closure of the nilpotent orbit. }
	\label{tab:HasseSU2branes}
\end{table}

\begin{table}[h]
	\centering
	\includegraphics{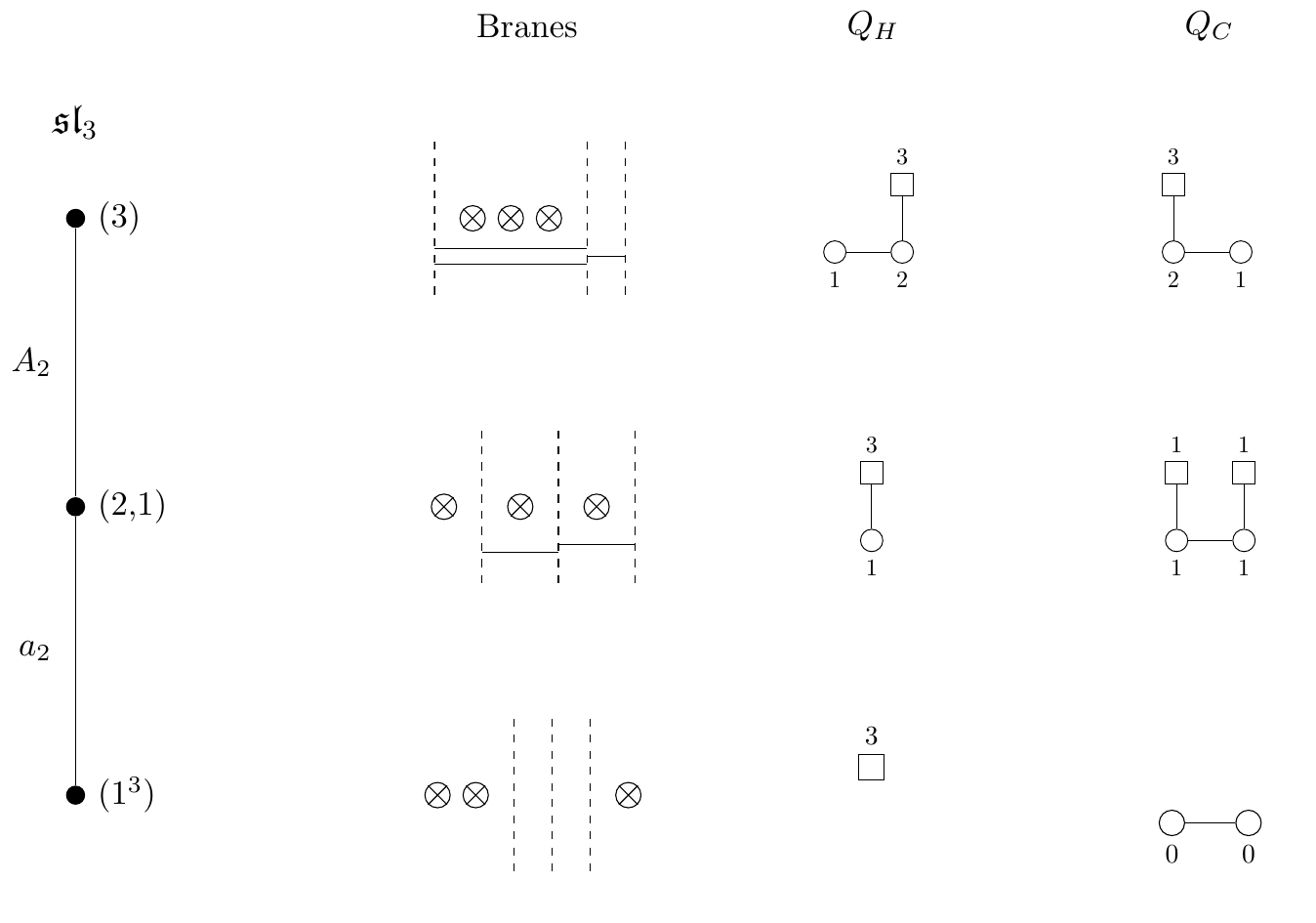}
	\caption{Hasse diagram with the partial order of all closures of nilpotent orbits of the algebra $\mathfrak{sl}_3$. The brane configurations can be obtained from the matrices that result from the matrix formalism computations. From each brane configuration we can obtain the quiver of the corresponding theory, labeled $Q_H$, for which the Higgs branch is the closure of the corresponding nilpotent orbit, and the quiver for the mirror model, denoted $Q_C$. For the mirrror model, the Coulomb branch is the closure of the nilpotent orbit. }
	\label{tab:HasseSU3branes}
\end{table}

\begin{table}[h]
	\centering
	\includegraphics{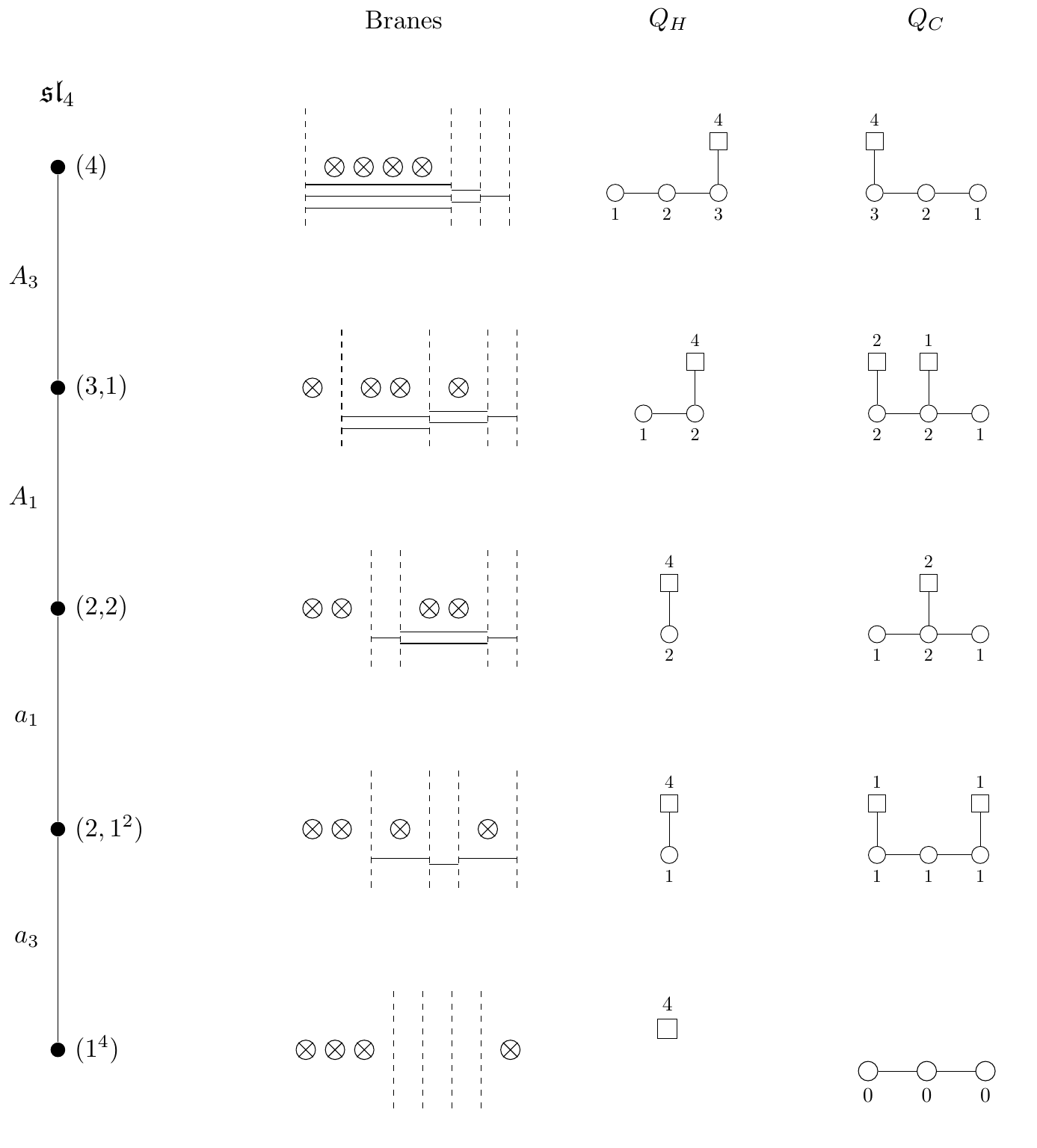}
	\caption{Hasse diagram with the partial order of all closures of nilpotent orbits of the algebra $\mathfrak{sl}_4$. The brane configurations can be obtained from the matrices that result from the matrix formalism computations. From each brane configuration we can obtain the quiver of the corresponding theory, labeled $Q_H$, for which the Higgs branch is the closure of the corresponding nilpotent orbit, and the quiver for the mirror model, denoted $Q_C$. For the mirrror model, the Coulomb branch is the closure of the nilpotent orbit. }
	\label{tab:HasseSU4branes}
\end{table}

\begin{table}[h]
	\centering
	\includegraphics{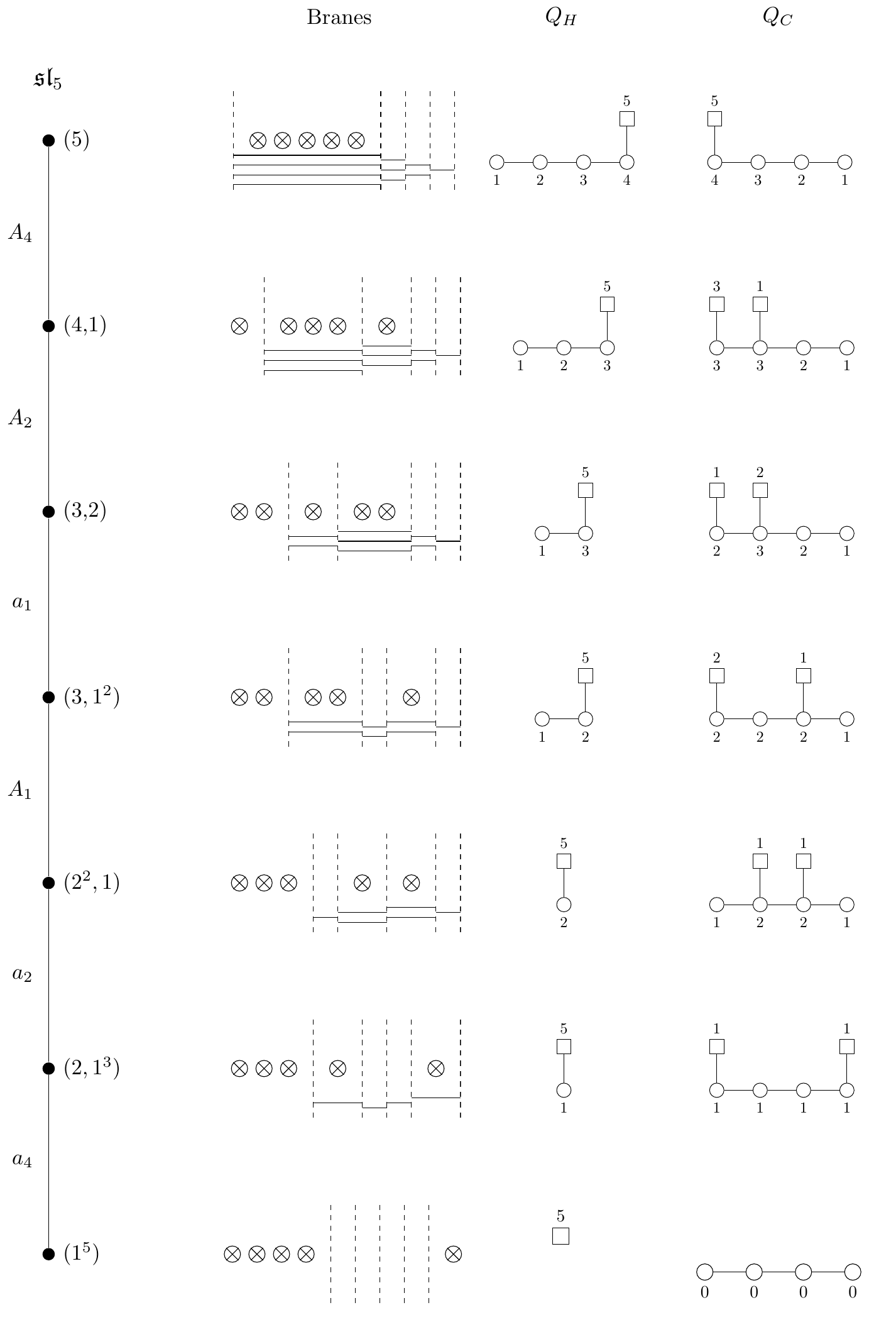}
	\caption{Hasse diagram with the partial order of all closures of nilpotent orbits of the algebra $\mathfrak{sl}_5$.}
	\label{tab:HasseSU5branes}
\end{table}

\pagebreak
\clearpage

\section{Conclusions and Outlook}\label{sec:C}

We want to recapitulate by emphasizing once more the extremely simple and yet powerful nature of the Kraft-Procesi transition. This is a physical process that has been developed during the study of moduli spaces that are closures of nilpotent orbits. However, it can now be applied to any generic model, as a way to systematically finding all minimal singularities in the moduli and establishing transitions to other models.\\

The next logical step in this research direction is to introduce O3-planes \cite{Feng} in the brane construction and apply the Kraft-Procesi transitions to models whose Higgs or Coulomb branch is the closure of a nilpotent orbit of the $\mathfrak{so}_n$ or the $\mathfrak{sp}_k$ algebra. We have already developed this approach and hope to be able to release a note on it soon. Many interesting mathematical features that are not present in nilpotent orbits of $\mathfrak{sl}_n$, like \emph{non-special} orbits or the \emph{collapse} of the partitions arise in this context.\\

Another natural application of the matrix formalism can be to Type IIB superstring brane configurations on a circle. These are very similar configurations to the ones we have seen here, with the difference that the spacial direction $x^6$ is considered to be a circle $S^1$. The computing algorithm can be straightforwardly modified to obtain a periodic pattern of KP transitions, starting for any given brane configuration with high enough number of D3-branes. We believe that the periodic Hasse diagrams that can be generated this way might be related to some notion of nilpotent orbits in affine Lie algebras\footnote{We want to thank Axel Kleinshmidt for discussions on nilpotent orbits of affine algebras that gave rise to these ideas.}.

\section*{Acknowledgments}

We are very grateful to Nipol Chaemjumrus, Stefano Cremonesi, Giulia Ferlito, Rudolph Kalveks, Axel Kleinshmidt, Noppadol Mekareeya, Claudio Procesi, Antonio Sciarappa, Edward Tasker and Alberto Zaffaroni for helpful conversations. S.C. is supported by an \mbox{EPSRC} DTP studentship. A. H. is supported by STFC Consolidated Grant ST/J0003533/1, and EPSRC Programme Grant EP/K034456/1.


\providecommand{\href}[2]{#2}\begingroup\raggedright\endgroup

\end{document}